\documentclass{aa}
\usepackage{graphicx}
\usepackage{txfonts}
\usepackage{longtable}
\usepackage{natbib}

\def\ltsima{$\; \buildrel < \over \sim \;$}
\def\simlt{\lower.5ex\hbox{\ltsima}}
\def\gtsima{$\; \buildrel > \over \sim \;$}
\def\simgt{\lower.5ex\hbox{\gtsima}}
\begin{document}
\title{An updated survey of globular clusters in M 31. III.}
\subtitle{
A spectroscopic metallicity scale for the Revised Bologna Catalog\thanks{Based 
on observations made at La Palma, at the Spanish Observatorio del
 Roque de los Muchachos of the IAC, with the William Herschel Telescope of the 
Isaac Newton Group and with the Italian Telescopio Nazionale Galileo (TNG) 
operated by the Fundaci\'on Galileo Galilei of INAF. 
Also based on observations made with the G.B.\ Cassini
Telescope at Loiano (Italy), operated by the Osservatorio Astronomico di 
Bologna (INAF).
}.}

   \author{S. Galleti\inst{1}, M. Bellazzini\inst{1},
           A. Buzzoni\inst{1}, L. Federici\inst{1}, 
          \and
       F. Fusi Pecci\inst{1}
}
         
      \offprints{S. Galleti}

   \institute{INAF - Osservatorio Astronomico di Bologna,
              Via Ranzani 1, 40127 Bologna, Italy\\
            \email{silvia.galleti@oabo.inaf.it, michele.bellazzini@oabo.inaf.it, 
          luciana.federici@oabo.inaf.it, \\
          alberto.buzzoni@oabo.inaf.it, flavio.fusipecci@oabo.inaf.it} }

     \authorrunning{S. Galleti et al.}
   \titlerunning{An updated survey of globular clusters in M 31. III.
   Metallicity.}

\date{Received 27 May 2009 / Accepted 17 September 2009}   

\abstract
% context
{}
% aims
{
We present a new homogeneous set of metallicity estimates 
based on Lick indices for the old globular clusters of the M31 galaxy. 
The final aim is to add homogeneous spectroscopic metallicities to as many 
entries as
possible of the Revised Bologna Catalog of M31 
clusters\thanks{RBC Version 4 available at: www.bo.astro.it/M31}, by 
reporting Lick
indices measurements from any source (literature, new observations, etc.) into
the same scale.}
%methods
{New empirical relations of [Fe/H] as a function of [MgFe] and Mg2 indices, as
defined by Trager et al. (1998), are based on well studied Galactic
Globular Clusters, complemented with theoretical model predictions for 
$-0.2\le [Fe/H]\le +0.5$. Lick indices for M31 clusters from various
literature sources (225 clusters) and from new observations by our team 
(71 clusters) have been transformed
into the Trager et al. (1998) system, yielding new metallicity 
estimates for  245 globular clusters of M31.  }
%results
{Our values are in good agreement with recent estimates based on detailed
spectral fitting and with those obtained from Color Magnitude Diagrams of
clusters imaged with the Hubble Space Telescope. The typical uncertainty on
individual estimates is  $\simeq \pm 0.25$ dex, as resulted from the comparison
with metallicities derived from Color Magnitude Diagrams of individual clusters.}
%conclusions
{ The metallicity distribution of M31 globular cluster is briefly discussed
and compared with that of the Milky Way. Simple parametric statistical tests
suggest that the distribution is likely not unimodal.  The strong correlation
between metallicity and kinematics found in previous studies is confirmed. The
most metal-rich GCs tend to be packed at the center of the system and to cluster
tightly around the galactic rotation curve defined by the HI disk, while the
velocity dispersion about the curve increases with decreasing metallicity.
However, also the clusters with $[Fe/H]<-1.0$ display a clear rotation pattern,
at odds with their Milky Way counterparts.}
%%%%%%%%%%%%%%%%%%%%%%%%%%%%%%%%%%%%%%%%%%%%%%%%%


\maketitle
%
%________________________________________________________________

\section{Introduction}

The concept of Simple Stellar Population (SSP) has proven to
be a very fruitful tool for the study of virtually 
any kind of stellar system \cite[][hereafter RFP88]{renbuz,renzfp}.
A SSP is completely characterized by only four ``parameters'': 
(a)  mass, (b)  chemical composition, (c)  age,
and (d) mass function, that determines the mass to light ratio (M/L)
of the SSP once fixed the age and the chemical composition
(see RFP88, for further possibly relevant variables that are not 
considered at zero-approximation). As a further simplification, the chemical composition is
typically represented with two main parameters, i.e. the Helium abundance (Y)
and the total abundance of elements heavier than He, usually parametrized by the
iron abundance [Fe/H] (see, for instance
\cite{zwscale} (ZW84),\cite{tc04mod,tc04alfa} and references therein). 
Even if the abundance of the so called $\alpha$-elements has been subject of an
increasing interest in the last two decades \citep{mcw,marmod,tc04alfa,grat}  
[Fe/H] remains the main parameter to rank stars and/or stellar populations
according to their abundance of heavy elements.     

In the study of globular clusters (GC), which are the best approximation 
of a SSP in nature, the metallicity is a key parameter that is also 
needed to infer ages and age differences 
\citep[see, for example RFP88 and][and references therein]{cg97}. 
The knowledge of the metallicity of a large
sample of globular clusters in a given galaxy allows one to search for
metallicity gradients, and the presence of distinct sub-populations of GCs
(as in the Milky Way (MW) \citet{Zinn}, and in many external galaxies 
\citet{bs06}), and, in general, to obtain
crucial information on the early phases of the formation and
chemical enrichment of the parent galaxy.

While modern instrumentation has allowed the determination of the
detailed abundance of several elements in single stars belonging to
GCs of the MW \cite[see][]{sneden,grat,euge}, the metallicities
of extragalactic GCs must be obtained from their integrated colors
and/or spectra, by comparison with Galactic templates and/or
theoretical models \citep{bs06}. 
Several broad-band integrated colors are fairly sensitive to metallicity and
relatively easy to measure for clusters out to very large distances.
However, they suffer from the well-known age-metallicity degeneracy (RFP88)
{\em and} they may be badly affected by the reddening due to extinction by
interstellar dust. While the foreground extinction associated with the dust
layers residing in our own Galaxy may be somehow constrained by observations and
modeled \citep{SFD}, the extinction intrinsic to external galaxies is largely
unknown. On the other hand, spectral indices based on the strength of an
absorption feature with respect to the surrounding continuum also
suffer from the age-metallicity degeneracy  
\citep[to different degrees, see][]{w94} but they are essentially unaffected by extinction 
\citep{MacArthur}, a very desirable characteristic.
The most widely used spectral indices were originally defined by \cite{bur}
and \cite{faber} at the Lick Observatory. These authors defined a set of indices 
that measure the
strength of the most pronounced absorption features that are seen in the
integrated low-resolution spectra of stellar systems  at optical wavelengths.
The use of Lick indices became widespread because they are easy to measure; as a
consequence, they were also included as standard predictions in all theoretical
models of SSP (see, for example, \cite{buz_mg2,buz94,w94,bruz} hereafter BC03, 
\citet{tc04mod,TMB}, hereafter TMB).

The M31 galaxy is an ideal target for studying GCs. It is nearby and it has a 
large cluster population ( $\sim$3 times larger than the MW).
The globular cluster system of M31 has been intensively studied in the past and
several authors have used Lick indices to constrain the age and 
the metallicity of clusters in the Andromeda Galaxy.

Integrated-light spectroscopy of M31 GCs was pioneered by \cite{syd69} who found
that the GC system of this galaxy extends to higher metallicities with respect
to the MW. In an important contribution, \cite{bur} comprehensively discussed
other interesting differences between GCs in M31 and in the MW. In particular
they showed that M31 clusters have significantly stronger H$\beta$ and CN
absorption indices at a given metallicity. In a series of papers
\cite{bh90,bh91} and \cite{huchra91} studied the metallicity distribution of M31 GCs using
an extensive sample of integrated spectra and line indices. They found that the
properties of the M31 GC system are broadly similar to the MW one, but they
confirmed the presence of a high-metallicity tail having no counterpart in our
Galaxy. They found that the mean metallicity  [Fe/H] was -1.2, and they
identified a weak metallicity  gradient as a function of projected radius.  From
the distribution of integrated colors \cite{bar00} found  evidences that the M31
GC system may  have a bimodal metallicity distribution 
\cite[like the Milky Way,][]{Zinn}, 
with peaks at [Fe/H]$\sim -1.4$  and [Fe/H]$\sim -0.6$.  Moreover,
they found that the $(V-K)_0$ color  distribution was best modeled assuming
three modes in the metallicity  distribution, instead of one or two.  Finally,
they found a small radial metallicity gradient and no  correlation between
cluster luminosity and metallicity  in M31 GCs. 
\cite{perr_cat}
produced a total sample of about 200  spectroscopic metallicities of M31 GCs,
calibrating Lick indices measured in their own system versus the metallicity of
M31 clusters in common with \cite{huchra91} 
(hence they used a set of {\em secondary} calibrators).
They confirmed the bimodality in the metallicity  distribution and  reported
that the metal--rich clusters have a higher rotation  amplitude with respect to
metal-poor ones, while both groups are known to  rotate faster than their
Galactic counterparts. Moreover, they found evidence for a radial metallicity 
gradient in the metal-poor population of M31 out to $\sim 60\arcmin$ from the 
galaxy center.

Metallicity (and age) estimates for various samples of
M31 clusters obtained by fitting spectra with theoretical SSP models 
have been
recently presented by  \cite{beasII} and \cite{puziam31}. 
\cite{beasII} studied a sample of 23 M31 GCs with very high Signal to Noise
(S/N) spectra,  seventeen of which were found to be old and to span a large range of
metallicity, while the remaining six were classified as intermediate  age 
clusters.
\cite{puziam31} presented the metallicity of 70  globular clusters
(including those studied by \cite{beasII}) finding a bimodal distribution with
peaks at [Z/H] $\sim -1.66$ ($\pm$0.05) and [Z/H] $\sim -0.45$  ($\pm$0.04) dex with dispersions 0.23 and
0.29 dex, respectively \footnote{[Z/H] is defined as 
[Fe/H] + 0.94[$\alpha$/Fe] taken 
from \citet[][see also Trager et al. 2000]{TMB}.}. 

More recently, \cite{LeeII} merged the metallicities from \cite{bar00} and
\cite{perr_cat}  with their own estimates from the  line indices measured from
WIYN/Hydra spectra.  They found that a bimodal and trimodal distribution are
statistically preferable to a unimodal metallicity distribution at a confidence
level of 99.8\%. \cite{Fan} assembled metallicities from the literature and with
estimates derived from integrated colors to obtain a global metallicity
distribution of M31 GCs. They found a bimodal distributions with  peaks at
[Fe/H]$\sim$-1.7 and $\sim$-0.7 dex with mean [Fe/H]=-1.21 dex, but  showed that
three-group fits are also statistically acceptable. They  found a metallicity
gradient as a function of projected radius for the metal-poor GCs, but no
gradient for  the metal-rich GCs.

The brief summary of modern studies above underlines the great degree of
heterogeneity of the available material. The various sets of estimates are
obtained from observables that are different in nature (integrated 
spectral indices or colors) and are based on different calibrations (empirical,
semi-empirical or theoretical; using primary or secondary calibrators). 
Even the actual definition of the same Lick indices varies from author to author;
thus, in general, the presented calibrations are valid only for a given
observational set-up and definition. In these circumstances it is clear that 
joining together different sets of metallicity estimates may be quite dangerous
as it may lead to a poor degree of self-consistency in the final merged set.

We have assembled and we continuously maintain and update a database of
parameters of confirmed and candidate clusters\footnote{We also keep lists of
targets previously suggested as candidate M31 globular clusters and later found
to be objects of different nature, like distant galaxies, foreground stars,
regions HII etc., see \cite{2mass,io_rv,io_est}.} in M31, the Revised Bologna Catalog
of M31 globular clusters \cite[RBC,][]{2mass,io_rv,io_est}. As we want to add a
reliable metallicity estimate to the confirmed clusters in the RBC, we need to 
devise the operational protocol to transform the actual measures provided in the
literature (as, for example, already available and future sets of Lick indices for M31
clusters) into a unique homogeneous metallicity scale.
In this paper, we describe the construction of this new homogeneous metallicity
scale for M31 GCs based on Lick line indices. 
Having set and tested
the new scale,  we present new metallicity estimates  for  245 M31 GCs.

The plan of the paper is as follows. In Section~\ref{scale}, we
describe the rationale and the procedure for the  construction of the new
metallicity scale.  In Section~\ref{data}, we report on the sample of M31 GC
spectra that we have obtained and reduced, and from which we have estimated Lick
indices. We also describe how we have collected Lick indices for M31 GCs from
the literature and how we have reported all of them to the same homogeneous
system. Finally, we derive the new values of the metallicities and, in  
Section~\ref{comp}, compare  our
scale with previous metallicity  estimates. In Section~\ref{analysis}, we 
present and discuss the metallicity distribution of the M31 GC system and on 
the correlations between metallicity and kinematics. 

\section{ An empirical metallicity scale \label{scale}}

The construction of a metallicity scale must be driven by a list of basic
requirements and a number of methodological/technical choices to achieve them,
as well as some trade-off between different possibilities. 
In particular, we identified the following ranked list of desiderata.

\begin{enumerate}

\item{} The scale must be consistent with at least one of the main 
metallicity scales currently used for the Galactic GCs, like 
\citet[][ZW84]{zwscale} or \cite{cg97}.

\item{} The scale must be calibrated on empirical templates. The agreement with
theoretical predictions is clearly desirable but it is not a {\em must}, as
theoretical models have problems and uncertainties on their own
\citep{tc04alfa,tc04mod}, while the chemical abundances of many Galactic GCs
are known in great detail \citep{euge}.

\item{} The observables that are used to settle the scale must be as sensitive
to the abundance of heavy elements as possible but also operationally 
well defined, and easy to measure out to large distances with
currently available instrumentation. 

\end{enumerate}

Several authors have studied and discussed in detail the sensitivity of the
various Lick indices to the abundance of various elements and to other
parameters 
\cite[see][and references therein]{gonz93,w94,wortsys,wo97,buz_mg2,buz95b,trager,marmod,TMB,tc04alfa,tc04mod}. 
Even if several indices are
fairly sensitive to metallicity, they are not necessarily suitable to serve as
the basis of a general purpose metallicity scale. As an extreme example, the
H$\beta$ index is very sensitive to metallicity but it is also very sensitive to
age, hence it would be a  misleading metallicity indicator, in general. Fe4648,
Fe5015, Fe5709, and Fe5782 have been indicated as very good metallicity
indicators, but none of these absorption features seems  ideal for reliable
metallicity determinations. Fe4648 was found to be sensitive to C, O, Mg, and
Si, hence it does not seem to trace any iron peak  element. Fe5015 is mostly
sensitive to iron, but it can  be affected by [$OIII$] emission. Fe5709 and
Fe5782 are weak features which require very high S/N spectra to be reliably
measured. There is general 
consensus that the
most reliable (and easy to measure) iron-sensitive Lick indices are Fe5270 and
Fe5335, as both measure predominantly strong iron lines. However all these
features are relatively weak in most old SSP spectra with respect to the Mg
features that are parametrized by the Mg2 and Mgb indices, both of which are
shown to correlate very well with [Fe/H] (\cite{w94,w96}).

As it become clear that old stellar populations (like GCs and classical
elliptical galaxies) are characterized by an enhancement of $\alpha$ elements 
(N, O, Mg, Ca, Na, Ne, S, Si, Ti), or, better, by an iron deficiency with respect
to the abundance pattern of the Sun, the impact of $[\alpha/Fe]$ on Lick indices
has been the subject of detailed study 
\cite[see][and references therein]{trager00,tc04alfa,marmod}.
 To reduce the influence 
of [${\alpha }$/Fe] variations on age and metallicity determinations, 
\cite{gonz93} introduced the [MgFe] index, 
$[MgFe]=\sqrt{Mgb \cdot <Fe>}$ with $<Fe>=(Fe5270+Fe5335)/2$, that appears to
be very sensitive to the total metallicity while minimizing the dependency
on [${\alpha }$/Fe] \citep[see][for discussion]{marmod,TMB}.

After many tests using several indices we decided to base our scale 
on four indices: three of them (Mgb, Fe5270, and Fe5335) are combined into 
[MgFe], the other is Mg2 (see Appendix \ref{correlazione}).
Mg2 has become a standard "metallicity" indicator for the 
integrated spectra (see i.e. \cite{buz_mg2}).
We found that [MgFe] and Mg2 provide the most consistent and strong 
correlations
with [Fe/H] in the ZW84 scale, once the 
\citet[][hereafter T98]{trager}
definitions of the Lick indices are adopted. 
Therefore, to obtain a metallicity estimate in a reliable and homogeneous scale,
the indices must be 
measured according to the T98 definitions and transformed into the T98 reference
frame
using a set of stars/stellar systems in common with T98 as standard calibrators
(see Sect.~3, below). 
Operationally, when the spectrum of a given cluster has a sufficient
signal-to-noise and wavelength coverage to allow a reliable measure of all the
involved indices, including Fe5270 and Fe5335, the estimates of [Fe/H] can be 
obtained from [MgFe]. A valid metallicity estimate can be
obtained even if reliable measures of Fe5270 and Fe5335 are lacking; the use of
both the assumed indicators is clearly preferable, but Mg2 alone is sufficient.

%%%%%%%%%%%%%%%%%%%%%%%%%%%%%%%%%%%%%%%%%%%%%%%%%%%%%%%%%%%###########
%------------------------FIG 1---S05-T98------------------------------
   \begin{figure}[]
   \centering
   \includegraphics[width=0.95\hsize, height=0.95\hsize,clip=]{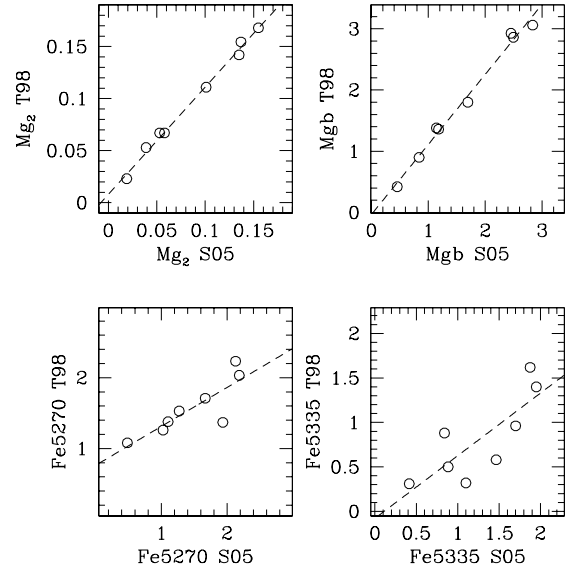}
     \caption{Comparison of passband measurements from S05 spectra and 
original Lick data for 8 Galactic Globular Clusters. 
The dashed is a least-square fit to the open circles. 
 }
        \label{lick_s05}
    \end{figure}
%
%%%%%%%%%%%%%%%%%%%%%%%%%%%%%%%%%%%%%%%%%%%%%%%%%%%%%%%%%%%%%%%%%
%-------------------------FIG 2----B04-T98/S05---------------------------
   \begin{figure}[]
   \centering
   \includegraphics[width=0.95\hsize, height=0.95\hsize,clip=]{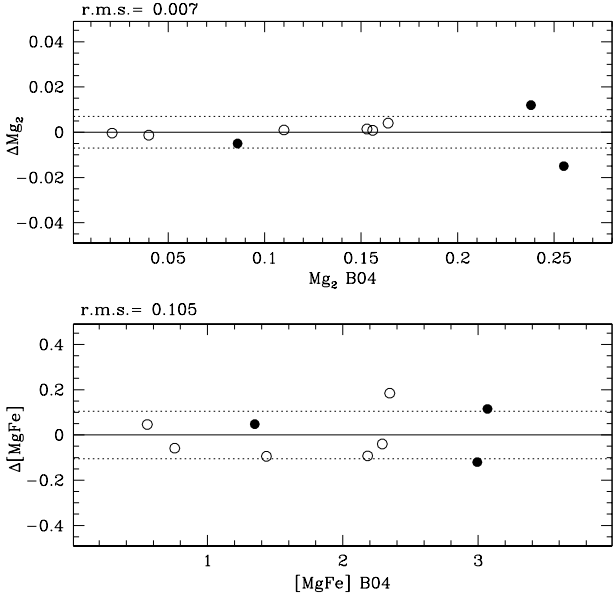}
     \caption{Comparison of the metallicity line indices adopted 
(open circles are T98, filled circles are S05 data) with the B04 values 
transformed to T98 system. 
The rms of the distributions are also reported.
 }
        \label{SA_B04}
    \end{figure}
%
%%%%%%%%%%%%%%%%%%%%%%%%%%%%%%%%%%%%%%%%%%%%%%%%%%%%%%%%%%%%%%%%%

\subsection{The calibrators}

To fulfill simultaneously the requirements \# 1 and \# 2 above, we decided to
adopt Galactic Globular clusters with metallicities in the ZW84 
scale as fundamental calibrators. All the metallicity values for MW GCs
adopted here are taken from ZW84 and \cite{AZ88}. It is clear 
 that the choice of the calibrators implies that the scale is valid
only for old populations having a similar degree of enhancement in the  abundance of
$\alpha$-elements with respect to the Sun (see \cite{alfaMW}).  We will use
theoretical SSP models to explore the effective range of ages and chemical
compositions in which our calibration can be considered valid. 

\begin{table}
  \begin{center}
  \caption{Linear fit coefficients for transformations of the Schiavon data
to the Lick system 
}\label{lick_sch}
  \begin{tabular}{|l|r|r|r|}
    \hline
    Index & a& b &r.m.s. \\
    \hline
\hline
Mg2     &  0.008  & 1.029  & 0.004 \\
Mgb	& -0.034  & 1.145  & 0.098 \\
Fe5270  &  0.761  & 0.551  & 0.199 \\ 
Fe5335  & -0.079  & 0.704  & 0.275 \\\hline
\end{tabular} \end{center}
 \end{table}

\begin{table}
  \begin{center}
  \caption{Linear fit coefficients for transformations of the Beasley data
to the Lick system 
}\label{lick_beasley}
  \begin{tabular}{|l|r|r|r|}
    \hline
    Index & a& b &r.m.s. \\
    \hline
\hline
Mg2     & -0.008  & 1.002 & 0.006 \\ 
Mgb	& -0.051  & 1.076 & 0.187\\
Fe5270  &  0.377  & 0.757 & 0.217 \\
Fe5335  & -0.009  & 0.872 & 0.205\\ \hline
\end{tabular} \end{center}
  \end{table}

We searched in the literature for assemble the
largest possible sample of Galactic GCs with well known metallicity and having
well measured Lick indices from high S/N spectra in the T98 system or that can
be easily transformed into this system.
First of all we took the data for 17 Galactic GCs provided by T98 themselves,
by definition in the T98 system.
The original spectra were obtained by \cite{bur} with the image 
dissector scanner (IDS) at the 3m Shane Telescope of the Lick Observatory, and 
the absorption-line indices were re-measured by T98.
Next, we incorporated the new data for 41 GCs 
from spectra obtained with the Blanco 4m telescope by 
\citet[][hereafter S05]{schiavon}\footnote{see http://www.noao.edu/ggclib.}.
We have measured the needed indices (Mg2, Mgb, Fe5270 and Fe5335)
from the Schiavon et al.'s spectra
as described in detail in  Section \ref{callick}, below. We used the 8
clusters in common with the T98 to derive a simple
least square fit converting indices from the S05 to the T98 system 
(using OLS(X$\mid$Y), according to \cite{OLS}; see Fig. \ref{lick_s05}). 
The derived values of the ($a, b$) coefficients for the
various indices are reported in Table\ref{lick_sch}.

%###########################-FIG 3 ####FIT MW##########################
%-----------------------------------------------------------
   \begin{figure}[!t]
   \centering
   \includegraphics[width=\hsize]{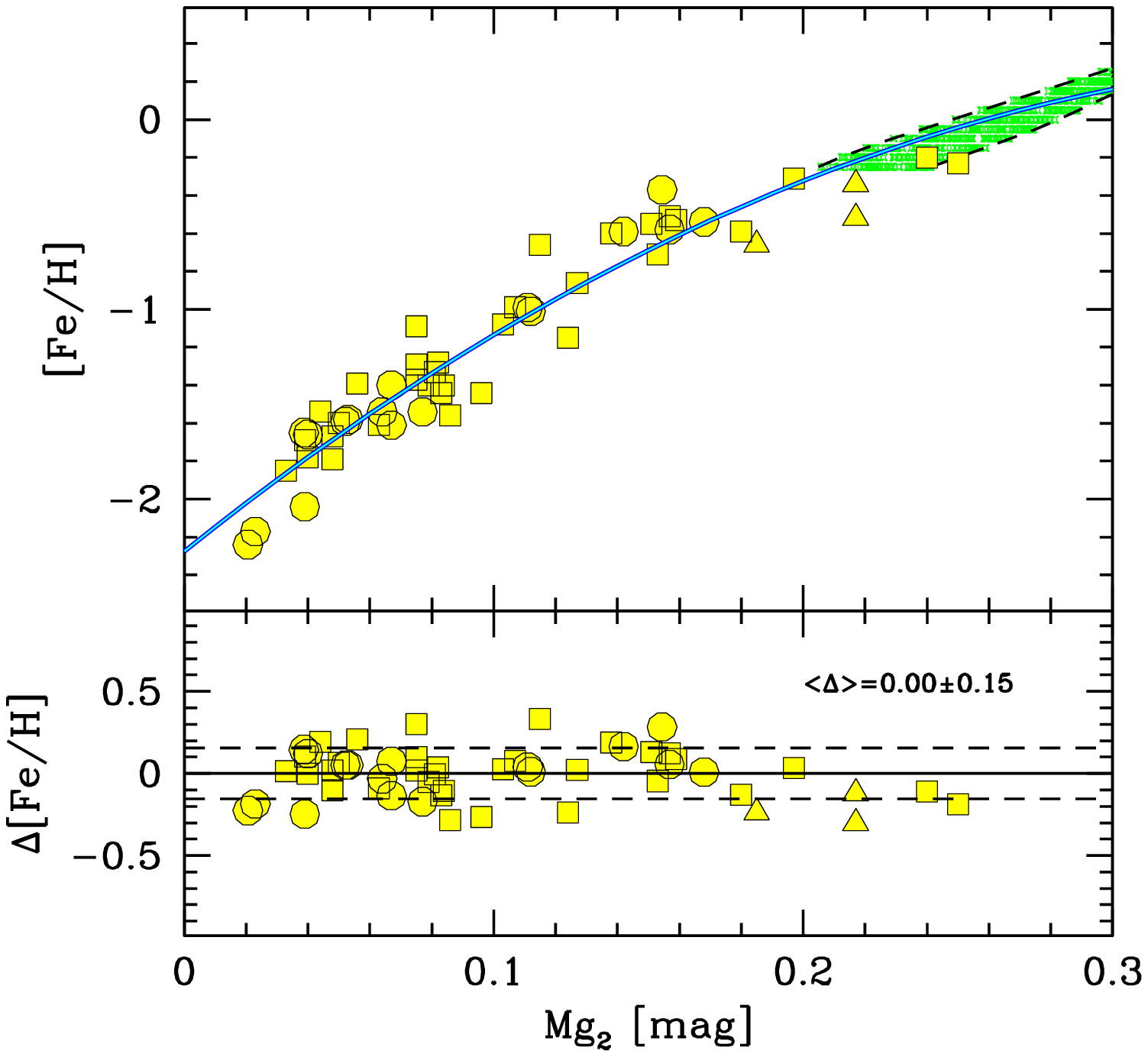}
   \includegraphics[width=\hsize]{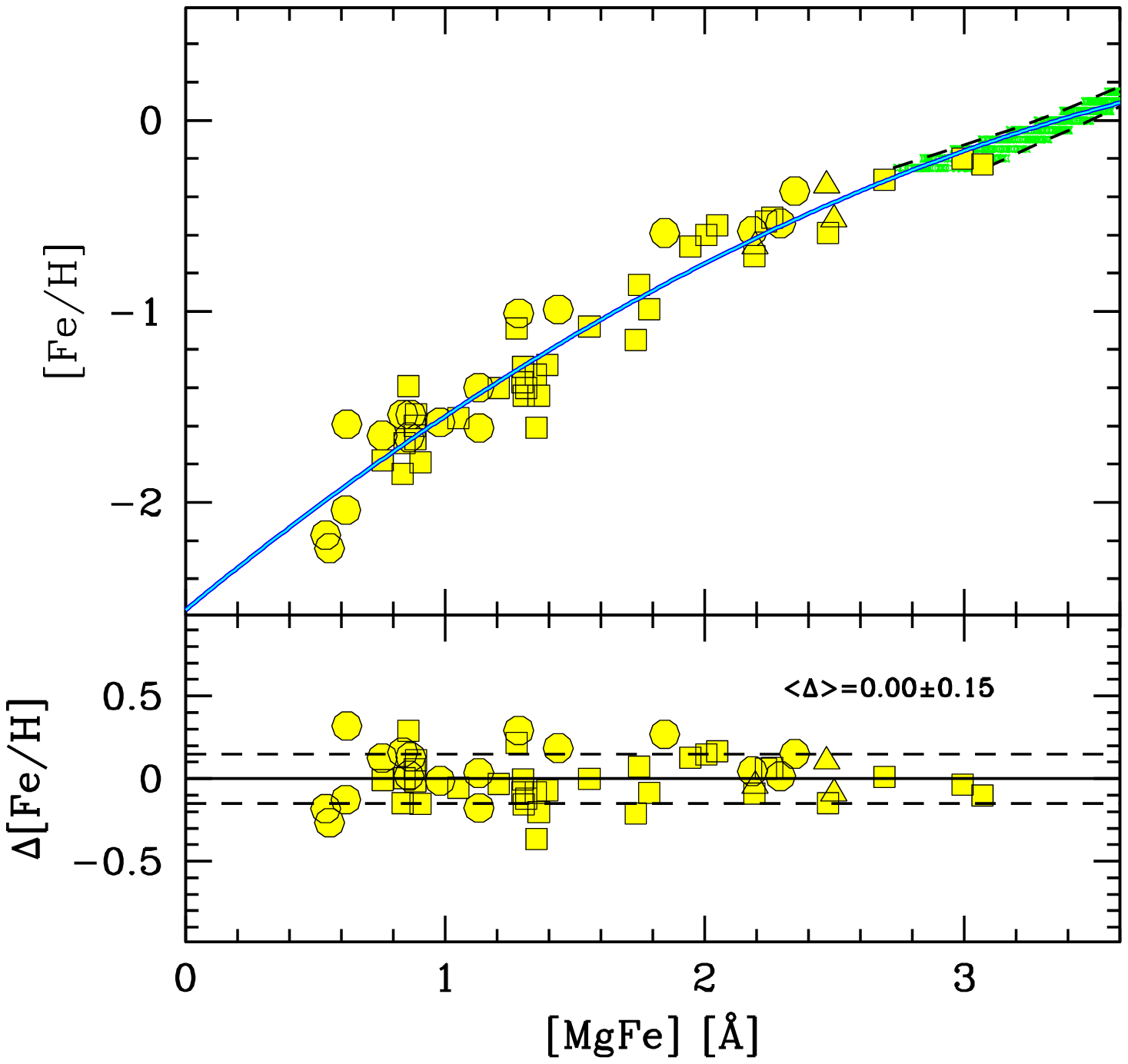}
     \caption{The adopted calibrations for Mg2 and [MgFe] vs. [Fe/H] 
(continuum lines) are superposed to the accounted set of Galactic globular
clusters from T98 (dots), S05 (squares), and B04 (triangles). The shaded
regions at super-solar metallicity are the envelope of the theoretical
predictions from several SSP models, as discussed in the text.
A comparison between the empirical metallicities and ZW84 with the mean 
differences and rms are also reported in the lower panels. The dashed 
lines enclose the rms. The color figures are available in the electronic
edition.}
    \label{fitmw}
    \end{figure}
%
%%%%%%%%%%%%%%%%%%%%%%%%%%%%%%%%%%%%%%%%%%%%%%%%%%%%%%%%%%%%%%%%%%%%%%%%%

Indices for further 3 GCs were taken from  \cite{beas} (hereafter B04)
who re-measured Lick indices for 12 GCs spectra obtained by
\cite{CohenGC}.  The coefficients given in Table \ref{lick_beasley},
derived by least-square fitting for clusters in common, as above, 
were used to convert  the B04 indices into the T98 system.  Fig.
\ref{SA_B04} compares the values of the indices from the sample
obtained by joining the data from T98 and S05 with the B04 sample,
after transformation to the T98 system. The plots show that the set of
measures we have assembled is {\em very} homogeneous: any residual
systematic and/or random scatter is of order of $\sim$ 5\% of the
range spanned by the index,  or smaller. 

\begin{table*}[!t]
        \begin{center}
  \caption{Lick indices for MW globular clusters. Indices were
computed with the passband definitions of \cite{trager} and were shifted in
this system.\label{mw_index}}
{\begin{footnotesize}
\begin{tabular}{lrrrrrrrrrr}
\hline\hline
Cluster &Mg$_2$ & eMg$_2$&Mg$b$ & eMg$b$ & Fe5270& eFe5270&Fe5335& eFe5335&  Source$^a$\\
        & mag & mag  &  \AA  &  \AA  &  \AA  &  \AA  &   \AA   &  \AA  &    & \\ 
\hline
NGC 104  & 0.153 & 0.007 & 3.022 & 0.013 & 1.925 & 0.015 & 1.256 & 0.017 & 2 \\
NGC 1851 & 0.075 & 0.009 & 1.396 & 0.024 & 1.606 & 0.026 & 0.820 & 0.030 & 2 \\
NGC 1904 & 0.039 & 0.009 & 0.788 & 0.035 & 1.297 & 0.039 & 0.516 & 0.046 & 2 \\
NGC 2298 & 0.033 & 0.012 & 0.907 & 0.063 & 1.154 & 0.071 & 0.391 & 0.081 & 2 \\
NGC 2808 & 0.075 & 0.007 & 1.374 & 0.015 & 1.613 & 0.017 & 0.838 & 0.019 & 2 \\
NGC 3201 & 0.063 & 0.009 & 1.836 & 0.032 & 1.407 & 0.036 & 0.583 & 0.041 & 2 \\
NGC 5024 & 0.039 & 0.010 & 0.830 & 0.295 & 0.370 & 0.268 & 0.550 & 0.281 & 1 \\
NGC 5272 & 0.040 & 0.008 & 1.010 & 0.235 & 1.060 & 0.220 & 0.410 & 0.213 & 1 \\
NGC 5286 & 0.048 & 0.009 & 0.903 & 0.023 & 1.287 & 0.026 & 0.526 & 0.030 & 2 \\
NGC 5904 & 0.067 & 0.010 & 1.380 & 0.300 & 1.530 & 0.282 & 0.320 & 0.257 & 1 \\
NGC 5927 & 0.197 & 0.009 & 3.885 & 0.035 & 2.183 & 0.039 & 1.549 & 0.044 & 2 \\
NGC 5946 & 0.056 & 0.009 & 0.844 & 0.058 & 1.337 & 0.063 & 0.410 & 0.072 & 2 \\
NGC 5986 & 0.048 & 0.009 & 0.900 & 0.026 & 1.278 & 0.029 & 0.465 & 0.034 & 2 \\
NGC 6121 & 0.081 & 0.009 & 1.677 & 0.023 & 1.467 & 0.026 & 0.705 & 0.030 & 2 \\
NGC 6171 & 0.111 & 0.015 & 1.800 & 0.447 & 1.710 & 0.420 & 0.580 & 0.427 & 1 \\
NGC 6205 & 0.039 & 0.006 & 0.725 & 0.140 & 0.976 & 0.131 & 0.609 & 0.136 & 1 \\
NGC 6218 & 0.067 & 0.011 & 1.360 & 0.338 & 1.380 & 0.317 & 0.500 & 0.316 & 1 \\
NGC 6235 & 0.079 & 0.012 & 1.188 & 0.068 & 1.599 & 0.074 & 0.855 & 0.084 & 2 \\
NGC 6254 & 0.050 & 0.009 & 0.861 & 0.023 & 1.315 & 0.026 & 0.495 & 0.029 & 2 \\
NGC 6266 & 0.082 & 0.007 & 1.638 & 0.015 & 1.555 & 0.017 & 0.821 & 0.019 & 2 \\
NGC 6284 & 0.084 & 0.009 & 1.562 & 0.037 & 1.475 & 0.041 & 0.730 & 0.047 & 2 \\
NGC 6229 & 0.077 & 0.013 & 1.100 & 0.398 & 0.980 & 0.372 & 0.390 & 0.356 & 1 \\
NGC 6304 & 0.180 & 0.009 & 3.545 & 0.035 & 2.056 & 0.039 & 1.399 & 0.044 & 2 \\
NGC 6316 & 0.151 & 0.009 & 2.896 & 0.045 & 1.864 & 0.049 & 1.033 & 0.055 & 2 \\
NGC 6333 & 0.040 & 0.009 & 0.748 & 0.028 & 1.213 & 0.032 & 0.336 & 0.036 & 2 \\
NGC 6342 & 0.115 & 0.012 & 2.737 & 0.081 & 1.663 & 0.091 & 1.099 & 0.102 & 2 \\
NGC 6352 & 0.157 & 0.009 & 3.198 & 0.044 & 1.867 & 0.049 & 1.329 & 0.056 & 2 \\
NGC 6341 & 0.021 & 0.005 & 0.800 & 0.176 & 0.448 & 0.161 & 0.323 & 0.151 & 1 \\
NGC 6356 & 0.168 & 0.009 & 3.060 & 0.251 & 2.034 & 0.236 & 1.400 & 0.263 & 1 \\
NGC 6362 & 0.103 & 0.009 & 2.099 & 0.034 & 1.567 & 0.038 & 0.741 & 0.044 & 2 \\
NGC 6388 & 0.138 & 0.007 & 2.393 & 0.017 & 2.029 & 0.018 & 1.335 & 0.021 & 2 \\
NGC 6440 & 0.217 & 0.010 & 3.443 & 0.080 & 2.131 & 0.130 & 1.412 & 0.130 & 3 \\
NGC 6441 & 0.159 & 0.009 & 2.921 & 0.020 & 2.043 & 0.022 & 1.383 & 0.025 & 2 \\
NGC 6522 & 0.096 & 0.009 & 1.543 & 0.032 & 1.597 & 0.035 & 0.803 & 0.039 & 2 \\
NGC 6528 & 0.250 & 0.009 & 4.377 & 0.042 & 2.421 & 0.046 & 1.884 & 0.051 & 2 \\
NGC 6539 & 0.185 & 0.010 & 2.970 & 0.100 & 1.995 & 0.150 & 1.246 & 0.150 & 3 \\
NGC 6544 & 0.086 & 0.009 & 0.950 & 0.046 & 1.564 & 0.049 & 0.764 & 0.055 & 2 \\
NGC 6553 & 0.240 & 0.009 & 4.300 & 0.042 & 2.418 & 0.045 & 1.751 & 0.051 & 2 \\
NGC 6569 & 0.127 & 0.009 & 2.329 & 0.049 & 1.654 & 0.054 & 0.967 & 0.061 & 2 \\
NGC 6624 & 0.154 & 0.008 & 2.860 & 0.240 & 2.233 & 0.226 & 1.618 & 0.253 & 1 \\
NGC 6626 & 0.083 & 0.009 & 1.554 & 0.025 & 1.445 & 0.027 & 0.743 & 0.031 & 2 \\
NGC 6637 & 0.142 & 0.010 & 2.926 & 0.288 & 1.369 & 0.269 & 0.961 & 0.292 & 1 \\
NGC 6638 & 0.124 & 0.009 & 2.184 & 0.035 & 1.736 & 0.038 & 1.021 & 0.043 & 2 \\
NGC 6652 & 0.107 & 0.009 & 2.381 & 0.030 & 1.722 & 0.034 & 0.957 & 0.039 & 2 \\
NGC 6712 & 0.112 & 0.015 & 1.570 & 0.431 & 1.410 & 0.405 & 0.690 & 0.423 & 1 \\
NGC 6723 & 0.075 & 0.009 & 1.554 & 0.031 & 1.462 & 0.035 & 0.636 & 0.040 & 2 \\
NGC 6752 & 0.044 & 0.009 & 0.924 & 0.025 & 1.264 & 0.029 & 0.445 & 0.033 & 2 \\ 
NGC 6760 & 0.217 & 0.010 & 3.496 & 0.080 & 2.116 & 0.140 & 1.455 & 0.130 & 3 \\
NGC 6838 & 0.157 & 0.010 & 2.628 & 0.295 & 1.774 & 0.277 & 1.855 & 0.314 & 1 \\
NGC 6981 & 0.064 & 0.014 & 0.810 & 0.421 & 1.280 & 0.397 & 0.450 & 0.389 & 1 \\
NGC 7006 & 0.052 & 0.012 & 0.665 & 0.352 & 0.529 & 0.326 & 0.633 & 0.344 & 1 \\
NGC 7078 & 0.023 & 0.007 & 0.420 & 0.211 & 1.080 & 0.201 & 0.310 & 0.182 & 1 \\
NGC 7089 & 0.053 & 0.008 & 0.900 & 0.242 & 1.260 & 0.228 & 0.880 & 0.245 & 1 \\
\hline
\hline					   
\end{tabular}
\end{footnotesize}
}
{	
\begin{flushleft}
$^a$ Dataset label: 1 -- \cite{trager}, 2 -- \cite{schiavon}, 3 -- \cite{beas}.
\end{flushleft}
}
\end{center}
\end{table*}
%%%%%%%%%%%%%%%%%%%%%%%%%%%%%%%%%%%%%%%%%%%%%%%%%%%%%%%%%%%%%%%%%

We merged all the sources described above into a global sample comprising
53 Galactic GCs with metallicities $-2.24 \leq [Fe/H] \leq -0.23$ in the 
ZW84 scale. In case of multiple measures we adopted one single
source according to the following ranking: T98, S05 and B04.
Lick indices, source and uncertainties for all sample adopted are given in 
Table \ref{mw_index}. 

 As recalled in Sect.~1, it is known since long time that M31, as well as
other giant galaxies \cite[see][]{harris5128}, hosts GCs that are significantly
more metal rich than found in the Milky Way, possibly up to super-solar
metallicities. To extend the range of applicability of our metallicity scale at
the super solar regime - where we lack observed calibrators - we complemented
our  sample with a suitable set of several models for old SSPs (i.e. 12-12.5
Gyr,  see \cite{grattonetamwgc}), with metallicity in the range $-0.3 \leq [Fe/H]
\leq +0.5$. In particular, the theoretical predictions by \citet{buz89},
\citet{w94},  \citet{bruz}, \citet{marmod}, and \citet{tc04mod} have been
considered. The simultaneous use of models from different authors provided a
confident estimate of the  internal uncertainty  of the theoretical framework,
intrinsic to the different input physics among the various theoretical synthesis
codes.  Therefore, to fit our calibrating relations we adopted a composite sample
made by the empirical set of 53 Galactic GC in the range  $-2.5\leq [Fe/H]
<-0.2$, plus the theoretically predicted index values  described above,
considered as observed points, in the range   $-0.2 \leq [Fe/H] \leq +0.50$.
While the agreement between the observed points and the models predictions is
quite good over the whole metallicity range covered by Galactic GCs (see Fig.~5,
below), the reader must be aware that the metallicity scale proposed here is not
constrained by empirical calibrators in the solar and super-solar regime.  

To avoid confusion resulting by plotting many different symbols in such a
restricted range of metallicities, in Fig.~3 we simply plot a sketch of the
envelope  enclosing all the theoretical points that were considered in the
calibration.

The two best-fit relations are shown in Fig.~\ref{fitmw}; they are the following
second order polynomials represented by:

\begin{equation}
\label{MgFe-Fe}
[Fe/H]_{[MgFe]} = -2.563 + 1.119[MgFe] - 0.106[MgFe]^{2}
\end{equation}
$\pm$ 0.15~dex, r.m.s.\\
\begin{equation}
\label{Mg2-Fe}
[Fe/H]_{Mg_2} = -2.276 + 13.053Mg_2 - 16.462Mg_2^2
\end{equation}
$\pm$ 0.15~dex, r.m.s\\

Equations \ref{MgFe-Fe} and \ref{Mg2-Fe} are the fundamental calibrating
relations of the proposed metallicity scale. When all the needed observables 
are available, the final metallicity value is obtained from Eq.~1; otherwise
one shall recur to Eq.~2.

The internal consistency of the adopted
scale is verified in Fig.~4, where the original metallicity 
values of the calibrating clusters are compared with those obtained 
with our calibrations. The r.m.s. scatter is $\simeq 0.15$ dex.
Fig.~3 and Fig.~4 suggest that our scale is less sensitive to metallicity
variations and more uncertain at the metal-poor end, for $[Fe/H]\la -1.9$.  
This is a general characteristic of scales based on integrated Lick indices 
\cite[see, for example][]{faber,cohen03} and must be taken into account when
very metal  poor clusters are considered.  

%--------------------------fig 4 comp-MW----------------------------- 
   \begin{figure}[!t]
   \centering
   \includegraphics[width=0.95\hsize, height=0.95\hsize,clip=]{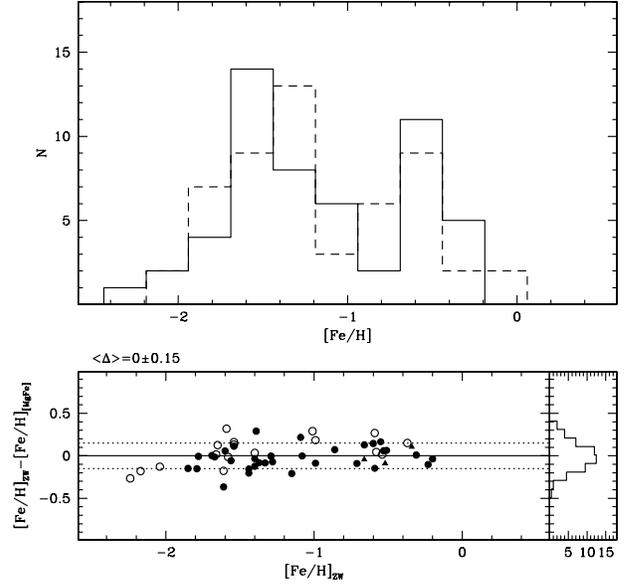}
     \caption{$\textit{ Upper Panel}$: Comparison of the input MW GC 
metallicities from ZW84 and \cite{AZ88} 
(continuous line) with those
obtained from the empirical metallicity calibration (dashed line).
$\textit{Lower Panels}$: Comparison of the metallicity observed 
for Galactic globular clusters with the estimates from the [MgFe] index.
Open circles are Galactic globulars from T98, filled  circle from S05 and
 triangles from B04. The dashed lines enclose the rms. 
                        }
        \label{dfitmw}
    \end{figure}
%%%%%%%%%%%%%%%%%%%%%%%%%%%%%%%%%%%%%%%%%%%%%%%%%%%%%%%%%%%%%%%%%%%%%%%%%

%%%%%%%%%%%%%%%%%%%%%%%%%%%%%%%fig 5 age %%%%%%%%%%%%%%%%%%%%%%%%%%%%%%%%%%
%----------------------------------------------------------- 
   \begin{figure}[!ht]
   \centering
  \includegraphics[width=0.85\hsize]{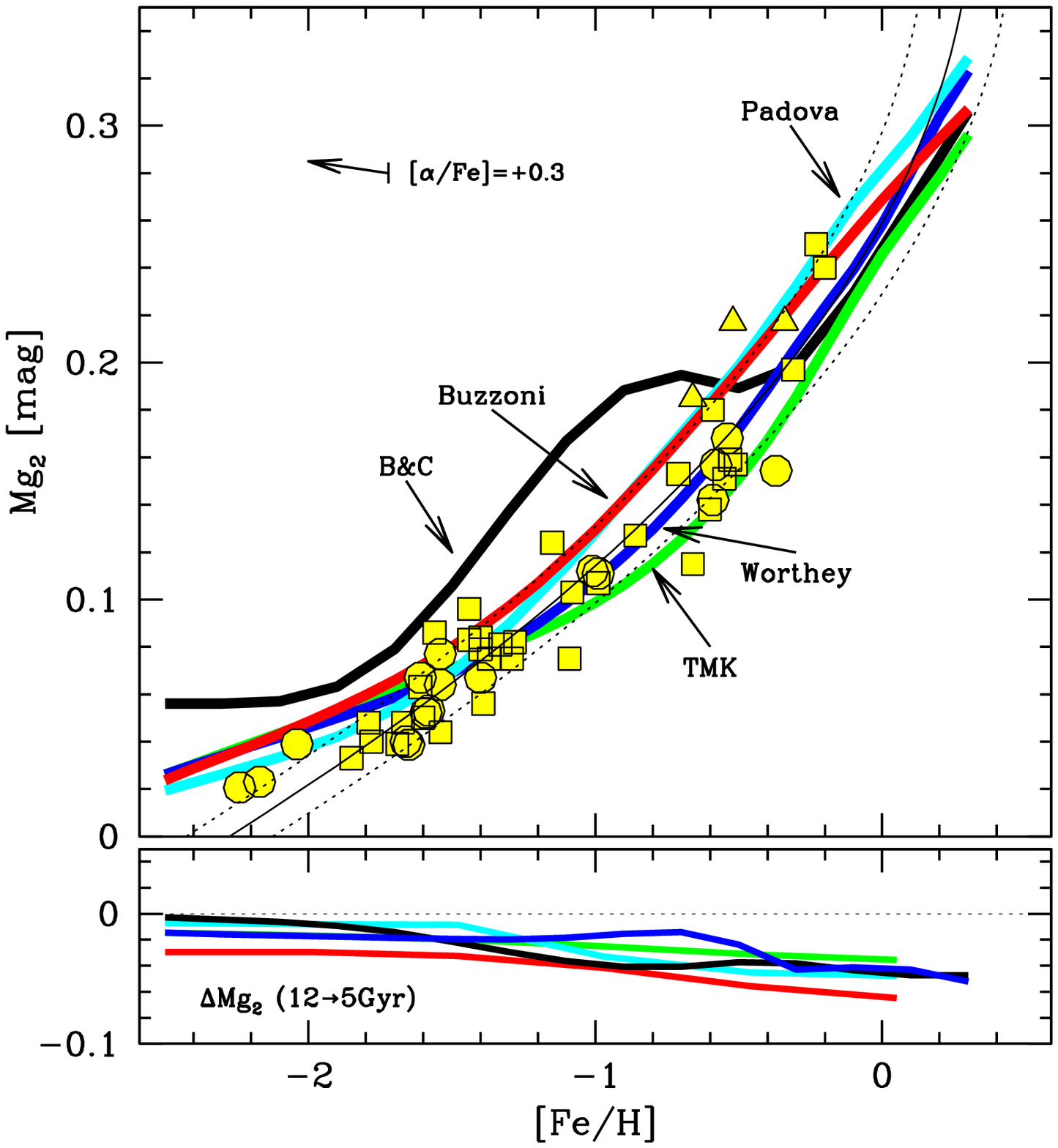}
     \caption{{\it (uppper panel)} - Mg2 index distribution of the MW GCs (symbols are 
the same as Fig. \ref{fitmw}) compared to the stellar population models of 
\cite{buz89,buz95a}, \cite{wortsys}, \cite{Girardi00} (labelled as ``Padova''
on the plot), \citet{bruz} (``B\&C''), \citet{tc04mod}  (``TMK''). 
An age of 12 Gyr is assumed throughout, with solar [$\alpha$/Fe]
element partition. The thin line indicates the empirical calibration 
with its 1$\sigma$ uncertainty. The effect of $\alpha$-element
enhancement is sketched by the arrow, for a change [$\alpha$/Fe] = +0.3.
In the {\it lower panel} we assess on the contrary the effect of a change in
age. For the same theoretical models we report the expected index variation 
($\Delta$~Mg$_2$) for a 5~Gyr stellar population along the full metallicity
range. The color figure is
 available in the electronic edition. }
        \label{ftmod}
    \end{figure}
%%%%%%%%%%%%%%%%%%%%%%%%%%%%%%%%%%%%%%%%%%%%%%%%%%%%%%%%%%%%%%%%%%%%%%%%%

The effects of age assumptions are explored in Fig.\ \ref{ftmod}, where
we compare our Mg$_2$ data with an illustrative set of theoretical models from several
population synthesis codes. In particular, we relied on the models
by \cite{buz89,buz95a}, \cite{wortsys}, \cite{Girardi00}, 
\citet{bruz}, \citet{tc04mod}. The upper panel of the
figure displays a collection of the 12 Gyr model predictions, while
the expected shift in the theoretical loci when moving to ages younger
 is estimated in the lower panel. One can see that
any change in age, say from 12 to 5~Gyr, reflects in a shallower slope
of the theoretical [Fe/H] vs.\ Mg$_2$ calibration, as a consequence of
a larger offset ($\Delta$~Mg$_2 \sim -0.05$~mag) at solar metallicity.
On the other hand, any enhancement in the [$\alpha$/Fe] element partition
results in a (roughly) solid shift of the curve shelf to correspondingly
lower values of [Fe/H], as sketched on the plot.

The comparison between our empirical calibrating relations and the model 
predictions reveal that the application of
our method to clusters as young as 5 Gyr, in the metallicity range $-2.0< [Fe/H]
<0.0$, would lead to systematic errors in the estimated metallicity as small as
$\le \pm 0.2$ dex, i.e. of the same order of the typical statistical
uncertainty. 
In any case, a good safety criterion would be to limit the
application to clusters older than 7-8 Gyr.

\section{The sample of M31 globular clusters \label{data}}

The M31 GCs Lick indices used in this study are taken from our
observations and from several sources available in literature.
In this section we describe the various data we
adopted and how we transformed the different sets of measures into the T98
system. 
In the following analysis we will consider only objects classified in the
RBC as {\em genuine old M31 clusters}, i.e. having {\em classification flag} f=1.
The possibility of contamination of the sample by spurious sources is discussed
in Sect.~3.4, below.

%%%%%%%%%%%%%%%%%%%%%%%%%%%%%%%fig 6 my- T98 %%%%%%%%%%%%%%%%%%%%%%%%%%%%%%%%5
   \begin{figure}[!ht]
   \centering
   \includegraphics[width=0.95\hsize, height=0.82\hsize,clip=]{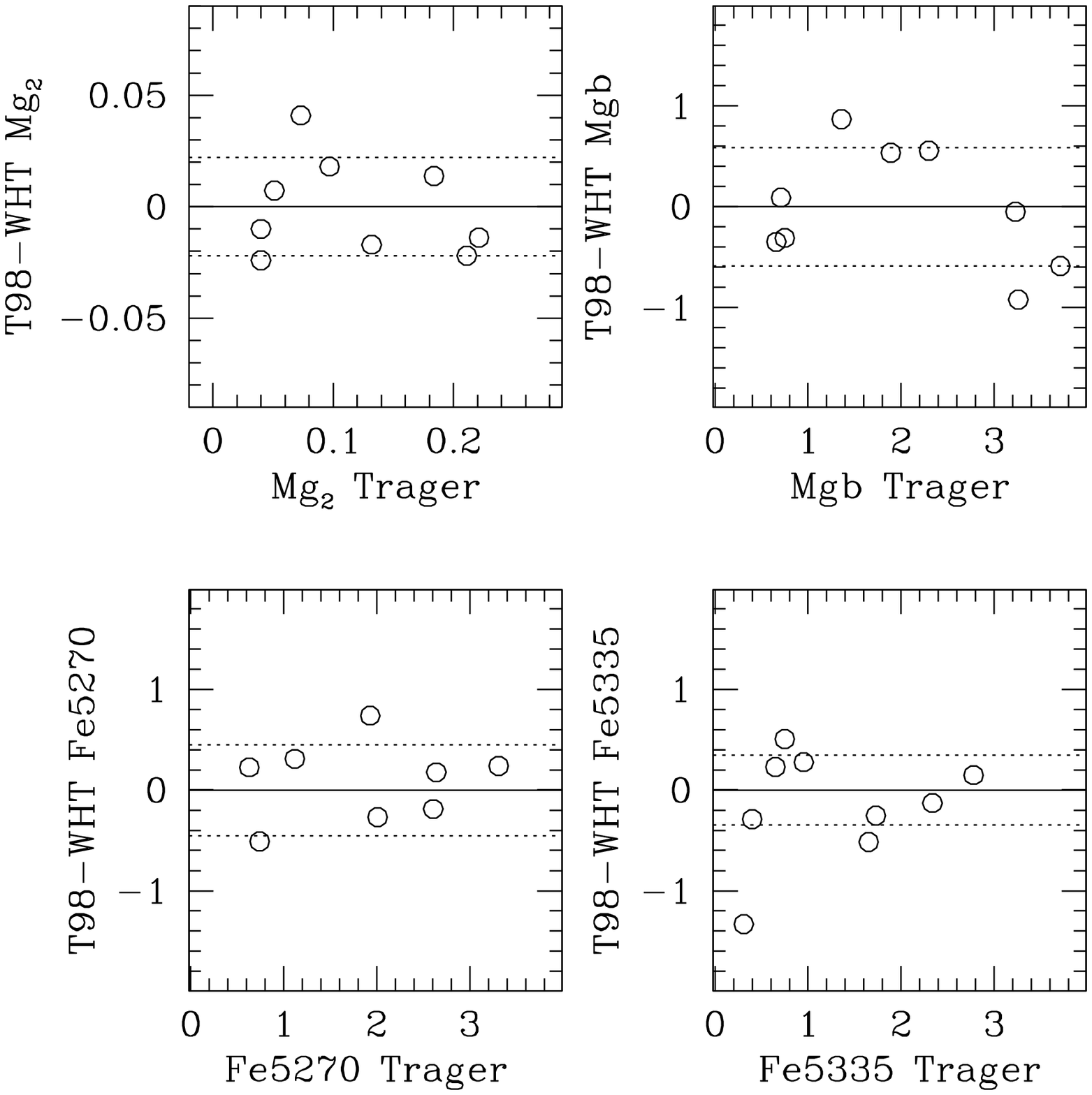}
   \includegraphics[width=0.95\hsize, height=0.82\hsize,clip=]{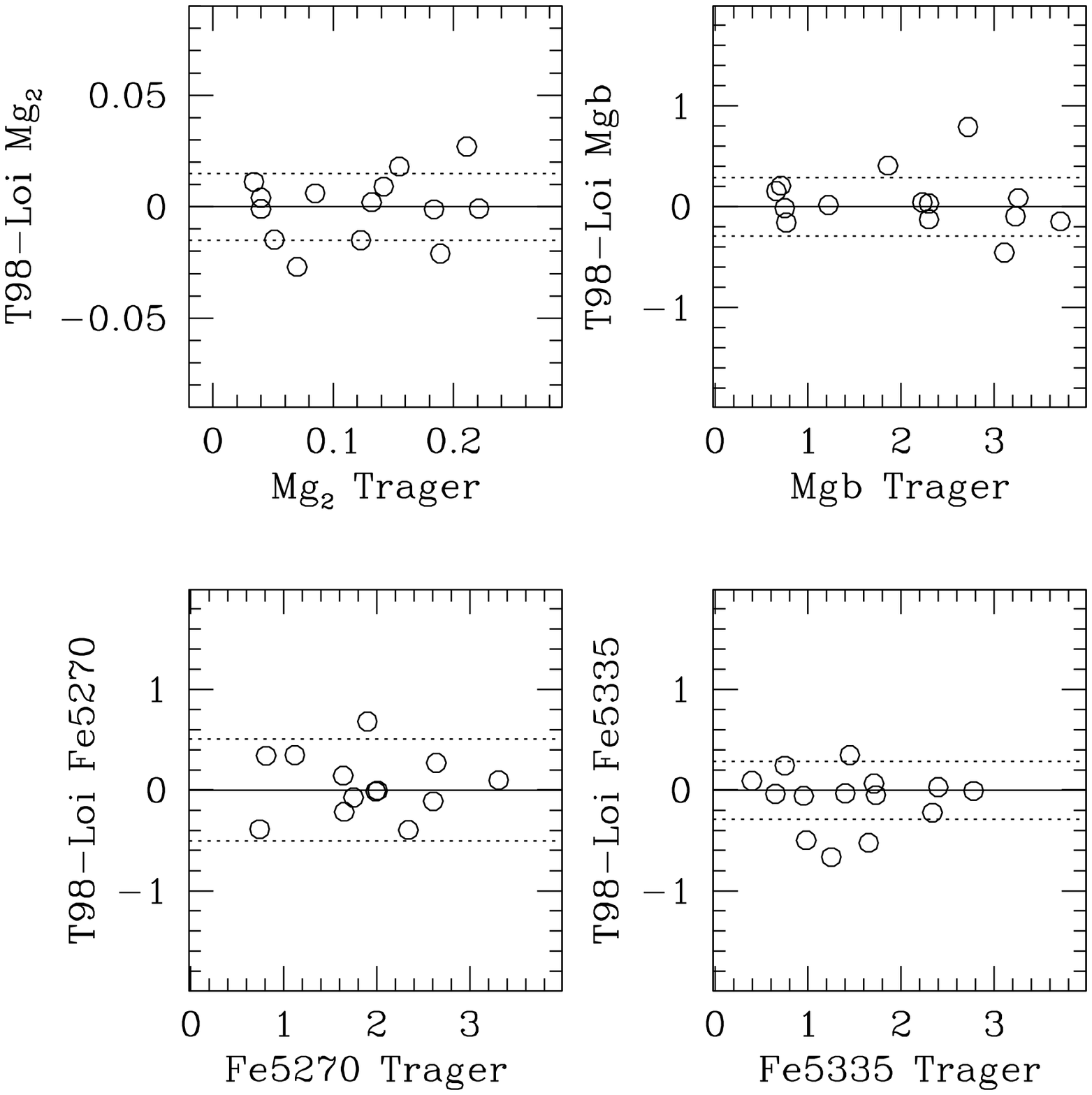}
   \includegraphics[width=0.95\hsize, height=0.82\hsize,clip=]{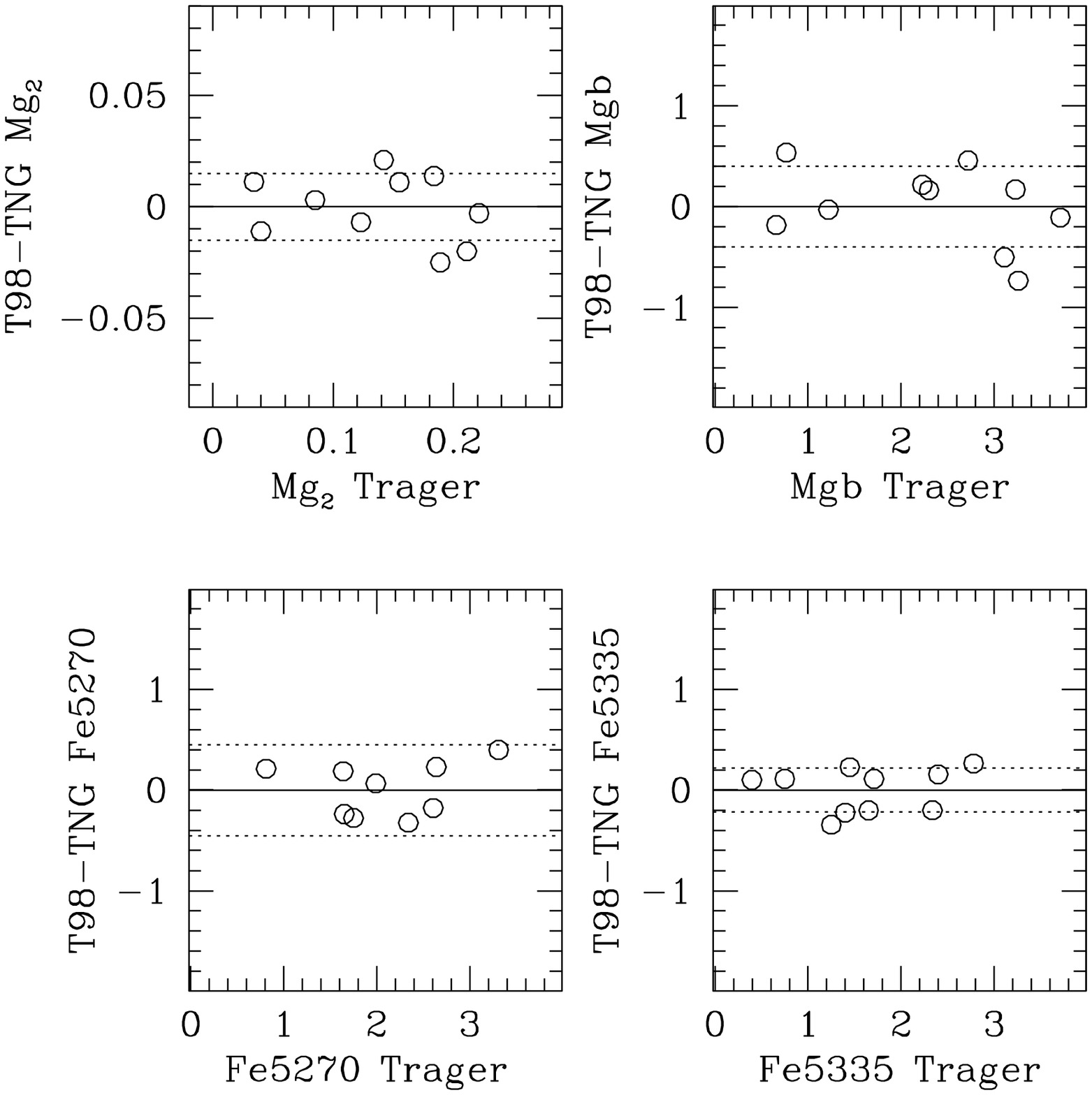}
     \caption{ Comparison of Lick indices measurements from WHT, Loiano, TNG 
after transformation and  T98, for the clusters in common. The dotted lines 
enclose the range of the rms. In AF2/WYFFOS data for Fe5335 we have not 
considered B178 in the fit data. In Loiano and TNG data we have not 
considered B193.
              }
        \label{indM}
    \end{figure}

\subsection{Indices from our own spectra \label{callick}}

First, we obtained new measures of Lick indices for 
the sample of M31 GCs described in \cite{io_rv} and \cite{io_est}. 
The spectra were taken with the AF2/WYFFOS multi-fiber spectrograph at
the 4.2m William Herschel telescope (WHT), with DoLoRes at the 3.5m 
Telescopio Nazionale Galileo (TNG),  Roque de los Muchachos (La
Palma, Spain), and with BFOSC at the Cassini 1.52m 
telescope of the Loiano Observatory (Bologna, Italy). 
The data acquisition, reduction and the resulting
radial velocities (and membership) are fully described in
\cite{io_rv}. 

All the spectra were flux-calibrated, using various spectrophotometric 
standard stars to convert counts into flux units.  
We selected 88 confirmed clusters having the best spectra, 
i.e. $S/N\ge 15$, 69 from the WHT set, 14 from the TNG set 
and 5 from the Loiano set. 
During each observing night  we also collected accurate ($S/N > 20$)  
observations of GCs in common with T98.

All the selected spectra span a wavelength range including indices 
from Fe4531 to Fe5406. 
Each spectrum was shifted to
zero radial velocity. 
Before measuring indices, one has carefully to degrade spectra of
higher resolution to the resolution of the Lick system. We strictly
followed the approach of \cite{wo97} and degraded our spectra to
the wavelength-dependent Lick resolution ($\sim 11.5 $ \AA~at 4000
\AA, 8.4 \AA~at 4900 \AA, and 9.8 \AA~ at 6000 \AA) with a variable-width 
Gaussian kernel. 

The derived indices were then compared with those provided by T98 for 9, 14, and
10 clusters in common for the WHT, Loiano and TNG sets, respectively. It was
found that all the considered indices can be reported into the T98 just by
adding the constant values listed in Table~\ref{corT98}. The comparison
between the {\em corrected} values from the various sets and the measures by T98
are shown in Fig.~\ref{indM}. 
Our Lick index measurements and index uncertainties are listed in 
Appendix \ref{rbc_new}, Table \ref{my_index}.
Errors were determined using 
photon statistics, following the formulae given in \cite{cardiel}.
They do not incorporate the uncertainty due to our
transformation to the Lick system, that is quantified by the r.m.s. scatter
reported in Table~\ref{corT98}.

%%%%%%%%%%%%%%%%%%%%%%%%%%%%%%%%%%%%%%%%%%%%%%%%%%%%%%%%%%%%%%%%%%
\begin{table}
  \begin{center}
  \caption{Correction terms of the transformation to match the Lick system for
WHT, Loiano and TNG data in the sense $I_{\rm Lick} = I_{\rm measured}+c$.}
\label{corT98}
  \begin{tabular}{|l|r|l||r|l||r|l|}
    \hline
Index & $c$  & rms & $c$  & rms  &  $c$  & rms \\
      &  WHT &  WHT &  Loi &  Loi &   TNG &  TNG\\
    \hline
\hline
Mg2    & 0.015 &0.022 & 0.018 &0.015 & -0.016 &0.015\\
Mgb    & 0.000 &0.588 &-0.122 &0.290 & -0.333 &0.399\\
Fe5270 &-0.210 &0.451 &-0.148 &0.504 & -0.108 &0.451\\
Fe5335 &-0.230 &0.346$^a$ &-0.188 &0.287 & -0.188 &0.218\\ 
\hline
\end{tabular} \end{center}
$^a$ We have not considered B178 in the fit for AF2/WYFFOS data.\\
  \end{table}
%%%%%%%%%%%%%%%%%%%%%%%%%%%%%%%%%%%%%%%%%%%%%%%%%%%%%%%%%%%%%%%

   \begin{figure*}[!ht]
   \centering
   \includegraphics[width=0.8\hsize, height=0.8\hsize,clip=]{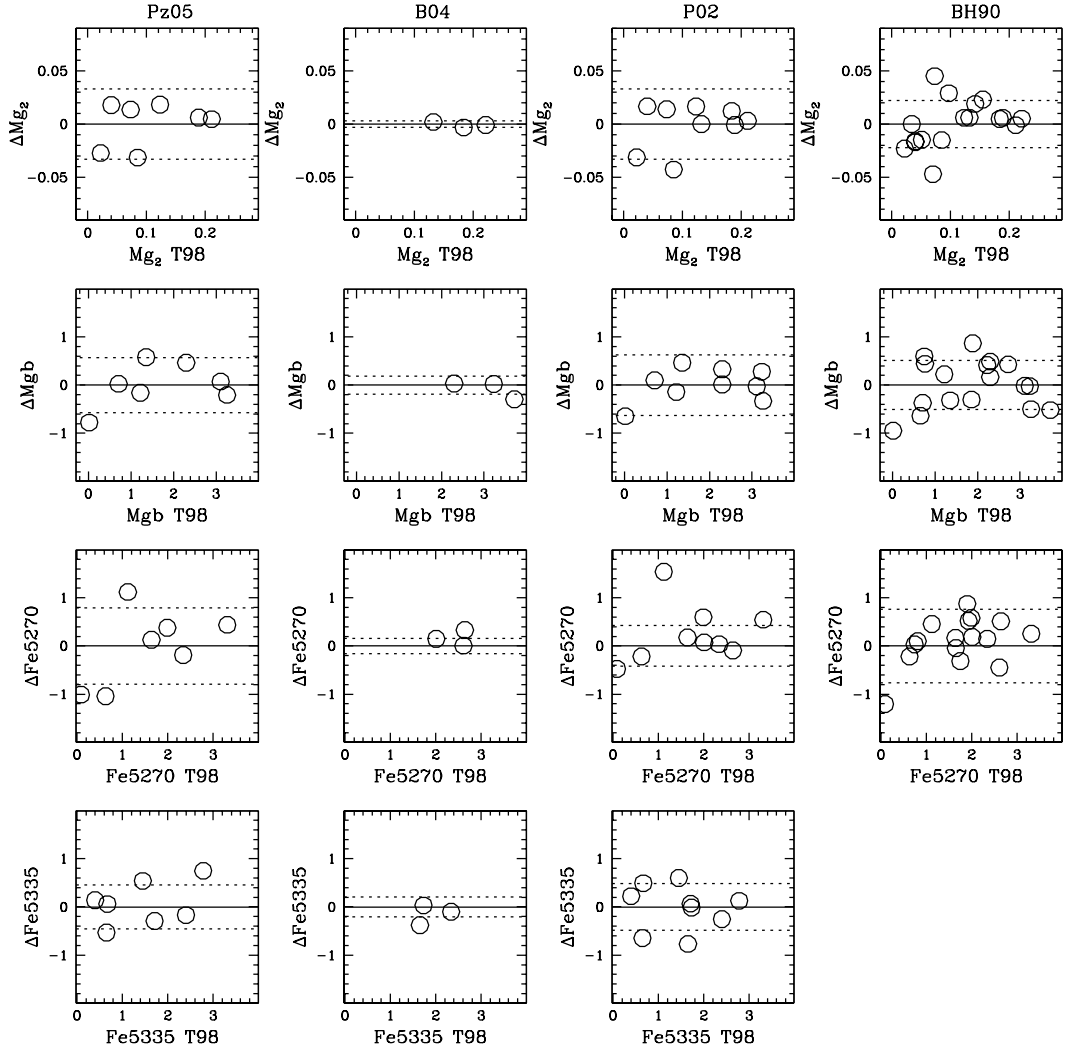}
     \caption{ A comparison of the common GCs between Trager and each sample 
 after the shifts according to Table \ref{allsh}. Dashed lines show 
the rms.
               }
        \label{pp}
    \end{figure*}

%%%%%%%%%%%%%%%%%%%%%%%%%%%%%%%%%%%%%%%%%%%%%%%%%%%%%%%%%%%%%%%%%%

\subsection{Data from literature}

To assemble as large as possible a dataset of metallicities in the new scale,
 we
complemented the measures described above with other clusters for which there
are published measures of the required Lick indices. In all cases we derived
the equations to transform the published values into the T98 system using
clusters in common between each considered set and T98, as done above for our
measures. The following sources were adopted:

\begin{table*}
  \begin{center}
  \caption{Coefficient for transformations to the Lick system 
of the literature data}\label{allsh}
  \begin{tabular}{|l||r|r|l||r|r|l||r|r|l||r|r|l|}
    \hline
Index & $a$ & $c$  & rms & $a$ & $c$  & rms  & $a$ &  $c$  & rms & $a$ & $c$  & rms \\
      &  Pz05 &&   & B04  & &  &   P02 && & BH90 && \\
\hline
\hline
Mg2    &0.731 & 0.017 &0.033 & 0.000 &-0.020&0.003 &0.730 &0.070&0.033    &0.000& 0.015 &0.022\\
Mgb    &0.764 & 0.424 &0.572 & 0.000 &0.000 &0.186 &0.681&0.240 &0.631$^a$&0.000&-0.27&0.508\\
Fe5270 &0.000 & 0.000 &0.790 & 0.000 &0.000 &0.161 &0.000&-0.353&0.422$^b$&0.639&0.731&0.762\\
Fe5335 &0.000 &0.000  &0.454 & 0.000 &0.000 &0.208 &0.000&0.000 &0.485    &     &     &     \\ 
\hline
\end{tabular} \end{center}
$^a$ we have not considered B015 in the fit data.\\
$^b$ we have not considered B012 in the fit data.
  \end{table*}

\begin{enumerate}

\item First, we have included the measures by T98 itself, that are
available for 18 M31 globular clusters. These spectra were obtained
by \cite{bur} with the image dissector scanner (IDS) at the 3m
Shane Telescope of the Lick Observatory. The absorption-line indices
are re-measured by T98 system and they define the standard-system.\\

\item Indices for 30 clusters have taken from
\cite{beas} (B04), who obtained high S/N spectra (S/N$>30$ pixel$^{-1}$) 
with the Low Resolution Imaging Spectrograph (LRIS) on the Keck I
telescope. The set of Lick line indices was measured with the
passband definitions of \cite{wortsys} and was not corrected to the
system of T98. However the three clusters in common with T98 show an excellent
agreement with the standard system (see Fig \ref{pp}).\\

\item We have incorporated the dataset of \cite{puziam31} (Pz05), who measured
Lick indices, with the T98 definition, for 29 M31 GCs from the best spectra 
(S/N $>25$ per $\AA$) of the 
\citet[][hereafter P02]{perr_cat} sample, 
obtained with the WYFOSS at the 4m WHT telescope.\\

\item We also have included the indices for further $\sim120$ clusters, 
from the lower
S/N spectra of the P02\footnote{Private communication.} sample. 
The absorption-line indices were
measured with the old passband definitions of \cite{faber73} and
\cite{bh}. 
When needed, we converted the value of
P02 to the commonly used $\AA$-scale for atomic
indices\footnote{The transformation between wavelength and magnitude
scale can be performed with the equations:
\begin{equation}
I_{{\AA}}=(\lambda_{max}-\lambda_{min})(1-10^{-0.4I_{m}})
\end{equation}
\begin{equation}
I_{m}= -2.5\log[1-(I_{{\AA}}/(\lambda_{max}-\lambda_{min}))]
\end{equation}
where $\lambda_{max}$ and $\lambda_{min}$ define the red and blue
boundaries of the feature passband.}.\\

\item The same procedure have applied to the \cite{bh90} (BH90) data obtained with the 
Multiple Mirror Telescope (MMT). BH90 have measured $\sim150$
absorption-line indices in their bandpass definitions from atmospheric 
cut-off at 3200 $\AA$ to NaI, not including thus the Fe5335 index.\\

\end{enumerate}

The various sets of indices ($I_{i}$) were transformed into the T98 system 
by the equation: 

$I_{T98}=I_{i}+aI_{i}+c$

The coefficients of the adopted transformations are reported in
Table~\ref{allsh}, and the
corrected indices are compared to the T98 values in Fig.\ref{pp}. 

\begin{figure}[t]
   \centering
   \includegraphics[width=9cm, height=9.5cm]{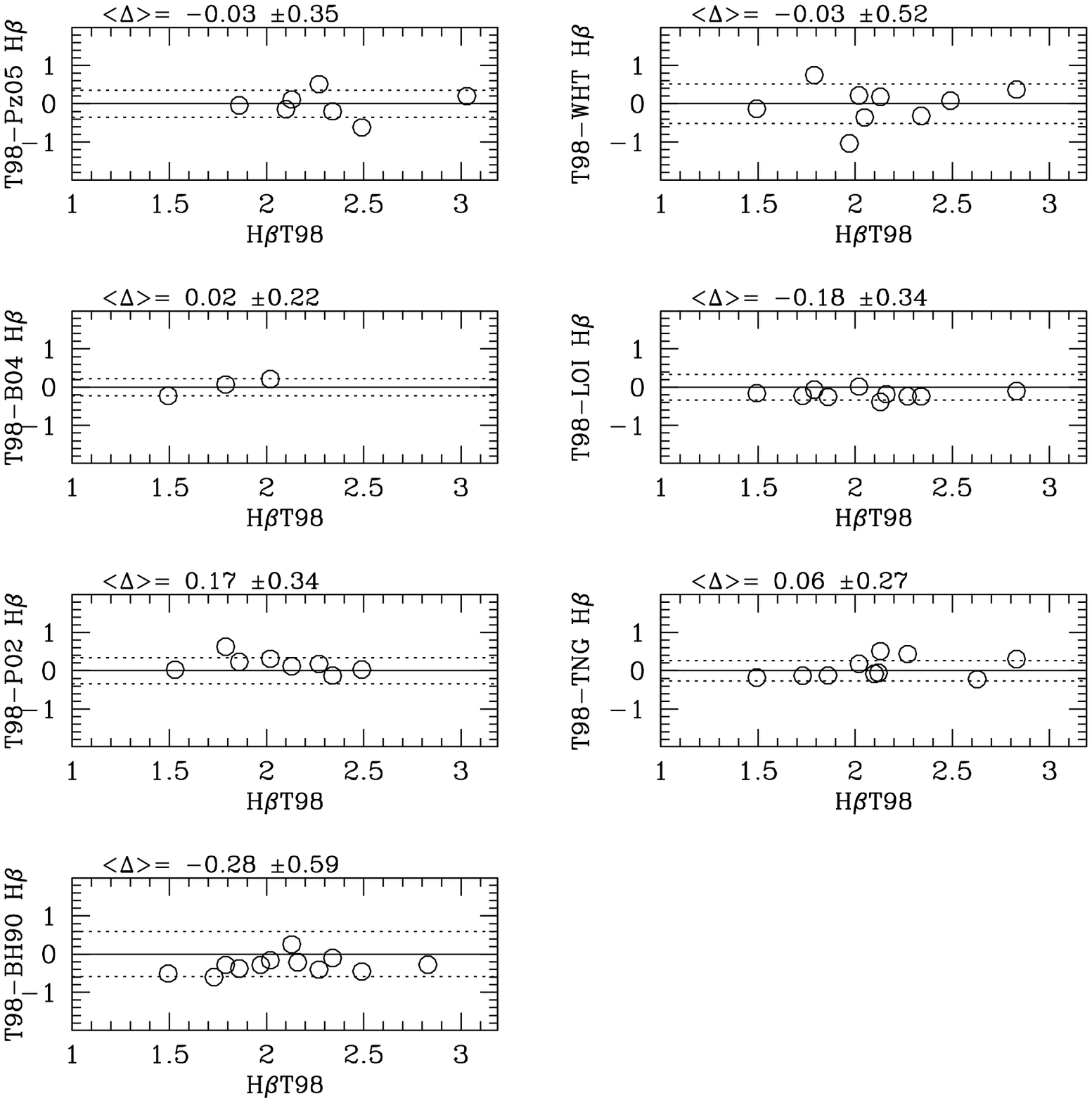}
     \caption{A comparison of the  $H\beta$ index measurements in common with 
T98. The mean differences and rms are also reported. The dashed lines 
encloses the rms.
 }
       \label{thb}
    \end{figure}

The same procedure has been adopted to transform also the $H\beta$
index into the T98 system, when available. In all the considered cases, a
constant shift appears to be an adequate transformation; the 
comparison of the corrected  $H\beta$ values and the adopted shifts and r.m.s.
are shown in  Fig.~\ref{thb}.

\subsection{Adopted metallicities}

The indices transformed into the T98 system were used to compute the
metallicities from Equations 1 and 2. In case of multiple measures of the same
spectral index for a given GC, we always choose the value obtained from the
spectrum with the highest S/N  (when available) and/or with the smallest error.
Individual indices estimates and the associated uncertainties 
for these datasets (296) are given in 
Appendix \ref{rbc_new}, Table \ref{all_index}.\\
Cluster metallicities and the associated uncertainties have been determined
from Eq.~1 when possible, and  from Mg2 in the other cases, i.e. when measures
of  Fe5270 and/or Fe5335 are lacking.
Since the index--metallicity relation used is valid only for old 
globular clusters, (see Sect.~\ref{scale}), we have removed all possibly young 
objects. 
The empirical metallicities for 245 M31 GCs (see Sect.~3.4, below) are 
reported in Appendix \ref{rbc_new}, Table \ref{meta}.

%%%%%%%%%%%%%%%%%%%%%%%%%%%%%%%%%%%%%%%%%%%%%%%%%%%%%%%%%%%%%%%%%%%%%%%%%%

\subsection{Possible contaminations}

All the M31 clusters comprised in our analysis  are class f=1 RBC entries; that
is, they are all classified as genuine M31  members whose nature has been
confirmed either spectroscopically and/or by means of  high-resolution imaging
\cite[see][for a detailed discussion about the classification of M31
GCs]{io_rv}. While the sample should be largely dominated by  bona-fide
clusters, some spurious object may always be present as a truly final word on
the nature of these objects can be obtained only by  resolving them (at least
partially) into individual stars by means of high resolution imaging. However,
the recently published large spectroscopic and imaging survey by
\citet{caldwell} allows us to extensively check the classifications adopted in
the RBC with fully independent and homogeneous data. 

Of the 252 class f=1 clusters that we originally considered in our analysis,
247 were also observed and classified by \cite{caldwell}, and only   {\em
seven} of them,  namely B025D, B026D, B043D, B046D, B215D, B248D and DAO25, 
were classified as non-clusters by these authors. For this reason, they have
been excluded from our sample, reducing the total number of clusters with
metallicity estimate from 252 to 245. Moreover, both B289D and B292D are
suspected  by \cite{caldwell} to be in fact stars. Since these cases are not
clear-cut, we mark  these objects as potentially misclassified but we keep them
in our globular  cluster sample. \cite{caldwell} {\em confirmed the RBC
classification for all the remaining 238 clusters in common between the two
samples}, i.e. all of them are classified as genuine globulars. We stress that
for the large majority of these clusters this classification has been
previously confirmed also by other authors. Finally, of the five clusters of
our sample that have not been observed by \cite{caldwell}, i.e. B514, MCGC1,
MCGC8, MCGC10, and B344D, the first four have been confirmed as genuine old
globulars from their HST CMDs \citep{acsb514,mc_acs}. Hence, according to the
above cross-check, we conclude that before the exclusion of the seven objects
re-classified by \cite{caldwell} the contamination of our sample by
non-clusters was $\la$~4\%, and should be significantly lower than this in the
final, cleaned sample.

As said in the Sect. 2.1, the derived calibrations are valid only  for old GCs
(age $>$ 7 Gyr), hence it would be wise to exclude possibly young clusters from
the final sample.  The most widely used age indicators among Lick indices are
the Balmer lines \cite[see][and references therein]{blcc,caldwell}. Here we
adopt the $H\beta$ index to clean our sample from possibly young objects on an
objective basis. In particular, we excluded since the beginning all the objects
with $H\beta$ $> 3.7$ $\AA$ \cite[see][for a detailed discussion]{blcc}
\footnote{In \citet{blcc}, to select clusters (possibly) younger than 2 Gyr the
selection criteria $H\beta$ $> 3.5$ $\AA$ was adopted, using $H\beta$ 
estimates from P02. However in the T98 system adopted here
$H\beta_{T98}$=$H\beta_{P02}$+0.17$\AA$ (Fig.~8), hence the limit has been
changed accordingly.}. 

To investigate the problem in deeper detail we have  checked any other cluster
that has been suggested by some author as  possibly young, irrespective of its
$H\beta$  value.  Of 245 clusters for which we have derived   metallicities,
there are 28 such clusters \cite[see][]{blcc}. Seven of these  have published
color-magnitude diagrams showing that they are  very likely old GCs: B311, B358
and B468 by \cite{rich05}, B008 by \cite{Perinacmd}, B083, B347 and NB16 by
\cite{Perinay}.  Additional 16 clusters have been recently classified as old
from their spectra by  \citet[][included in the cross-check described
above]{caldwell}, B015, B030, B047, B060, B070, B090, B117, B146, B154, B164,
B197, B214, B232, B292, B328, B486. We did not find additional information for
the remaining five (B018, B316, B431, B240D, DAO30): conservatively, we
maintain them in the list of clusters for which we provide a metallicity
estimate, but we exclude them from the cleaned sample used in the following
analysis (Sect.~5).

In conclusion, the degree of contamination from any kind of spurious object
(non-cluster or young cluster) in the sample considered in Sect.~5, should be
extremely low. It should be recalled that the adopted selection in $H\beta$
would not exclude from our sample intermediate-age clusters (2 Gyr $\la$ age
$\la$ 7 Gyr) that may be included with a wrong metallicity (lower than the true
value). However these clusters should be quite rare in M31, if any, as among
the several tens of M31 clusters having a CMD from HST, none has been found in
that age range.

\section{Comparisons with other sets of metallicities \label{comp}}

In the following sections we compare our metallicity estimates for 
M31 globular clusters with those already available in the literature.
We will discuss separately the comparison with (a) estimates obtained from
empirical calibrations of spectral indices or colors, (b) estimates obtained
from the fit of observed spectra with theoretical SPSS models, and (c) estimates
obtained from the analysis of the Color Magnitude Diagrams of individual
clusters \cite[see][and references therein]{Fusipecci96,holl97,jablonka00,
b379cmd,rich05,mc_acs,Perinacmd}. 

\subsection{Comparison with [Fe/H] from empirical calibrations}

Before proceeding to with the comparise our new scale with previous 
analysis,
it is worth having a look at the degree of consistency between already
existing sets. The comparison between H91 and P02 is particularly relevant in
this context, as (a) they are the largest sets of empirical metallicities for 
M31GCs
available in the literature, (b) they should be consistent {\em by definition},
as P02 used the same definitions of the indices as H91  \citep[see][hereafter
BH90]{bh90}, and used metallicities by H91 for clusters in common between the
two sets to calibrate [Fe/H] vs. indices. To the original set of H91 we added
the metallicities for further 35 M31 GCs obtained by \cite{bar00} with the same
method and strictly in the same system as H91.

\begin{figure}[t]
   \centering
   \includegraphics[width=9cm,height=8cm]{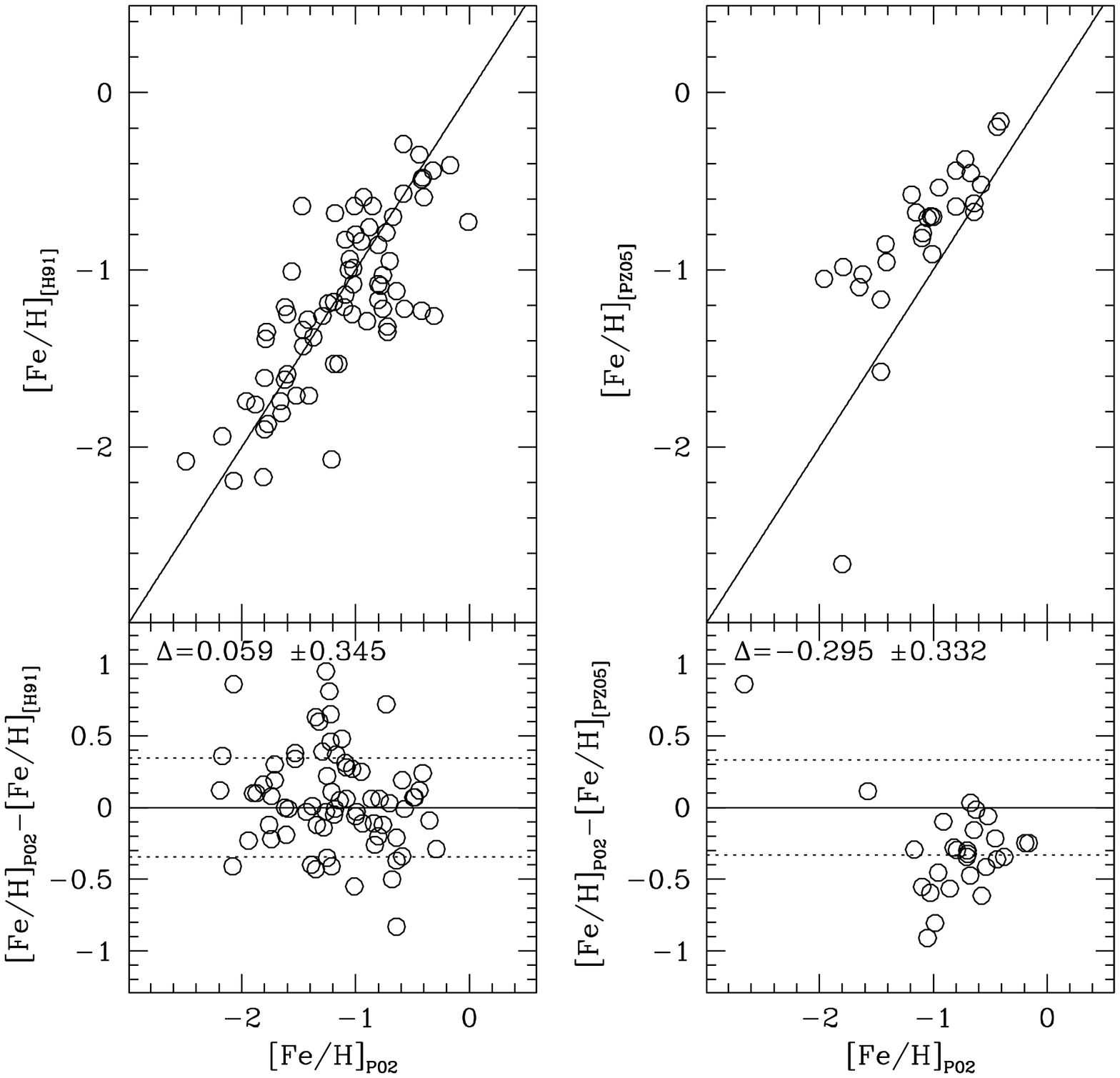}
     \caption{Comparison of the metallicities from
P02 with H91 and B00 in the left panel
and from P02 with PZ05 in the right panel. The solid line indicates 
the one-to-one relation.
The dotted lines in the lower panel mark the $\sigma$ value. 
                      }
       \label{difph}
    \end{figure}

Fig.\ref{difph} left panel reveals that there is a considerable 
scatter between the H91
and P02 sets of measures: the r.m.s. is 0.34 dex but differences up to
$\sim 1$ dex
are also present. This can be taken as a reference of the typical degree of
agreement between independent sets. 
In the right panel we compare the P02 metallicities with those Pz05 where the 
dataset is the same. A large spread is also evident in this case.

It is important to recall here that to obtain the metallicity of M31 GCs from
line  indices, BH90 and H91 calibrated a relation between [Fe/H] and the
infrared colors ((V-K), (J-K)) using Galactic GCs. 
Then, they used the infrared photometry of 40 M31 GCs by
\cite{Frogel} and \cite{Bonolia} to obtain their metallicity from that 
relation, and merged this
sample  with i) a sample of Galactic GCs for which they measured the same line
indices, and ii) with the average of the indices measured in several individual
stars in the open cluster NGC188, that was adopted as a template for solar
metallicity populations lacking among MW globulars. They used the merged
sample to calibrate various indices against [Fe/H]. Finally, they used the
relations to  obtain a metallicity estimate for each index, and they adopted
the  weighted average of the values obtained from the various indices  as their
final metallicity estimate (H91). The complex procedure outlined above was
dictated by the requirement to obtain the largest possible sample from the data
available at the time, and to average out the errors by using the information
from all the available indicators. On the other hand, our aim is to provide a
clean and easily repeatable process to obtain metallicities from few selected
spectral indices, as it is nowadays relatively easy to obtain high S/N
spectra for most M31 GCs with 4m telescopes, and the 10m class telescopes are
entering the game.  

%%-----------------------------fig 9 p02------------------------------ S_vib

   \begin{figure}[th]
   \includegraphics[width=9cm, height=9cm]{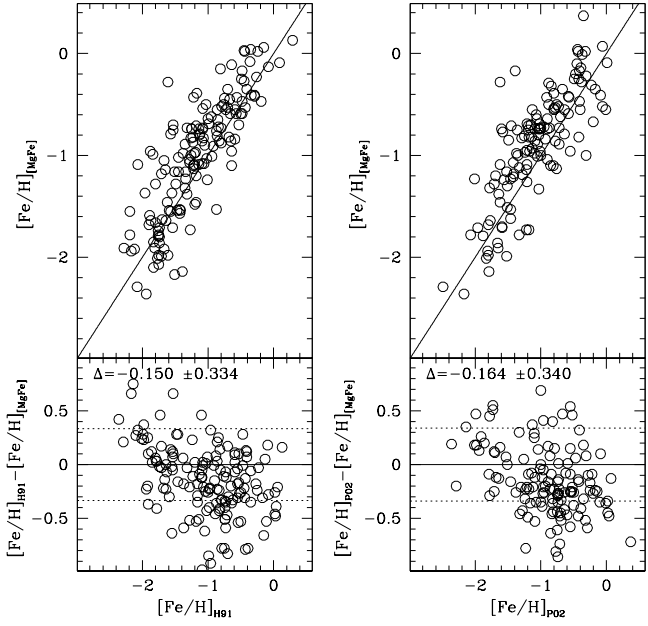}
     \caption{Comparison of our empirical metallicities of M31 GC with 
    P02 and H91. The solid line indicates the one-to-one relation.
 The dotted lines in the lower panel enclose the rms. Error bars have been 
omitted in the panels for clarity.
               }
        \label{fe_oldl}
    \end{figure}

   \begin{figure}[!h]
   \includegraphics[width=9cm, height=9cm]{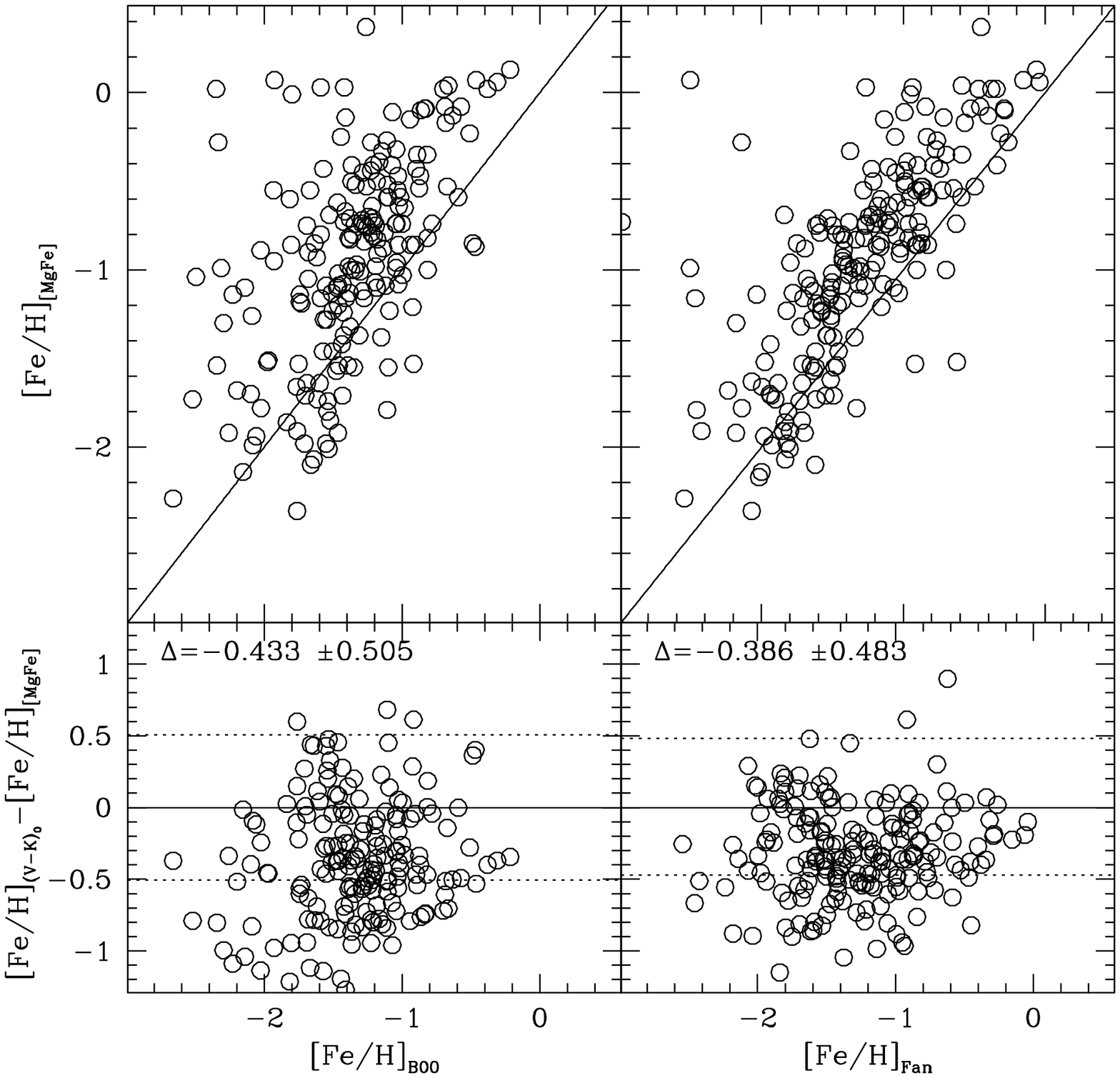}
     \caption{Comparison of our empirical metallicities of M31 GC with those
obtained from $(V-K)_0$ colors, using the calibrations by \citet{bar00}
and adopting two different sets of reddening estimates: 
\citet{bar00} and \citet{Fan}. 
The solid line indicates the one-to-one relation and the dotted lines 
in the lower panel enclose the rms . 
               }
        \label{fe_vk}
    \end{figure}
%%%%%%%%%%%%%%%%%%%%%%%%%%%%%%%%%%%%%%%%%%%%%%%%%%%%%%%%%%%%%%%%%

Fig.~\ref{fe_oldl} shows that the difference between our metallicity estimates
and those from H91 and P02 presents a scatter of the same amplitude as 
that existing between H91 and P02. Moreover, metallicities from our scale are
systematically larger by up to $\sim 0.3$ dex for $[Fe/H]\ga -1.4$ and
systematically smaller  for $[Fe/H]\la -2.0$. Fig.~\ref{fe_oldl} provides a
strong warning on the reliability of empirical metallicity scales based
on Lick indices from integrated spectra. While our metallicities and those by H91
and P02 present strong correlations, estimates for individual clusters
can well differ by as much as $\pm 0.5$ dex (or more) because of statistic
or systematic effects. This is the fundamental reason that convinced us to avoid
assembling metallicities from different sources for the RBC, trying instead to
reach the maximum degree of homogeneity at the ``index level'' (Sect.~2 and 3)
and the highest degree of internal consistency by converting the indices into
metallicity with the same calibrating relations. 
The choice of using just the  Mg2,
Mgb, Fe5270 and Fe5335 indices is also intended 
to minimize the effects on the final 
metallicity estimate of variations/anomalies in the abundance of other
elements, like for instance C, N etc., or age effects, that may affect other 
indices (see, for example,  \citet{burst04}, \citet{blcc} and references 
therein). It is worth noting, in this context, that we have no particular
a-posteriori reason to claim that our scale is {\em superior} to other existing empirical
scales based on spectra. 2, we feel that we have made all the
possible efforts to construct a very homogeneous and internally consistent scale
for the RBC, designed for the easy and safe inclusion of any new set of indices
that will be published in the future\footnote{This will be possible a 
the condition that the
considered set of indices have a sufficient number of clusters in common
with our sample to obtain a good transformation of the indices into the 
T98 system}.

Finally, Fig.~\ref{fe_vk} shows the comparison between our metallicities and
those obtained from $(V-K)_0$ colors, using the calibrations by \citet{bar00},
taking the V-K colors from the RBC, and adopting two different sets of
reddening estimates, i.e. those from 
\citet{bar00}\footnote{Private communication.} and from \citet{Fan}. 
This comparison reveals the critical role of the (uncertain) reddening estimates on
any metallicity scale based on colors: the two sets considered here differ only
in the adopted reddening, yet the r.m.s scatter in the final metallicity is as
large as $\pm 0.30$ dex. The overall
behavior in comparison with our scale is relatively similar to that of the H91
and P02 sets. This may be due to the fact that at the origin of these scales
there is also a calibration of metallicity vs. integrated $(V-K)_0$  colors.

It is interesting to note that, independently of the adopted set of
reddening, the metallicities obtained from $(V-K)_0$ are systematically lower
than our spectroscopic estimates by a large amount, i.e. $\simeq 0.4$ dex, in
average. We do not have a straightforward explanation for this remarkable
systematic difference, we can just put forward some hypothesis for its origin.
The observed effect can arise if the reddening values are systematically
overestimated: using the [Fe/H] - $(V-K)_0$ calibration by \citet{bar00} and
assuming $E(V-K)=2.75E(B-V)$, according to \citet{cardelli}, an overestimate of
E(B-V) by 0.09, in average, is sufficient to account for the whole 0.4 dex
difference between the metallicity scales. While the required systematic in
E(B-V) is probably too  large to be realistic, an overestimate of the reddening
may provide a relevant (possibly the largest) contribution to the observed
systematic difference in the metallicity. Systematic differences in the age
distribution and/or in the abundance pattern between MW and M31 globulars can
also contribute to the effect. In particular, Fig.~9 of \citet{bar00} seems to
suggest that the clusters of the two galaxies may not share the same [Fe/H] -
$(V-K)_0$ relation.

\subsection{Comparison with [Fe/H] from SED fitting}

\begin{figure}[t]
   \centering
   \includegraphics[width=9cm, height=9cm]{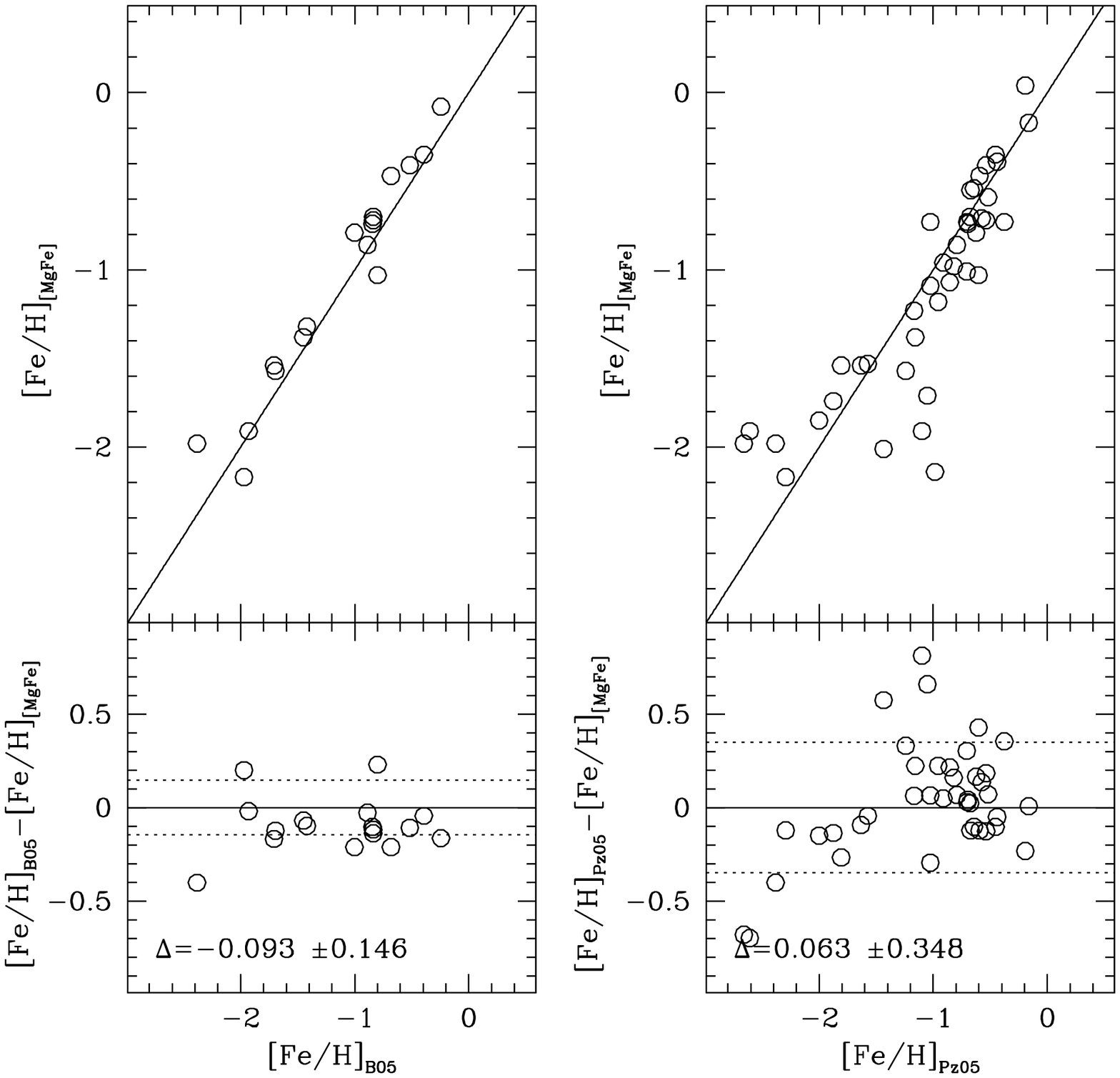}
     \caption{A comparison of the common GCs between this work, Puzia and 
Beasley data using [Fe/H] definitions by TMB.  
            }
        \label{ssp}

   \centering
   \includegraphics[width=9cm, height=9cm]{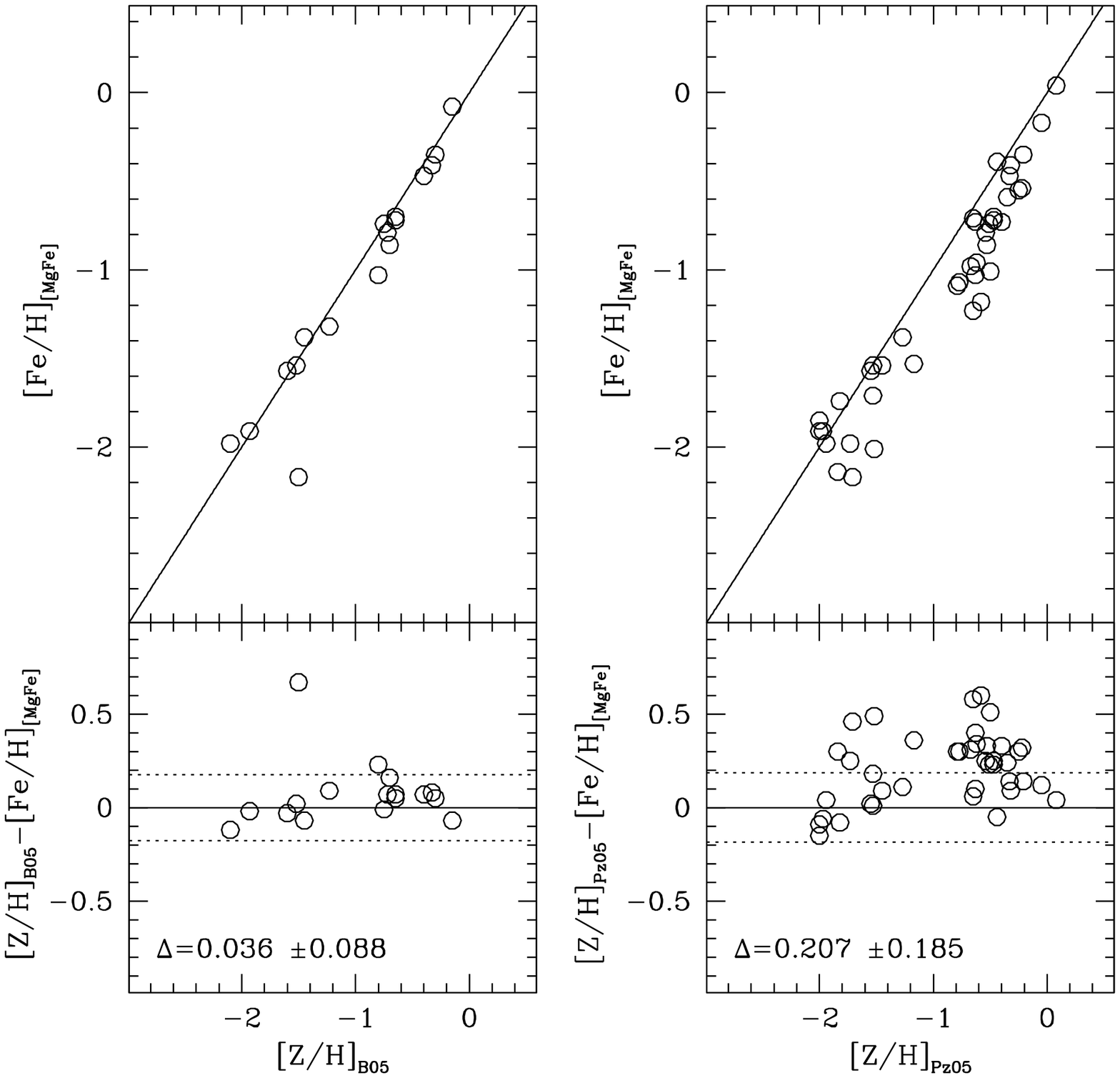}
     \caption{The same as Fig. \ref{ssp} but using [Z/H].
            }
        \label{ssp2}
    \end{figure}

We compare the metallicities of the M31 clusters derived from our
empirical calibrations 
with those derived from SSP
model fitting by \cite{puziam31} and \cite{beasII} (using the TMB
models) in Fig. \ref{ssp}.  The model report the
metallicities in [Z/H] scale and a transformation to [Fe/H]  has been
done through the equation:  [Fe/H]= [Z/H] - 0.94[$\alpha$/Fe] taken
from \citet[][see also Trager et al. 2000]{TMB}.  The clusters with
derived age $<$8 Gyr are excluded  because our empirical calibrations
are valid only for old GCs. To first order, the agreement between the
literature values and the metallicities from our empirical relations
is clearly satisfactory with B05, and acceptable with Pz05, as shown
in Fig.~\ref{ssp}.  The two sets of measures show systematics of
opposite sign with respect to our scale: B05 finds values slightly
lower  that ours (by $\sim 0.1$dex), Pz05 estimates are larger by
$\sim 0.2$ dex, on average. 
It may be worthwhile to check if part of
these differences may be due to the fact that these authors consider
separately the metallicity from Iron peaks elements and the abundance
of $\alpha$ elements, while our scale neglects this potentially
relevant discrimination\footnote{It is important to recall that our scale is
not expected to strictly trace the abundance of {\em Iron}. In fact it is based
on the ZW84 scale that, in turn, is based on the metallicities derived  by
\citet{Cohen} from lines of various elements, including Mg \cite[see][for a
detailed discussion and references]{mendel}. Therefore it is likely a better
proxy for the {\em total} metallicity than for the actual Iron abundance.}. 

Fig.~\ref{ssp2} shows that this may be the
case for B05: if our [Fe/H] estimates are compared with B05 estimates
of [Z/H] the offset is reduced to zero and even the r.m.s. scatter is
slightly reduced (having excluded the outlier B328). 
On the other hand the comparison with [Z/H]
exacerbates the systematic difference with Pz05, while significantly
reducing the r.m.s. scatter. 
In conclusion, our new metallicity scale
seems in much better agreement with scales derived from the detailed
fitting of spectra with SPSS models than with other empirical scales.

\subsection{Comparison with [Fe/H] from CMDs}

The estimates that can be obtained from CMDs of individual clusters by
comparing the observed Red Giant Branch (RGB) with the RGB templates
of well studied Galactic GCs  probably provides one of the most
reliable metallicities currently available for M31 GCs, of course
under the hypothesis that the basic properties of 
the two GC systems are the same. For clusters that are not too compact
and are  not immersed in exceedingly crowded fields, HST photometry
(either from the WFPC2, e.g.  \citet{rich05}, or the ACS, e.g. see
\cite{acsb514} and \cite{mc_acs})  can provide  clean and
well defined CMDs of the RGB. In addition the Horizontal Branch
morphology and the lack of bright Main Sequence stars give the best
sanity check on the actual age of the cluster that can be currently
achieved (see, in particular  \cite{b379cmd}).  Therefore, the
comparison of our estimates with those obtained from good CMDs from
HST is a compelling test of the reliability and accuracy of our new
metallicity scale.

\begin{figure}[t]
   \centering
   \includegraphics[width=8cm, height=8cm]{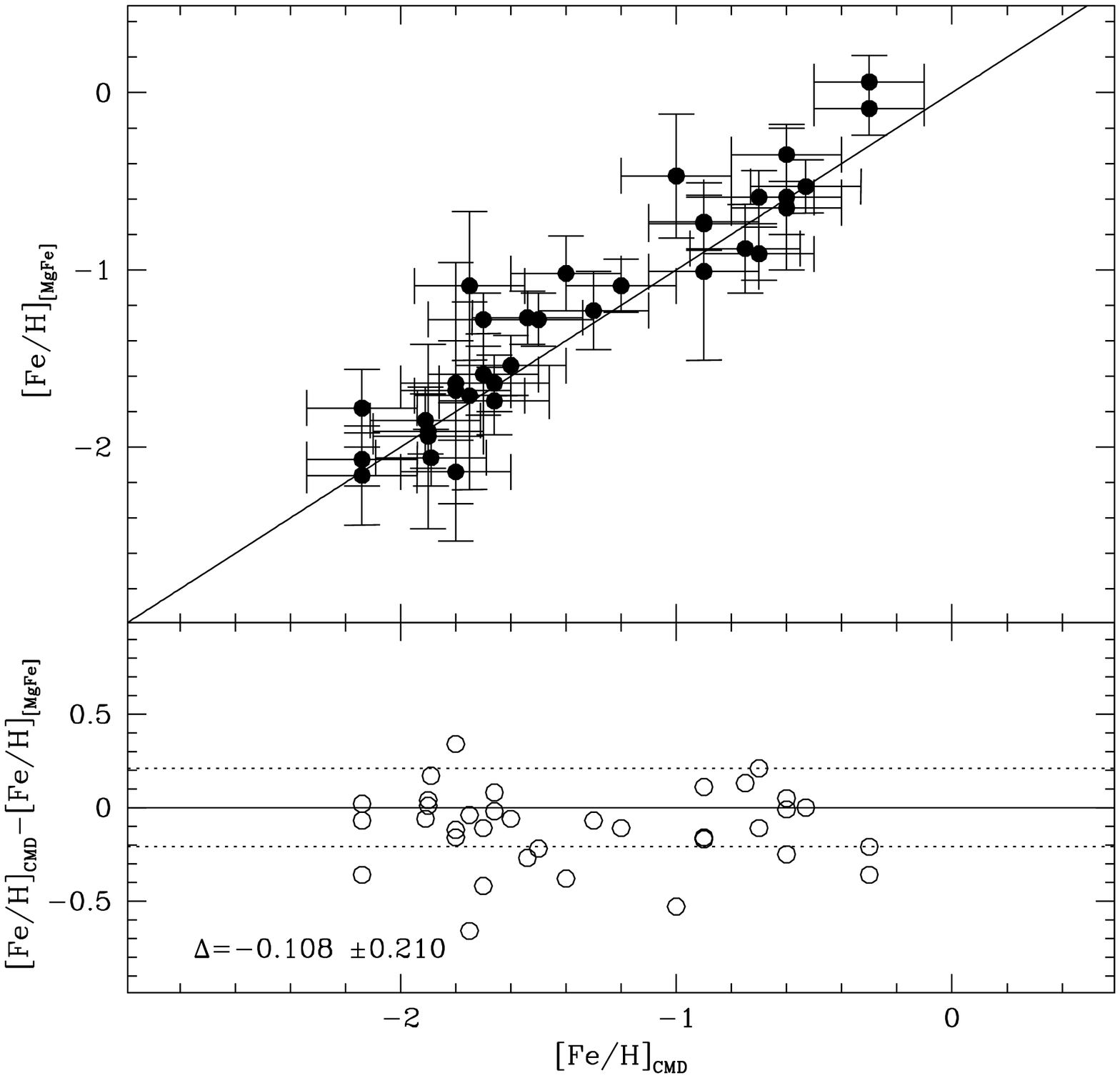}
     \caption{[Fe/H] estimates from CMDs of individual clusters (from 
     good quality HST photometry) are compared with the metallicities derived in
     the present analysis. The mean offset $\pm$ the standard deviation are
     reported in the lower panel.}
        \label{fecmd}
    \end{figure}

We collected metallicity from CMDs for 35 clusters in common with our
list, from the following sources: \cite{rich05}, that comprise the 
largest sample of published CMD of M31 GCs; \cite{jablonka00} that
analyzed  three GC in the bulge of M31; \cite{b379cmd}, that studied
in great detail B379; \cite{acsb514} and \cite{mc_acs} that considered
clusters located in the outskirts of the galaxy and \cite{Perinacmd}. 
The comparison between our estimates and those obtained from the CMDs
is presented in Fig.~\ref{fecmd}. The agreement is quite
satisfying over the whole considered range. 
However, there is a small systematic offset between the metallicity
estimates  obtained from the spectra or from the CMDs, in the sense that the
former are  larger than than latter by $\sim 0.1$ dex, in average. This points to
a real difference between the two independent scales, possibly related to how 
[$\alpha$/Fe] is included in the two calibrations.
We take the r.m.s. of the
difference computed over the whole sample as the typical accuracy of
our metallicity estimates ($\pm 0.25$ dex).

   \begin{figure}[t]
   \centering
   \includegraphics[width=9cm, height=9cm]{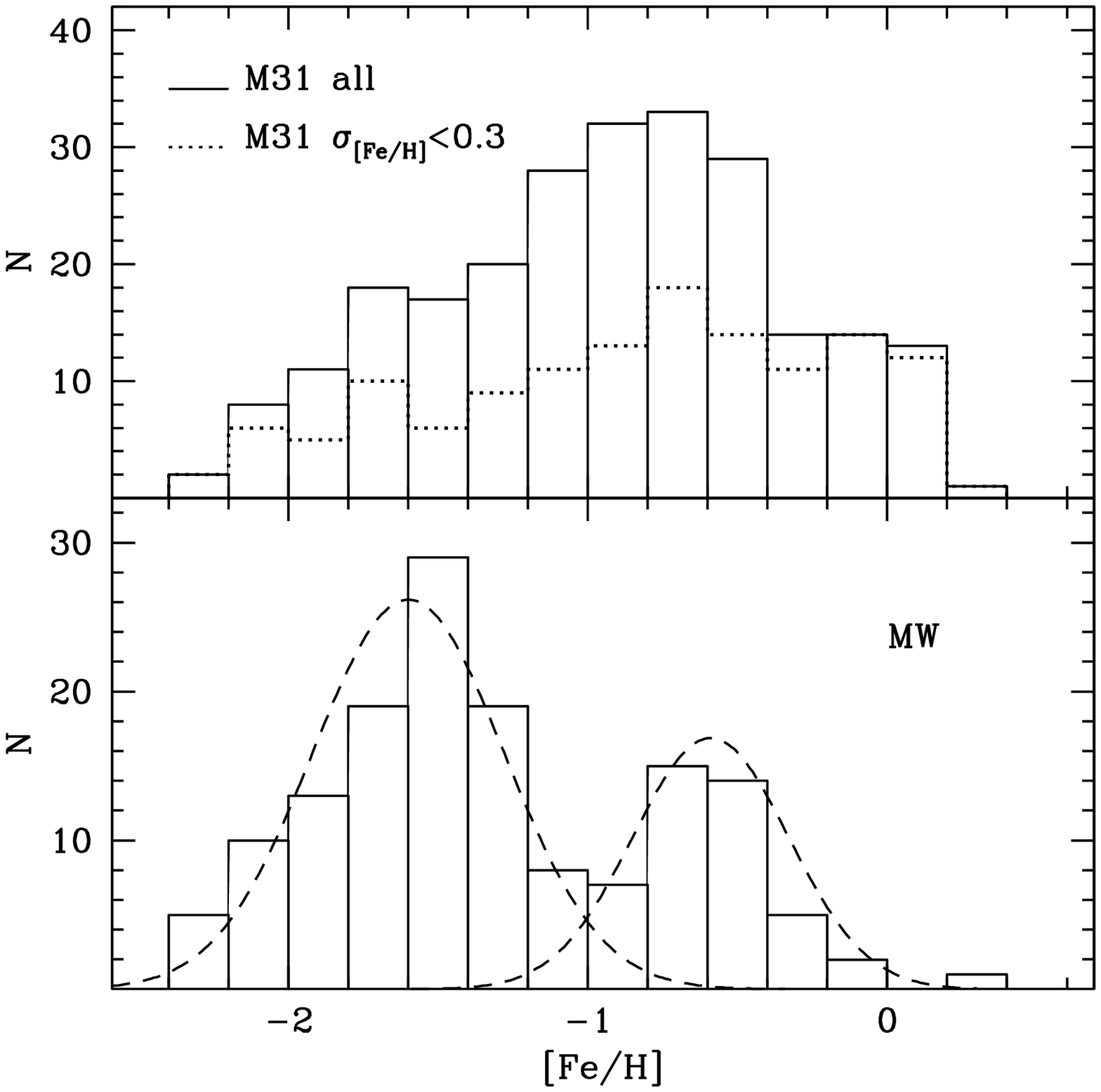}
     \caption{Metallicity histogram for the M31 globular cluster system 
(top) and the MW GC system (bottom), reported for comparison. The dashed lines
in the lower plot are the gaussian curves in the best fit models as found 
by the KMM algorithm for two subpopulations (Fe/H]=-1.60 and -0.59).              
\label{m31md1} }
        
   \includegraphics[width=8cm, height=8cm]{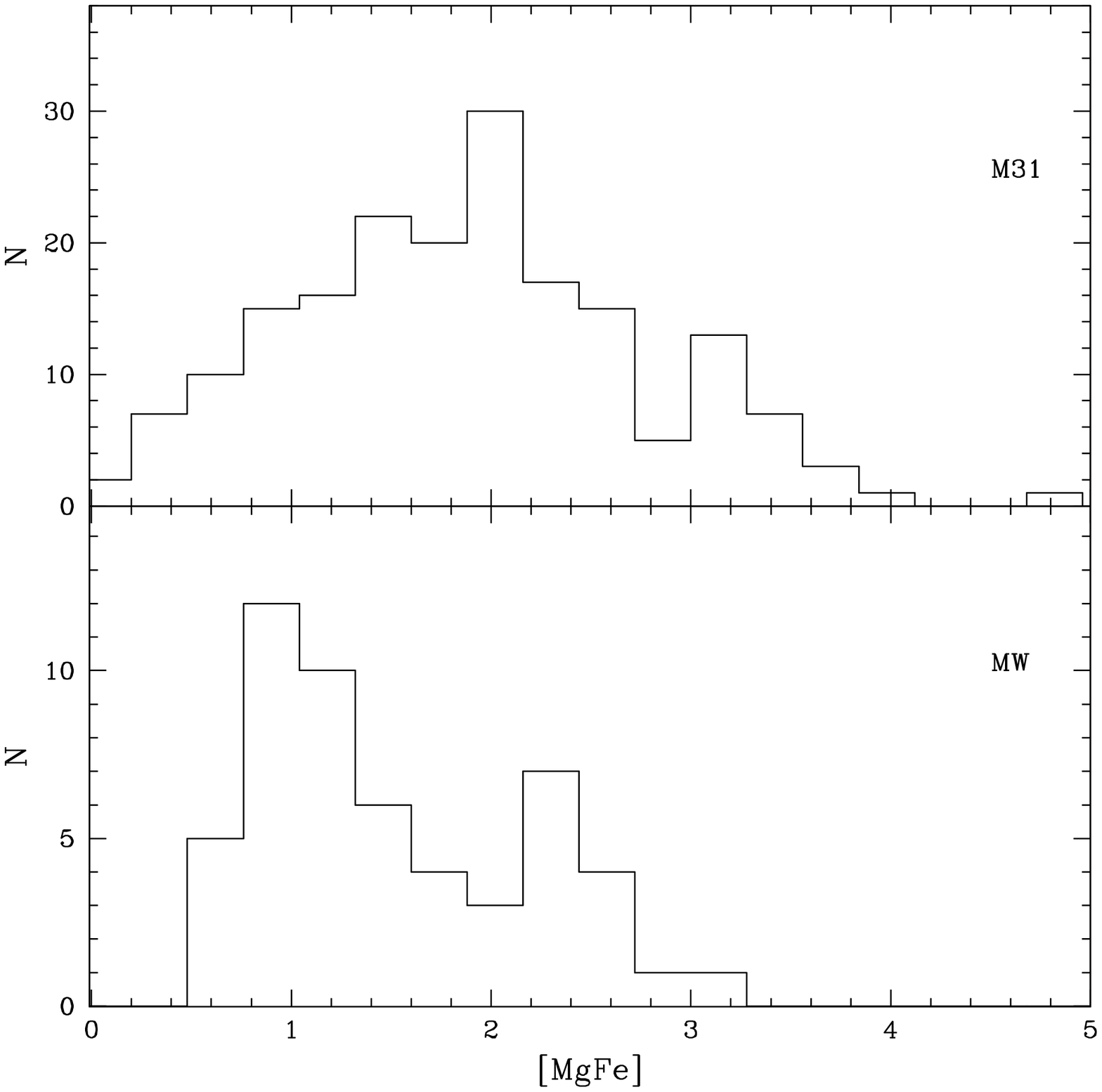}
     \caption{Distribution of [MgFe] of the M31 GCs. The lower panel
shows the distribution for Milky Way GCs 1 is clearly bimodal.
        \label{m31md2}
}
    \end{figure}

\section{Discussion and Conclusions \label{analysis}}

Using our own data as well as datasets available in the literature,
we have established a new homogeneous metallicity scale for M31 GCs.
The scale is based on the Lick index Mg2 and on the combination of
Mgb and Fe indices [MgFe], that have been calibrated against well
studied Galactic globulars (for $[Fe/H]<-0.2$) and a variety of
old-age SSP theoretical models for $-0.2 \leq [Fe/H] \leq 0.50$. 
Our scale has been shown to be self-consistent within $\pm 0.25$ dex, and it 
should be applied only to classical, old globular clusters.

In the following we  briefly describe a few natural applications
of the newly derived metallicity scale. In particular, (a) we 
derive and discuss the metallicity distribution of M31 GCs, (b)
we explore the correlations
between metallicity and kinematics,  for the sample of 240 bona-fide old GCs
described in Sect.~3.4, above.

   \begin{figure*}[t]
   \centering
   \includegraphics[width=14cm, height=14cm]{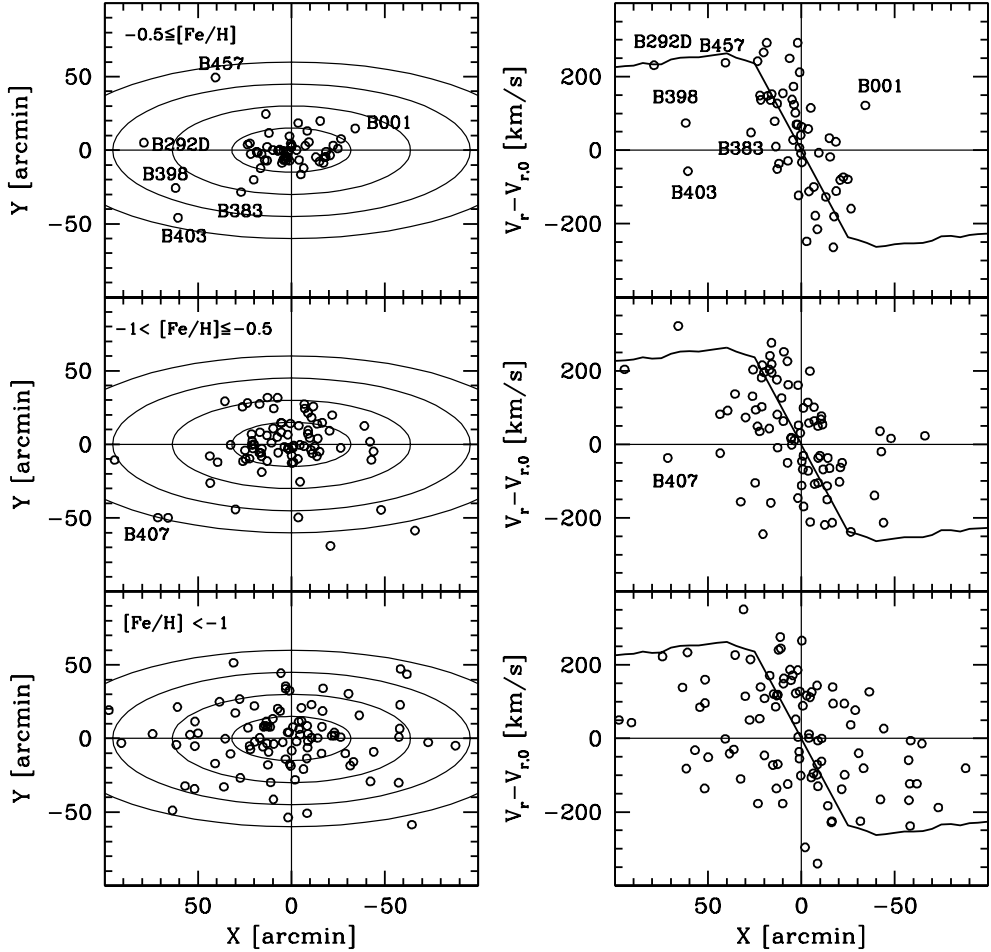}
     \caption{Left Panels: Spatial distribution of three metallicity groups 
 GCs in M31. 
The ellipses have a semimajor axis of 15, 30, 45, 60 arcmin.
  Right Panels:  Radial velocities vs. the projected distances along the 
major axis (X). 
The solid line shows a HI rotation curve from \cite{hI}. 
               }
        \label{m31k3}
    \end{figure*}

\subsection{Metallicity distribution}

In Fig.~\ref{m31md1} the Metallicity Distribution (MD) of our sample of 
M31 GCs is compared with its Milky Way counterpart.

The highest peak in the M31 MD
occurs at [Fe/H]$\sim -0.9$, coinciding with the overall average of the sample 
$<[Fe/H]> =-0.94$, significantly more metal rich than in the MW case, where the
maximum is at [Fe/H]$\sim -1.5$ and the overall mean is $<[Fe/H]>=-1.30$ 
\cite[based on data from][that are in the ZW84 scale]{h96}. 
The M31 system appears also to have a much
larger fraction of clusters having [Fe/H]$>-0.5$ (23\%
of the total sample) with respect to the Milky Way  (7\%).
It should be considered that the individual metallicity estimates for M31
clusters have much larger uncertainties with respect to their MW counterparts and
this may produce some spurious widening of the MD for M31. However the shape of
the distribution is essentially unchanged if we limit the analysis to the subset
of clusters having errors in metallicity lower than $\pm 0.3$ dex (132 clusters;
dotted histogram in the upper panel of Fig.~\ref{m31md1}). Fig.~\ref{m31md2}
shows that the difference between the MDs of the two galaxies cannot be ascribed
to spurious effects due to our calibration, as it can be re-conduced to genuine
differences in the observable [MgFe].

 The MD of M31 GCs do not present any obvious structure like the bimodality
encountered in the GC system of the Milky Way. Nevertheless the distribution for
M31 clusters does not seem to be well represented by a single Gaussian
distribution. Note that large errors on individual metallicities should
contribute to wipe out real structures, not to produce spurious ones.
The hypothesis of a multimodal underlying distribution has been compared with a
unimodal representation using the parametric KMM test \citep{KMM}, that compares
the fits to the MD made with one or more Gaussian distributions. A two component
model with modes at [Fe/H]=-1.54 and [Fe/H]=-0.64 is preferred to the unimodal
case at the 99.1\% confidence level (homoscedatic case) and at the 98.7\% c.l. in
the  heteroscedastic case with peaks at [Fe/H]=-1.79 and [Fe/H]=-0.76.  A three
component model with modes at $[Fe/H]=-0.25$,  $-0.89$ and $-1.72$ is also
preferable to the unimodal one (99.8\% c.l., in the homoscedastic case, and
99.6\% c.l., in the heteroscedastic case with peaks at [Fe/H]=-1.77, 
[Fe/H]=-0.80 and [Fe/H]=-0.01), and nearly equivalent to the bimodal
representation, from a statistical point of view.  
 The preference of bi- and three- modal models over the unimodal case remains
even if we  consider the subset of clusters with the lowest metallicity errors
described above. While clearly not conclusive, the above analysis suggests that
there may be real structures in the MD of M31 globular clusters, in good
agreement with the conclusions reached by \citet{bar00}, P02, Pz05, \cite{Fan}
and \cite{LeeII}.

\subsection{Metallicity and kinematics}

Fig.~\ref{m31k3} shows the positional and kinematical properties of M31 GCs
divided into three groups according to their metallicity, i.e. a Metal Poor (MP)
group ($[Fe/H]\le -1.0$), a Metal Intermediate (MI) group ($-1.0<[Fe/H]<-0.5$),
and a Metal Rich (MR) group ($[Fe/H]\ge -0.5$). 
The left panels of Fig.~\ref{m31k3} show the
spatial distribution of the considered clusters in the canonical X,Y projected
coordinate system \cite[see][and references therein]{2mass}, with X along the
major axis of the galaxy. In the right panels the radial velocity of the
clusters (in the reference frame of M31) is plotted versus the X coordinate and
compared with the rotation curve of the HI disk from \cite{hI}.

It results quite clear from the inspection of Fig.~\ref{m31k3} that  
the MR and MI subsamples display a significant rotation pattern, much
similar to the rotation curve of neutral Hydrogen disk of M31. The MR clusters
are more densely packed near the center of the galaxy and appear to follow more
closely the HI curve, whereas the MI clusters display a larger dispersion. MR
clusters are likely associated with the prominent Bulge of M31.

The MP clusters show a much larger velocity dispersion at any distance from the
center of the galaxy; in spite of that, they follow a significant rotation
pattern in the same sense as the other clusters. Dividing the MP sample at X=0,
we find (M31-centric) average velocities of $\langle V_{M31}\rangle=+59$ km/s and
$\langle V_{M31}\rangle=-48$ km/s for the clusters with $X>0$ and $X<0$,
respectively; the difference in the median velocities is even larger, as
$V_{M31}^{med}=+86$ km/s and $V_{M31}^{med}=-59$ km/s, for the two subsets.
Finally if the $V_{M31}$ distributions of the $X>0$ and $X<0$ MP clusters are
compared with a Kolmogorov-Smirnov test, it turns out that the probability that
the two samples are drawn from the same distribution of $V_{M31}$ is just 0.2\%.
In Sect.~3.4 we have shown that our sample should be reasonably clean from
spurious sources \cite[as for instance young massive clusters, that may be
misclassified as metal-poor GCs and would follow the rotation pattern of the thin
disk they belong to, see][]{blcc}, hence we conclude that the rotation pattern of
MP clusters is probably real. However an ultimate conclusion on this (relevant)
issue could be achieved only when the actual nature of a significant subsample of
MP clusters will be confirmed beyond any doubt from the CMD of their individual
stars. 

The above results are in good agreement with what previously found by P02 and
\cite{LeeII}, among  others.  A more detailed discussion of these correlations
between kinematics and metallicity is beyond the scope of the present paper. We
address the interested reader to the thorough discussion by \cite{LeeII}. Here we
just want to draw the attention of the reader on five of the six MR clusters
lying at $R>30\arcmin$ (labeled in Fig.~\ref{m31k3}). B001, B398 and B403 show no
correlation with the overall rotation pattern.  On the other hand B292D, and
B457  lie straight  on the flat branch of the HI rotation curve in spite of the
fact that they are more than $\sim 7$ kpc away from all the other MR clusters
(except B398 and B403, of course).  These five objects clearly deserve new
observations with high S/N spectra to verify both their metallicity and their
radial velocity. If confirmed, their odd positions and kinematics would require
an interpretation. Moreover, B403 and B407 (labeled in the MI panel of
Fig.~\ref{m31k3}, and having  $[Fe/H]=-0.65\pm 0.15$) have very similar position
and velocity (differing by  $\sim20$~km/s). The case of these two relatively
metal-rich clusters in the outer halo of M31 is discussed in more detail in
\cite{Perinacmd}.\footnote{{\em Note Added in Proofs:} After the acceptance of the 
present paper, a work appeared presenting metallicities of a few M31 GCs from high 
resolution integrated spectra \citep{colucci}.
It is interesting to note that for four of the five clusters in common, our [Fe/H] 
estimates agree with those of \citet{colucci} to within $<0.1$ dex. Our estimate 
for the fifth cluster (B358, which is in the deep metal-poor regime in which our 
observable are less sensitive to metallicity, see Sect.~2, above) is 0.36 dex 
higher than what found by \citet{colucci}, i.e.
 $[Fe/H]=-1.85\pm 0.19$ vs.\ $[Fe/H]=-2.21\pm 0.03$; however the difference is 
less than $2\sigma$. It appears that our own scale and that based on high-resolution 
spectroscopy introduced by these authors are in {\em excellent} agreement.
We note also that our metallicity estimate for the remote M31 cluster MGC1, 
$[Fe/H]=-2.16\pm 0.28$, is in excellent agreement with what recently obtained by
\citet{macGC1} from a high-quality CMD, $[M/H]\simeq -2.3$. }   

\begin{acknowledgements}

M.B. acknowledge the financial support of INAF through the PRIN2007 grant  CRA
1.06.10.04. We are very grateful to P. Barmby and K. Perrett for providing 
their unpublished data for M31 GCs. This research has made use of NASA's 
Astrophysics Data System Bibliographic Services.
We would like to thank the anonymous referee for their careful reading of the 
manuscript and helpful comments. 
\end{acknowledgements}

\bibliographystyle{aa}

%%%%%%%%%%%%%%%%%%%%%%%%%%%%%%%%%%%%%%%%%%%%%%%%%%%%%%%%
\newpage
\begin{appendix}

\section{Homogeneous Lick indices in the T98 system for M31 globular clusters 
 \label{rbc_new}}

%%%%%%%%%%%%%%%%%%%%%%%%%%%%%%%%%%%%%%%%%%%%%%%%%%%%%%%%%%%%%%%%%%%%%%%%%%%%%%%%%%%%%%%%%%%%%%%
\onecolumn
%\clearpage
\begin{small}

\begin{longtable}{l r r r r r r r r r r r r r r }
  \caption{Lick indices for M31 globular clusters from new observations
(Sect.~3.1). 
\label{my_index}}\\
 \hline\hline
Cluster&Mg$_2$&eMg$_2$&Mg$b$&eMg$b$&Fe5270&eFe5270&Fe5335&eFe5335&Fe5406&eFe5406&H$\beta$ &eH$\beta$ & y$^1$&Set\\
        & mag & mag  &  \AA  &  \AA  &  \AA  &  \AA  &   \AA   &  \AA  &  \AA &  \AA  & \AA &  \AA \\ 
\hline
\endfirsthead
\caption{continued.}\\
\hline\hline
 Cluster&Mg$_2$&eMg$_2$&Mg$b$&eMg$b$&Fe5270&eFe5270&Fe5335&eFe5335&Fe5406&eFe5406&H$\beta$ &eH$\beta$ & y$^1$&Set\\
        & mag & mag  &  \AA  &  \AA  &  \AA  &  \AA  &   \AA   &  \AA  &  \AA &  \AA  & \AA &  \AA \\ 
\hline
\endhead
\hline
\endfoot
   B003    & 0.086&  0.013&  1.554&  0.547&   1.906&   0.648&  1.676&	0.774&   0.643&     0.603&    2.72&   0.39&0& WHT\\ 
   B006    & 0.210&  0.005&  3.051&  0.221&   2.112&   0.262&  1.942&	0.305&   1.389&     0.230&    1.83&   0.17&0& WHT\\ 
   B012    & 0.064&  0.004&  0.619&  0.188&   0.811&   0.223&  0.421&	0.266&   0.303&     0.202&    2.63&   0.13&0& WHT\\ 
   B017    & 0.159&  0.007&  1.979&  0.291&   1.798&   0.338&  1.862&	0.393&   0.957&     0.300&    1.73&   0.22&0& WHT\\ 
   B019    & 0.159&  0.005&  2.313&  0.187&   1.790&   0.222&  1.704&	0.259&   0.916&     0.199&    1.73&   0.14&0& WHT\\ 
   B020    & 0.120&  0.002&  1.962&  0.082&   1.929&   0.092&  1.710&	0.105&   1.406&     0.078&    1.98&   0.07&0& TNG\\ 
   B022    & 0.061&  0.014&  1.339&  0.597&   1.371&   0.712&  1.109&	0.853&   0.209&     0.664&    3.09&   0.42&0& WHT\\ 
   B023    & 0.137&  0.004&  1.929&  0.171&   1.824&   0.186&  1.432&	0.211&   1.205&     0.154&    1.96&   0.17&0& LOI\\ 
   B032    & 0.210&  0.016&  4.270&  0.619&   2.693&   0.730&  2.805&	0.840&   1.221&     0.642&    1.73&   0.53&0& WHT\\ 
   B034    & 0.201&  0.013&  3.339&  0.533&   2.450&   0.631&  1.916&	0.744&   1.656&     0.555&    1.15&   0.41&0& WHT\\ 
   B039    & 0.176&  0.008&  2.630&  0.320&   1.920&   0.370&  1.742&	0.429&   0.880&     0.326&    1.40&   0.26&0& WHT\\ 
   B042    & 0.161&  0.008&  2.204&  0.325&   1.629&   0.374&  1.458&	0.432&   0.997&     0.323&    1.87&   0.27&0& WHT\\ 
   B051    & 0.170&  0.010&  2.565&  0.397&   1.608&   0.467&  1.054&	0.553&   0.746&     0.416&    1.37&   0.31&0& WHT\\ 
   B058    & 0.097&  0.007&  1.452&  0.294&   1.220&   0.352&  1.478&	0.402&   0.812&     0.312&    2.34&   0.24&0& LOI\\ 
   B060    & 0.134&  0.012&  2.067&  0.518&   1.474&   0.622&  0.691&	0.744&   0.916&     0.560&    2.53&   0.37&5& WHT\\ 
   B070    & 0.123&  0.010&  1.358&  0.406&   0.976&   0.481&  0.944&	0.564&   0.292&     0.436&    2.53&   0.29&5& WHT\\ 
   B071    & 0.275&  0.013&  4.838&  0.499&   2.710&   0.598&  2.279&	0.702&   1.725&     0.532&    2.04&   0.38&0& WHT\\ 
   B073    & 0.207&  0.012&  3.623&  0.465&   2.767&   0.546&  2.572&	0.635&   1.235&     0.490&    2.21&   0.35&0& WHT\\ 
   B082    & 0.193&  0.012&  2.608&  0.467&   2.111&   0.526&  1.955&	0.604&   0.835&     0.459&    1.63&   0.40&0& WHT\\ 
   B083    & 0.037&  0.011&  0.787&  0.442&   0.977&   0.494&  0.677&	0.575&   0.826&     0.423&    1.72&   0.42&2& WHT\\ 
   B095    & 0.186&  0.017&  2.754&  0.710&   1.149&   0.846&  1.592&	0.973&   2.058&     0.700&    1.59&   0.55&0& WHT\\ 
   B099    & 0.166&  0.010&  2.216&  0.412&   1.680&   0.485&  1.374&	0.571&   0.824&     0.437&    1.74&   0.30&0& WHT\\ 
   B110    & 0.183&  0.009&  2.655&  0.359&   1.738&   0.424&  1.788&	0.491&   1.302&     0.370&    1.64&   0.28&0& WHT\\ 
   B111    & 0.144&  0.019&  2.342&  0.774&   1.845&   0.923&  1.084&	1.079&   0.804&     0.819&    1.60&   0.60&0& WHT\\ 
   B117    & 0.067&  0.005&  0.897&  0.206&   0.630&   0.232&  0.621&	0.265&   0.253&     0.196&    2.06&   0.20&4& WHT\\ 
   B131    & 0.279&  0.005&  4.067&  0.204&   2.389&   0.243&  2.118&	0.285&   1.485&     0.218&    1.57&   0.15&0& WHT\\ 
   B147    & 0.242&  0.002&  3.952&  0.082&   3.032&   0.091&  2.816&	0.103&   2.162&     0.076&    1.66&   0.08&0& TNG\\ 
   B148    & 0.240&  0.008&  4.036&  0.299&   2.946&   0.326&  2.743&	0.368&   0.911&     0.283&    3.74&   0.28&0& WHT\\ 
   B151    & 0.199&  0.006&  3.179&  0.227&   1.981&   0.269&  1.879&	0.312&   1.347&     0.235&    1.75&   0.18&0& WHT\\ 
   B153    & 0.247&  0.011&  4.057&  0.432&   2.427&   0.513&  2.228&	0.598&   1.860&     0.450&    1.55&   0.34&0& WHT\\ 
   B155    & 0.212&  0.010&  2.732&  0.405&   3.711&   0.420&  3.685&	0.469&   1.221&     0.364&    3.24&   0.38&0& WHT\\ 
   B156    & 0.078&  0.009&  1.259&  0.376&   1.400&   0.410&  1.224&	0.465&   0.571&     0.345&    1.99&   0.38&0& WHT\\ 
   B158    & 0.149&  0.012&  1.744&  0.499&   2.278&   0.570&  1.982&	0.670&   1.575&     0.507&    1.01&   0.38&0& WHT\\ 
   B162    & 0.270&  0.015&  4.720&  0.602&   2.198&   0.738&  2.727&	0.843&   1.623&     0.646&    1.91&   0.46&0& WHT\\ 
   B163    & 0.235&  0.005&  4.301&  0.187&   2.797&   0.206&  2.466&	0.231&   1.555&     0.170&    1.59&   0.20&0& WHT\\ 
   B169    & 0.280&  0.013&  5.389&  0.486&   2.657&   0.543&  1.977&	0.624&   2.404&     0.447&    1.65&   0.52&0& WHT\\ 
   B171    & 0.214&  0.002&  3.610&  0.091&   2.663&   0.102&  2.245&	0.116&   1.979&     0.085&    1.89&   0.09&0& TNG\\ 
   B174    & 0.103&  0.006&  1.950&  0.243&   1.417&   0.270&  1.194&	0.306&   0.892&     0.225&    1.89&   0.25&0& WHT\\ 
   B178    & 0.079&  0.006&  1.354&  0.226&   1.191&   0.249&  1.639&	0.280&   0.420&     0.209&    2.98&   0.23&0& WHT\\ 
   B179    & 0.116&  0.008&  1.933&  0.317&   1.408&   0.350&  1.520&	0.396&   0.648&     0.299&    1.10&   0.32&0& WHT\\ 
   B180    & 0.125&  0.006&  2.589&  0.216&   1.924&   0.244&  1.146&	0.283&   0.603&     0.215&    1.57&   0.21&0& WHT\\ 
   B182    & 0.146&  0.014&  2.776&  0.554&   1.593&   0.667&  0.878&	0.799&   0.328&     0.607&    1.92&   0.42&0& WHT\\ 
   B183    & 0.182&  0.006&  3.134&  0.221&   2.361&   0.244&  1.423&	0.283&   0.922&     0.211&    1.92&   0.22&0& WHT\\ 
   B185    & 0.159&  0.007&  2.834&  0.289&   2.258&   0.316&  1.745&	0.358&   0.893&     0.261&    1.70&   0.31&0& WHT\\ 
   B187    & 0.123&  0.012&  1.370&  0.494&   0.892&   0.547&  0.653&	0.624&   0.447&     0.461&    2.52&   0.45&0& WHT\\ 
   B193    & 0.233&  0.004&  4.182&  0.167&   3.071&   0.183&  2.631&	0.207&   1.558&     0.155&    1.92&   0.17&0& WHT\\ 
   B204    & 0.139&  0.005&  2.642&  0.204&   2.256&   0.223&  2.733&	0.249&   1.415&     0.185&    1.94&   0.21&0& WHT\\ 
   B206    & 0.082&  0.004&  1.494&  0.143&   1.516&   0.157&  1.293&	0.179&   0.837&     0.131&    2.23&   0.14&0& WHT\\ 
   B212    & 0.050&  0.005&  1.009&  0.219&   0.183&   0.252&  0.245&	0.288&   0.358&     0.212&    2.43&   0.21&0& WHT\\ 
   B215    & 0.196&  0.007&  3.493&  0.288&   2.306&   0.320&  1.758&	0.364&   0.861&     0.273&    1.92&   0.29&0& WHT\\ 
   B218    & 0.130&  0.002&  2.137&  0.065&   1.923&   0.073&  1.598&	0.083&   1.331&     0.062&    2.05&   0.06&0& TNG\\ 
   B219    & 0.157&  0.009&  3.211&  0.339&   2.070&   0.386&  1.241&	0.449&   1.057&     0.344&    2.41&   0.32&0& WHT\\ 
   B224    & 0.042&  0.006&  0.909&  0.263&   0.448&   0.299&  1.162&	0.339&   0.177&     0.258&    2.17&   0.24&0& WHT\\ 
   B225    & 0.170&  0.002&  3.279&  0.077&   2.464&   0.085&  2.167&	0.098&   1.021&     0.074&    1.77&   0.08&0& WHT\\ 
   B228    & 0.129&  0.007&  2.104&  0.300&   1.759&   0.328&  1.473&	0.375&   1.594&     0.272&    2.19&   0.29&0& WHT\\ 
   B230    & 0.057&  0.007&  0.107&  0.286&   0.461&   0.315&  0.174&	0.363&   0.267&     0.268&    2.53&   0.26&0& WHT\\ 
   B232    & 0.032&  0.005&  0.493&  0.214&   0.404&   0.236&  0.690&	0.267&   0.236&     0.197&    2.38&   0.21&4& WHT\\ 
   B233    & 0.099&  0.005&  1.709&  0.192&   1.883&   0.210&  1.629&	0.240&   0.825&     0.181&    2.04&   0.18&0& WHT\\ 
   B235    & 0.133&  0.006&  2.380&  0.222&   1.690&   0.246&  1.773&	0.279&   0.973&     0.207&    1.96&   0.22&0& WHT\\ 
   B236    & 0.052&  0.012&  0.767&  0.486&   0.318&   0.534&  0.009&	0.613&   0.320&     0.441&    4.40&   0.46&0& WHT\\ 
   B238    & 0.137&  0.006&  3.478&  0.231&   1.714&   0.267&  1.862&	0.301&   0.873&     0.225&    1.78&   0.24&0& WHT\\ 
   B240    & 0.044&  0.005&  1.060&  0.185&   1.252&   0.206&  0.677&	0.239&   0.212&     0.179&    2.38&   0.18&0& WHT\\ 
   B318    & 0.027&  0.004&  0.112&  0.165&   0.586&   0.190&  0.231&	0.222&   0.601&     0.165&    5.49&   0.12&1& TNG\\ 
   B338    & 0.082&  0.002&  1.250&  0.078&   1.454&   0.088&  1.223&	0.102&   1.194&     0.075&    2.24&   0.07&0& TNG\\ 
   B344    & 0.109&  0.007&  1.669&  0.270&   2.161&   0.291&  2.320&	0.330&   1.114&     0.249&    1.85&   0.26&0& WHT\\ 
   B347    & 0.052&  0.007&  1.175&  0.274&   0.447&   0.311&  0.994&	0.351&   0.179&     0.265&    2.46&   0.26&4& WHT\\ 
   B348    & 0.136&  0.008&  2.169&  0.303&   2.221&   0.327&  1.443&	0.383&   1.061&     0.284&    2.52&   0.28&0& WHT\\ 
   B356    & 0.075&  0.008&  0.877&  0.325&   1.133&   0.353&  0.817&	0.405&   0.378&     0.297&    2.44&   0.31&0& WHT\\ 
   B358    & 0.023&  0.003&  0.230&  0.122&   0.599&   0.137&  0.296&	0.159&   0.649&     0.116&    2.91&   0.11&0& TNG\\ 
   B373    & 0.167&  0.016&  2.451&  0.619&   2.169&   0.832&  1.940&	0.810&   1.435&     0.611&    1.55&   0.53&0& WHT\\ 
   B381    & 0.075&  0.006&  1.372&  0.257&   1.766&   0.283&  1.650&	0.323&   0.722&     0.242&    1.67&   0.25&0& WHT\\ 
   B399    & 0.043&  0.004&  0.817&  0.178&   0.854&   0.201&  1.164&	0.229&   0.752&     0.170&    2.89&   0.16&0& TNG\\ 
   B457    & 0.265&  0.000&  4.029&  0.013&   3.851&   0.014&  3.496&	0.016&   2.891&     0.012&    1.44&   0.01&0& TNG\\ 
   B468    & 0.113&  0.007&  2.583&  0.277&   1.472&   0.320&  1.095&	0.367&   0.657&     0.274&    2.50&   0.25&4& TNG\\ 
   B472    & 0.080&  0.004&  3.214&  0.142&   1.378&   0.168&  1.266&	0.192&   0.514&     0.143&    2.23&   0.15&0& WHT\\ 
   G001    & 0.133&  0.003&  2.187&  0.104&   1.866&   0.115&  1.915&	0.131&   0.936&     0.098&    2.37&   0.10&0& LOI\\ 
   B020D   & 0.092&  0.016&  1.713&  0.653&   0.340&   0.805&  0.897&	0.917&   0.602&     0.696&    3.22&   0.49&0& WHT\\ 
   VDB0    & 0.031&  0.002&  0.186&  0.088&   0.598&   0.101&  0.568&	0.116&   0.366&     0.087&    4.50&   0.07&1& TNG\\ 
   B025D$^{\star}$   & 0.250&  0.024&  4.182&  0.955&   1.875&   1.094&  2.463&	1.258&   0.035&     1.008&    0.08&   0.86&0& WHT\\ 
   B041D   & 0.126&  0.020&  0.685&  0.851&   1.416&   0.945&  1.897&	1.075&   1.019&     0.817&    1.97&   0.65&0& WHT\\ 
   B046D$^{\star}$   & 0.230&  0.026&  3.245&  1.091&   2.712&   1.255&  3.013&	1.454&   1.154&     1.160&    2.12&   0.76&0& WHT\\ 
   B090D   & 0.291&  0.007&  4.122&  0.298&   2.604&   0.349&  2.241&	0.408&   1.732&     0.310&    1.51&   0.23&0& WHT\\ 
   B215D$^{\star}$   & 0.187&  0.009&  2.449&  0.392&   1.676&   0.463&  1.450&	0.547&   1.061&     0.414&    1.79&   0.28&0& WHT\\ 
   B344D   & 0.121&  0.001&  2.517&  0.022&   2.054&   0.024&  1.679&	0.028&   1.372&     0.021&    2.66&   0.02&0& TNG\\ 
   B514    & 0.062&  0.003&  0.300&  0.137&   0.176&   0.154&  1.279&	0.169&   0.282&     0.159&    2.32&   0.13&0& LOI\\ 
   MCGC1   & 0.041&  0.007&  0.566&  0.290&   0.489&   0.327&  0.005&	0.379&   0.397&     0.276&    1.84&   0.29&0& LOI\\ 
   MCGC8   & 0.093&  0.003&  1.334&  0.115&   1.400&   0.128&  1.206&	0.147&   1.041&     0.109&    1.98&   0.10&0& TNG\\ 
   MCGC10  & 0.031&  0.003&  0.395&  0.104&   0.633&   0.118&  0.427&	0.136&   0.677&     0.100&    2.93&   0.09&0& TNG\\ 
\end{longtable}
\end{small}
{\begin{flushleft}{
$^1$ y=BLCC (young cluster) from \cite{blcc} 0-- old cluster; 1-- color selected; 2-- $H_{\beta}$ selected; 
3-- color and $H_{\beta}$ selected;  4-- reportedly young objects by other authors and candidates BLCC (table 2)\\ 
}
\end{flushleft}}
{\begin{flushleft}{
$^{\star}$ B025D, B046D and B215D are classified by \cite{caldwell} to be not-clusters
and in the following analysis are not considered.} 
\end{flushleft}}

%%%%%%%%%%%%%%%%%%%%%%%%%%%%%%%%%%%%%%%%%%%%%%%%%%%%%%%%%%%%%%%%%%%%%%%%%%%%%%%%%%%%%%%%%%%%%%%

\onecolumn
%\clearpage
\begin{small}

\begin{longtable}{l r r r r r r r r r r r r r r }
  \caption{Lick indices M31 globular clusters taken from literature
sources and reported into the T98 system (Sect.~3.2). \label{all_index}}\\
 \hline\hline
Cluster&Mg$_2$&eMg$_2$&Mg$b$&eMg$b$&Fe5270&eFe5270&Fe5335&eFe5335&Fe5406&eFe5406&H$\beta$ &eH$\beta$ & y$^1$&Set$^2$\\
        & mag & mag  &  \AA  &  \AA  &  \AA  &  \AA  &   \AA   &  \AA  &  \AA &  \AA  & \AA &  \AA \\ 
\hline
\endfirsthead
\caption{continued.}\\
\hline\hline
 Cluster&Mg$_2$&eMg$_2$&Mg$b$&eMg$b$&Fe5270&eFe5270&Fe5335&eFe5335&Fe5406&eFe5406&H$\beta$ &eH$\beta$ & y$^1$&Set$^2$\\
        & mag & mag  &  \AA  &  \AA  &  \AA  &  \AA  &   \AA   &  \AA  &  \AA &  \AA  & \AA &  \AA \\ 
\hline
\endhead
\hline
\endfoot
     B001  &   0.160&  0.033&  2.485&  0.631&  2.366&  0.422&  2.753&  0.485& 99.999& 99.999&	2.17 &  0.30 & 0 &4\\
     B003  &   0.086&  0.013&  1.554&  0.547&  1.906&  0.648&  1.676&  0.774&  0.643&  0.603&	2.75 &  0.39 & 0 &3\\
     B004  &   0.081&  0.033&  2.094&  0.631&  0.772&  0.422&  1.846&  0.485& 99.999& 99.999&	3.16 &  0.30 & 0 &4\\
     B005  &   0.153&  0.033&  2.379&  0.631&  2.033&  0.422&  1.021&  0.485& 99.999& 99.999&	2.05 &  0.30 & 0 &4\\
     B006  &   0.144&  0.016&  2.449&  0.620&  2.060&  0.630&  2.010&  0.630&  1.163&  0.630&	2.00 &  0.56 & 0 &1\\
     B008  &   0.144&  0.033&  2.005&  0.631&  2.893&  0.422&  2.984&  0.485& 99.999& 99.999&	3.67 &  0.30 & 2 &4\\
     B009  &   0.060&  0.021&  1.246&  0.035&  1.160&  0.031& 99.999& 99.999& 99.999& 99.999&	3.31 &  0.03 & 0 &5\\
     B010  &   0.073&  0.033&  0.628&  0.631&  1.561&  0.422&  1.021&  0.485& 99.999& 99.999&	3.00 &  0.30 & 0 &4\\
     B011  &   0.046&  0.021&  1.081&  0.035&  1.649&  0.032& 99.999& 99.999& 99.999& 99.999&	1.79 &  0.04 & 0 &5\\
     B012  &   0.064&  0.004&  0.619&  0.188&  0.811&  0.223&  0.421&  0.266&  0.303&  0.202&	2.66 &  0.13 & 0 &3\\
     B013  &   0.185&  0.033&  1.085&  0.631&  4.366&  0.422&  3.050&  0.485& 99.999& 99.999&	2.89 &  0.30 & 0 &4\\
     B015  &   0.362&  0.010&  6.430&  0.292&  4.160&  0.275&  3.300&  0.317&  1.990&  0.237&	1.53 &  0.27 & 4 &0\\
     B016  &   0.130&  0.033&  2.952&  0.631&  1.899&  0.422&  1.778&  0.485& 99.999& 99.999&	1.72 &  0.30 & 0 &4\\
     B017  &   0.159&  0.007&  1.979&  0.291&  1.798&  0.338&  1.862&  0.393&  0.957&  0.300&	1.76 &  0.22 & 0 &3\\
     B018  &   0.090&  0.033&  1.733&  0.631&  1.999&  0.422&  2.453&  0.485& 99.999& 99.999&	2.40 &  0.30 & 7 &4\\
     B019  &   0.159&  0.005&  2.313&  0.187&  1.790&  0.222&  1.704&  0.259&  0.916&  0.199&	1.76 &  0.14 & 0 &3\\
     B020  &   0.120&  0.002&  1.962&  0.082&  1.929&  0.092&  1.710&  0.105&  1.406&  0.078&	1.98 &  0.07 & 0 &6\\
     B021  &   0.109&  0.033&  2.659&  0.631&  1.254&  0.422&  1.778&  0.485& 99.999& 99.999&	1.91 &  0.30 & 0 &4\\
     B022  &   0.061&  0.014&  1.339&  0.597&  1.371&  0.712&  1.109&  0.853&  0.209&  0.664&	3.12 &  0.42 & 0 &3\\
     B023  &   0.137&  0.004&  1.929&  0.171&  1.824&  0.186&  1.432&  0.211&  1.205&  0.154&	1.96 &  0.17 & 0 &7\\
     B024  &   0.163&  0.019&  2.741&  0.032&  1.997&  0.028& 99.999& 99.999& 99.999& 99.999&	1.22 &  0.03 & 0 &5\\
     B025  &   0.088&  0.019&  0.997&  0.870&  1.480&  0.880&  0.610&  0.880&  0.253&  0.880&	3.24 &  0.78 & 0 &1\\
     B026  &   0.213&  0.033&  4.461&  0.631&  1.832&  0.422&  2.620&  0.485& 99.999& 99.999&	1.48 &  0.30 & 0 &4\\
     B027  &   0.052&  0.014&  0.860&  0.022&  0.776&  0.020& 99.999& 99.999& 99.999& 99.999&	2.39 &  0.02 & 0 &5\\
     B028  &   0.092&  0.033&  1.347&  0.631&  1.764&  0.422& -0.395&  0.485& 99.999& 99.999&	3.98 &  0.30 & 2 &4\\
     B029  &   0.171&  0.033&  3.607&  0.631&  3.988&  0.422&  2.486&  0.485& 99.999& 99.999&	0.50 &  0.30 & 0 &4\\
     B030  &   0.228&  0.033&  2.694&  0.631&  3.797&  0.422&  3.083&  0.485& 99.999& 99.999&	1.79 &  0.30 & 4 &4\\
     B031  &   0.111&  0.033&  2.572&  0.631& -0.346&  0.422&  0.847&  0.485& 99.999& 99.999&	0.80 &  0.30 & 0 &4\\
     B032  &   0.210&  0.016&  4.270&  0.619&  2.693&  0.730&  2.805&  0.840&  1.221&  0.642&	1.76 &  0.53 & 0 &3\\
     B033  &   0.079&  0.033&  1.217&  0.631&  1.798&  0.422&  1.948&  0.485& 99.999& 99.999&	3.30 &  0.30 & 0 &4\\
     B034  &   0.122&  0.011&  1.830&  0.490&  1.490&  0.500&  1.680&  0.500&  0.863&  0.500&	2.18 &  0.47 & 0 &1\\
     B035  &   0.080&  0.033&  1.825&  0.631&  1.254&  0.422&  3.669&  0.485& 99.999& 99.999&	2.45 &  0.30 & 0 &4\\
     B037  &   0.100&  0.033&  1.825&  0.631&  3.477&  0.422&  1.914&  0.485& 99.999& 99.999&  99.99 &  9.99 & 0 &4\\
     B038  &   0.090&  0.033&  0.782&  0.631&  1.289&  0.422& -0.143&  0.485& 99.999& 99.999&	2.89 &  0.30 & 0 &4\\
     B039  &   0.176&  0.008&  2.630&  0.320&  1.920&  0.370&  1.742&  0.429&  0.880&  0.326&	1.43 &  0.26 & 0 &3\\
     B040  &   0.019&  0.033&  0.743&  0.631& -0.321&  0.422&  1.982&  0.485& 99.999& 99.999&	7.58 &  0.30 & 3 &4\\
     B041  &   0.047&  0.033&  1.217&  0.631&  1.595&  0.422&  1.982&  0.485& 99.999& 99.999&	2.73 &  0.30 & 0 &4\\
     B042  &   0.161&  0.008&  2.204&  0.325&  1.629&  0.374&  1.458&  0.432&  0.997&  0.323&	1.90 &  0.27 & 0 &3\\
     B043  &   0.040&  0.033&  0.512&  0.631&  0.807&  0.422&  0.567&  0.485& 99.999& 99.999&	5.70 &  0.30 & 3 &4\\
     B044  &   0.105&  0.031&  4.037&  0.049&  2.148&  0.046& 99.999& 99.999& 99.999& 99.999&  99.99 &  9.99 & 0 &5\\
     B045  &   0.138&  0.014&  1.417&  0.640&  2.290&  0.640&  1.530&  0.640&  1.323&  0.650&	2.34 &  0.56 & 0 &1\\
     B046  &   0.115&  0.039& -0.126&  0.065&  1.071&  0.061& 99.999& 99.999& 99.999& 99.999&	1.98 &  0.06 & 0 &5\\
     B047  &   0.207&  0.036&  2.427&  0.061& -2.290&  0.056& 99.999& 99.999& 99.999& 99.999&	2.69 &  0.06 & 2 &5\\
     B048  &   0.151&  0.019&  2.219&  0.860&  2.370&  0.860&  2.380&  0.880&  0.493&  0.880&	2.90 &  0.77 & 0 &1\\
     B049  &   0.052&  0.033&  1.272&  0.631&  1.629&  0.422& -0.793&  0.485& 99.999& 99.999&	9.48 &  0.30 & 3 &4\\
     B050  &   0.089&  0.033&  1.951&  0.631&  0.144&  0.422&  1.846&  0.485& 99.999& 99.999&	1.86 &  0.30 & 0 &4\\
     B051  &   0.170&  0.016&  1.983&  0.730&  2.530&  0.740&  1.630&  0.740&  1.393&  0.740&	1.80 &  0.67 & 0 &1\\
     B054  &   0.207&  0.033&  3.391&  0.631&  3.024&  0.422&  2.786&  0.485& 99.999& 99.999&	1.77 &  0.30 & 0 &4\\
     B055  &   0.161&  0.033&  3.408&  0.631&  1.561&  0.422&  2.719&  0.485& 99.999& 99.999&	2.80 &  0.30 & 0 &4\\
     B056  &   0.229&  0.033&  3.755&  0.631&  3.829&  0.422&  2.819&  0.485& 99.999& 99.999&	1.77 &  0.30 & 0 &4\\
     B057  &   0.076&  0.033&  1.291&  0.631&  2.795&  0.422& -0.539&  0.485& 99.999& 99.999&	5.73 &  0.30 & 2 &4\\
     B058  &   0.070&  0.009&  1.860&  0.270&  1.900&  0.254&  0.980&  0.275&  0.910&  0.220&	2.16 &  0.25 & 0 &0\\
     B059  &   0.144&  0.033&  3.391&  0.631&  1.865&  0.422&  0.496&  0.485& 99.999& 99.999&	1.21 &  0.30 & 0 &4\\
     B060  &   0.134&  0.012&  2.067&  0.518&  1.474&  0.622&  0.691&  0.744&  0.916&  0.560&	2.56 &  0.37 & 1 &3\\
     B061  &   0.187&  0.033&  2.728&  0.631&  2.598&  0.422&  1.436&  0.485& 99.999& 99.999&	3.12 &  0.30 & 0 &4\\
     B063  &   0.141&  0.020&  1.735&  0.032&  1.317&  0.029& 99.999& 99.999& 99.999& 99.999&	1.20 &  0.03 & 0 &5\\
     B064  &   0.077&  0.020&  0.357&  0.033&  1.802&  0.030& 99.999& 99.999& 99.999& 99.999&	1.77 &  0.03 & 0 &5\\
     B065  &   0.080&  0.033&  1.402&  0.631&  1.663&  0.422&  0.952&  0.485& 99.999& 99.999&	2.22 &  0.30 & 0 &4\\
     B066  &   0.055&  0.033&  0.991&  0.631&  0.319&  0.422& -0.467&  0.485& 99.999& 99.999&	4.84 &  0.30 & 3 &4\\
     B068  &   0.187&  0.024&  4.062&  0.041&  1.428&  0.034& 99.999& 99.999& 99.999& 99.999&	0.49 &  0.04 & 0 &5\\
     B069  &   0.135&  0.033&  1.384&  0.631&  1.730&  0.422&  2.386&  0.485& 99.999& 99.999&	7.34 &  0.30 & 3 &4\\
     B070  &   0.123&  0.010&  1.358&  0.406&  0.976&  0.481&  0.944&  0.564&  0.292&  0.436&	2.56 &  0.29 & 1 &3\\
     B071  &   0.275&  0.013&  4.838&  0.499&  2.710&  0.598&  2.279&  0.702&  1.725&  0.532&	2.06 &  0.38 & 0 &3\\
     B072  &   0.164&  0.033&  2.641&  0.631&  3.284&  0.422&  2.520&  0.485& 99.999& 99.999&	2.15 &  0.30 & 0 &4\\
     B073  &   0.207&  0.012&  3.623&  0.465&  2.767&  0.546&  2.572&  0.635&  1.235&  0.490&	2.24 &  0.35 & 0 &3\\
     B074  &   0.078&  0.033&  0.609&  0.631&  1.152&  0.422&  0.426&  0.485& 99.999& 99.999&	4.09 &  0.30 & 2 &4\\
     B075  &   0.048&  0.033&  1.697&  0.631& -0.784&  0.422&  2.620&  0.485& 99.999& 99.999&	2.17 &  0.30 & 0 &4\\
     B076  &   0.133&  0.033&  2.641&  0.631&  0.249&  0.422&  2.218&  0.485& 99.999& 99.999&	3.21 &  0.30 & 0 &4\\
     B081  &   0.036&  0.033&  0.648&  0.631&  0.945&  0.422&  1.778&  0.485& 99.999& 99.999&	8.15 &  0.30 & 3 &4\\
     B082  &   0.193&  0.012&  2.608&  0.467&  2.111&  0.526&  1.955&  0.604&  0.835&  0.459&	1.66 &  0.40 & 0 &3\\
     B083  &   0.037&  0.011&  0.787&  0.442&  0.977&  0.494&  0.677&  0.575&  0.826&  0.423&	1.75 &  0.42 & 2 &3\\
     B085  &   0.014&  0.021&  0.916&  0.034&  1.160&  0.031& 99.999& 99.999& 99.999& 99.999&	3.31 &  0.03 & 0 &5\\
     B086  &   0.038&  0.015& -0.443&  0.025&  1.183&  0.022& 99.999& 99.999& 99.999& 99.999&	2.55 &  0.03 & 0 &5\\
     B088  &   0.043&  0.033&  0.820&  0.631&  0.108&  0.422&  0.742&  0.485& 99.999& 99.999&	2.77 &  0.30 & 0 &4\\
     B090  &   0.218&  0.033&  3.308&  0.631&  3.381&  0.422&  1.982&  0.485& 99.999& 99.999&	3.45 &  0.30 & 4 &4\\
     B091  &   0.116&  0.033&  1.384&  0.631&  1.932&  0.422&  1.880&  0.485& 99.999& 99.999&	7.47 &  0.30 & 3 &4\\
     B092  &   0.100&  0.024&  1.843&  0.039&  0.410&  0.035& 99.999& 99.999& 99.999& 99.999&	1.79 &  0.04 & 0 &5\\
     B093  &   0.123&  0.033&  1.933&  0.631&  2.167&  0.422&  2.016&  0.485& 99.999& 99.999&	3.14 &  0.30 & 0 &4\\
     B094  &   0.131&  0.033&  2.797&  0.631&  2.598&  0.422&  2.386&  0.485& 99.999& 99.999&	2.31 &  0.30 & 0 &4\\
     B095  &   0.186&  0.017&  2.754&  0.710&  1.149&  0.846&  1.592&  0.973&  2.058&  0.700&	1.62 &  0.55 & 0 &3\\
     B096  &   0.215&  0.031&  2.974&  0.055&  3.055&  0.042& 99.999& 99.999& 99.999& 99.999&	1.44 &  0.05 & 0 &5\\
     B097  &   0.124&  0.033&  1.569&  0.631&  2.400&  0.422&  1.402&  0.485& 99.999& 99.999&	2.87 &  0.30 & 0 &4\\
     B098  &   0.182&  0.027&  4.602&  0.045&  1.649&  0.040& 99.999& 99.999& 99.999& 99.999&	1.53 &  0.04 & 0 &5\\
     B099  &   0.166&  0.010&  2.216&  0.412&  1.680&  0.485&  1.374&  0.571&  0.824&  0.437&	1.77 &  0.30 & 0 &3\\
     B103  &   0.184&  0.013&  3.333&  0.021&  1.250&  0.018& 99.999& 99.999& 99.999& 99.999&	1.56 &  0.02 & 0 &5\\
     B105  &   0.120&  0.033&  1.513&  0.631&  2.033&  0.422&  2.016&  0.485& 99.999& 99.999&	1.67 &  0.30 & 0 &4\\
     B106  &   0.135&  0.033&  0.943&  0.056&  1.932&  0.050& 99.999& 99.999& 99.999& 99.999&	0.67 &  0.05 & 0 &5\\
     B107  &   0.093&  0.014&  1.219&  0.023&  1.824&  0.020& 99.999& 99.999& 99.999& 99.999&	2.02 &  0.02 & 0 &5\\
     B109  &   0.250&  0.033&  2.711&  0.631&  2.828&  0.422&  2.719&  0.485& 99.999& 99.999&  99.99 &  9.99 & 0 &4\\
     B110  &   0.183&  0.009&  2.655&  0.359&  1.738&  0.424&  1.788&  0.491&  1.302&  0.370&	1.67 &  0.28 & 0 &3\\
     B111  &   0.144&  0.019&  2.342&  0.774&  1.845&  0.923&  1.084&  1.079&  0.804&  0.819&	1.63 &  0.60 & 0 &3\\
     B112  &   0.291&  0.038&  5.997&  0.063&  3.727&  0.054& 99.999& 99.999& 99.999& 99.999&	2.09 &  0.06 & 0 &5\\
     B115  &   0.273&  0.013&  3.862&  0.022&  1.976&  0.018& 99.999& 99.999& 99.999& 99.999&	0.12 &  0.02 & 0 &5\\
     B116  &   0.171&  0.033&  2.326&  0.631&  2.598&  0.422&  2.252&  0.485& 99.999& 99.999&	2.73 &  0.30 & 0 &4\\
     B117  &   0.067&  0.005&  0.897&  0.206&  0.630&  0.232&  0.621&  0.265&  0.253&  0.196&	2.09 &  0.20 & 4 &3\\
     B119  &   0.226&  0.033&  3.524&  0.631&  2.532&  0.422&  1.982&  0.485& 99.999& 99.999&	1.65 &  0.30 & 0 &4\\
     B122  &   0.171&  0.033&  1.679&  0.631&  2.100&  0.422&  0.249&  0.485& 99.999& 99.999&	2.61 &  0.30 & 0 &4\\
     B125  &   0.063&  0.033&  1.440&  0.631& -0.367&  0.422&  0.777&  0.485& 99.999& 99.999&	3.21 &  0.30 & 0 &4\\
     B126  &   0.046&  0.014&  1.280&  0.240&  1.020&  0.170&  0.810&  0.200&  0.470&  0.140&	3.65 &  0.14 & 0 &2\\
     B127  &   0.189&  0.004&  2.716&  0.210&  3.020&  0.210&  0.950&  0.210&  1.203&  0.210&	1.68 &  0.20 & 0 &1\\
     B129  &   0.183&  0.033&  2.449&  0.631&  3.024&  0.422&  0.532&  0.485& 99.999& 99.999&	2.73 &  0.30 & 0 &4\\
     B130  &   0.070&  0.033&  1.458&  0.631&  0.876&  0.422&  1.880&  0.485& 99.999& 99.999&	3.43 &  0.30 & 0 &4\\
     B131  &   0.279&  0.005&  4.067&  0.204&  2.389&  0.243&  2.118&  0.285&  1.485&  0.218&	1.60 &  0.15 & 0 &3\\
     B134  &   0.109&  0.014&  2.220&  0.240&  1.790&  0.170&  1.590&  0.200&  0.990&  0.150&	1.78 &  0.16 & 0 &2\\
     B135  &   0.076&  0.033&  1.160&  0.631&  1.323&  0.422&  0.777&  0.485& 99.999& 99.999&	2.26 &  0.30 & 0 &4\\
     B137  &   0.099&  0.033&  0.915&  0.631&  1.999&  0.422&  1.914&  0.485& 99.999& 99.999&	2.84 &  0.30 & 0 &4\\
     B140  &   0.247&  0.033&  3.706&  0.631&  3.251&  0.422&  0.812&  0.485& 99.999& 99.999&	0.18 &  0.30 & 0 &4\\
     B141  &   0.072&  0.033&  0.686&  0.631&  0.529&  0.422&  1.470&  0.485& 99.999& 99.999&	2.93 &  0.30 & 0 &4\\
     B143  &   0.241&  0.015&  4.136&  0.026&  2.256&  0.023& 99.999& 99.999& 99.999& 99.999&	1.53 &  0.03 & 0 &5\\
     B144  &   0.187&  0.011&  2.647&  0.470&  2.430&  0.470&  1.570&  0.480&  1.143&  0.480&	1.76 &  0.46 & 0 &1\\
     B146  &   0.171&  0.042&  5.672&  0.066&  3.985&  0.059& 99.999& 99.999& 99.999& 99.999&	0.20 &  0.07 & 4 &5\\
     B147  &   0.242&  0.002&  3.952&  0.082&  3.032&  0.091&  2.816&  0.103&  2.162&  0.076&	1.66 &  0.08 & 0 &6\\
     B148  &   0.145&  0.010&  2.151&  0.390&  1.960&  0.390&  2.040&  0.390&  1.263&  0.390&	2.01 &  0.37 & 0 &1\\
     B149  &   0.090&  0.033&  1.328&  0.631&  2.200&  0.422&  1.948&  0.485& 99.999& 99.999&	2.93 &  0.30 & 0 &4\\
     B151  &   0.199&  0.006&  3.179&  0.227&  1.981&  0.269&  1.879&  0.312&  1.347&  0.235&	1.78 &  0.18 & 0 &3\\
     B152  &   0.062&  0.026&  2.295&  0.042&  2.170&  0.039& 99.999& 99.999& 99.999& 99.999&	1.22 &  0.04 & 0 &5\\
     B153  &   0.247&  0.011&  4.057&  0.432&  2.427&  0.513&  2.228&  0.598&  1.860&  0.450&	1.58 &  0.34 & 0 &3\\
     B154  &   0.265&  0.029&  4.963&  0.047&  3.034&  0.044& 99.999& 99.999& 99.999& 99.999&	1.65 &  0.05 & 4 &5\\
     B155  &   0.212&  0.010&  2.732&  0.405&  3.711&  0.420&  3.685&  0.469&  1.221&  0.364&	3.27 &  0.38 & 0 &3\\
     B156  &   0.078&  0.009&  1.259&  0.376&  1.400&  0.410&  1.224&  0.465&  0.571&  0.345&	2.02 &  0.38 & 0 &3\\
     B158  &   0.130&  0.013&  2.270&  0.210&  1.860&  0.100&  1.700&  0.130&  1.060&  0.090&	1.74 &  0.10 & 0 &2\\
     B159  &   0.177&  0.033&  2.113&  0.631&  2.300&  0.422&  1.367&  0.485& 99.999& 99.999&	2.57 &  0.30 & 0 &4\\
     B161  &   0.180&  0.033&  2.237&  0.631&  2.167&  0.422&  1.436&  0.485& 99.999& 99.999&	1.98 &  0.30 & 0 &4\\
     B162  &   0.270&  0.015&  4.720&  0.602&  2.198&  0.738&  2.727&  0.843&  1.623&  0.646&	1.94 &  0.46 & 0 &3\\
     B163  &   0.222&  0.013&  4.010&  0.190&  2.600&  0.070&  2.440&  0.090&  1.570&  0.060&	1.74 &  0.07 & 0 &2\\
     B164  &   0.216&  0.033&  3.308&  0.631&  1.561&  0.422&  2.319&  0.485& 99.999& 99.999&	1.65 &  0.30 & 4 &4\\
     B165  &   0.050&  0.025& -0.328&  0.041&  1.693&  0.038& 99.999& 99.999& 99.999& 99.999&	2.51 &  0.04 & 0 &5\\
     B167  &   0.180&  0.033&  3.037&  0.631&  3.154&  0.422&  2.083&  0.485& 99.999& 99.999&	2.03 &  0.30 & 0 &4\\
     B169  &   0.280&  0.013&  5.389&  0.486&  2.657&  0.543&  1.977&  0.624&  2.404&  0.447&	1.68 &  0.52 & 0 &3\\
     B170  &   0.116&  0.033&  3.424&  0.631&  0.354&  0.422&  2.050&  0.485& 99.999& 99.999&	4.69 &  0.30 & 2 &4\\
     B171  &   0.189&  0.011&  3.110&  0.314&  2.340&  0.296&  2.400&  0.338&  1.550&  0.256&	2.27 &  0.29 & 0 &0\\
     B171  &   0.214&  0.002&  3.610&  0.091&  2.663&  0.102&  2.245&  0.116&  1.979&  0.085&	1.89 &  0.09 & 0 &6\\
     B174  &   0.103&  0.006&  1.950&  0.243&  1.417&  0.270&  1.194&  0.306&  0.892&  0.225&	1.92 &  0.25 & 0 &3\\
     B178  &   0.097&  0.009&  1.890&  0.270&  1.930&  0.254&  0.310&  0.229&  0.730&  0.220&	1.97 &  0.25 & 0 &0\\
     B179  &   0.116&  0.008&  1.933&  0.317&  1.408&  0.350&  1.520&  0.396&  0.648&  0.299&	1.13 &  0.32 & 0 &3\\
     B180  &   0.125&  0.006&  2.589&  0.216&  1.924&  0.244&  1.146&  0.283&  0.603&  0.215&	1.60 &  0.21 & 0 &3\\
     B182  &   0.076&  0.011&  1.700&  0.460&  1.740&  0.470&  1.660&  0.470&  0.803&  0.470&	2.39 &  0.42 & 0 &1\\
     B183  &   0.182&  0.006&  3.134&  0.221&  2.361&  0.244&  1.423&  0.283&  0.922&  0.211&	1.95 &  0.22 & 0 &3\\
     B184  &   0.219&  0.033&  4.398&  0.631&  2.532&  0.422&  2.553&  0.485& 99.999& 99.999&	1.79 &  0.30 & 0 &4\\
     B185  &   0.159&  0.007&  2.834&  0.289&  2.258&  0.316&  1.745&  0.358&  0.893&  0.261&	1.73 &  0.31 & 0 &3\\
     B187  &   0.123&  0.012&  1.370&  0.494&  0.892&  0.547&  0.653&  0.624&  0.447&  0.461&	2.55 &  0.45 & 0 &3\\
     B188  &   0.094&  0.033&  1.624&  0.631&  0.038&  0.422&  1.298&  0.485& 99.999& 99.999&	2.43 &  0.30 & 0 &4\\
     B190  &   0.094&  0.033&  1.915&  0.631&  2.066&  0.422&  1.812&  0.485& 99.999& 99.999&	2.52 &  0.30 & 0 &4\\
     B193  &   0.233&  0.004&  4.182&  0.167&  3.071&  0.183&  2.631&  0.207&  1.558&  0.155&	1.95 &  0.17 & 0 &3\\
     B197  &   0.234&  0.033&  3.998&  0.631&  2.991&  0.422&  2.786&  0.485& 99.999& 99.999&	1.31 &  0.30 & 4 &4\\
     B198  &   0.160&  0.033&  2.255&  0.631&  2.532&  0.422&  2.184&  0.485& 99.999& 99.999&	2.33 &  0.30 & 0 &4\\
     B199  &   0.068&  0.033&  0.877&  0.631&  0.494&  0.422&  1.091&  0.485& 99.999& 99.999&	3.16 &  0.30 & 0 &4\\
     B200  &   0.111&  0.033&  2.728&  0.631&  1.391&  0.422&  3.181&  0.485& 99.999& 99.999&	2.17 &  0.30 & 0 &4\\
     B201  &   0.106&  0.017&  2.216&  0.028&  2.019&  0.027& 99.999& 99.999& 99.999& 99.999&	1.88 &  0.03 & 0 &5\\
     B203  &   0.175&  0.033&  2.745&  0.631&  2.400&  0.422&  0.987&  0.485& 99.999& 99.999&	1.04 &  0.30 & 0 &4\\
     B204  &   0.139&  0.005&  2.642&  0.204&  2.256&  0.223&  2.733&  0.249&  1.415&  0.185&	1.97 &  0.21 & 0 &3\\
     B205  &   0.097&  0.008&  1.789&  0.013&  1.272&  0.012& 99.999& 99.999& 99.999& 99.999&	1.58 &  0.01 & 0 &5\\
     B206  &   0.082&  0.004&  1.494&  0.143&  1.516&  0.157&  1.293&  0.179&  0.837&  0.131&	2.26 &  0.14 & 0 &3\\
     B207  &   0.078&  0.033&  1.422&  0.631&  1.932&  0.422&  1.333&  0.485& 99.999& 99.999&	3.23 &  0.30 & 0 &4\\
     B208  &   0.220&  0.033&  2.606&  0.631&  3.638&  0.422&  1.914&  0.485& 99.999& 99.999&	3.21 &  0.30 & 0 &4\\
     B209  &   0.090&  0.033&  2.077&  0.631&  1.561&  0.422&  1.160&  0.485& 99.999& 99.999&	2.03 &  0.30 & 0 &4\\
     B210  &   0.052&  0.033&  0.782&  0.631&  1.186&  0.422&  1.229&  0.485& 99.999& 99.999&	7.00 &  0.30 & 3 &4\\
     B211  &   0.028&  0.024&  1.546&  0.039&  2.617&  0.037& 99.999& 99.999& 99.999& 99.999&	3.05 &  0.03 & 0 &5\\
     B212  &   0.050&  0.005&  1.009&  0.219&  0.183&  0.252&  0.245&  0.288&  0.358&  0.212&	2.46 &  0.21 & 0 &3\\
     B213  &   0.159&  0.033&  2.397&  0.631&  2.233&  0.422&  1.402&  0.485& 99.999& 99.999&	1.26 &  0.30 & 0 &4\\
     B214  &   0.071&  0.033&  1.217&  0.631&  1.932&  0.422&  2.586&  0.485& 99.999& 99.999&	3.41 &  0.30 & 4 &4\\
     B215  &   0.196&  0.007&  3.493&  0.288&  2.306&  0.320&  1.758&  0.364&  0.861&  0.273&	1.95 &  0.29 & 0 &3\\
     B217  &   0.095&  0.033&  2.184&  0.631&  1.595&  0.422&  1.607&  0.485& 99.999& 99.999&	1.98 &  0.30 & 0 &4\\
     B218  &   0.123&  0.009&  2.300&  0.261&  1.990&  0.245&  1.710&  0.276&  1.460&  0.213&	1.86 &  0.24 & 0 &0\\
     B219  &   0.157&  0.009&  3.211&  0.339&  2.070&  0.386&  1.241&  0.449&  1.057&  0.344&	2.44 &  0.32 & 0 &3\\
     B220  &   0.092&  0.033&  1.752&  0.631&  1.391&  0.422&  1.333&  0.485& 99.999& 99.999&	2.26 &  0.30 & 0 &4\\
     B221  &   0.135&  0.033&  1.897&  0.631&  2.532&  0.422&  1.229&  0.485& 99.999& 99.999&	2.08 &  0.30 & 0 &4\\
     B222  &   0.101&  0.015&  1.300&  0.390&  1.870&  0.370&  1.170&  0.430&  0.970&  0.310&	4.46 &  0.31 & 2 &2\\
     B224  &   0.042&  0.006&  0.909&  0.263&  0.448&  0.299&  1.162&  0.339&  0.177&  0.258&	2.20 &  0.24 & 0 &3\\
     B225  &   0.187&  0.013&  3.210&  0.190&  2.310&  0.070&  2.030&  0.090&  1.310&  0.060&	1.83 &  0.07 & 0 &2\\
     B228  &   0.129&  0.007&  2.104&  0.300&  1.759&  0.328&  1.473&  0.375&  1.594&  0.272&	2.21 &  0.29 & 0 &3\\
     B230  &   0.057&  0.007&  0.107&  0.286&  0.461&  0.315&  0.174&  0.363&  0.267&  0.268&	2.56 &  0.26 & 0 &3\\
     B231  &   0.102&  0.033&  2.077&  0.631&  1.629&  0.422&  1.710&  0.485& 99.999& 99.999&	2.59 &  0.30 & 0 &4\\
     B232  &   0.032&  0.005&  0.493&  0.214&  0.404&  0.236&  0.690&  0.267&  0.236&  0.197&	2.41 &  0.21 & 4 &3\\
     B233  &   0.061&  0.015&  0.554&  0.025&  1.736&  0.022& 99.999& 99.999& 99.999& 99.999&	2.16 &  0.03 & 0 &5\\
     B234  &   0.113&  0.014&  2.370&  0.240&  2.030&  0.180&  1.500&  0.210&  0.990&  0.150&	1.72 &  0.16 & 0 &2\\
     B235  &   0.133&  0.006&  2.380&  0.222&  1.690&  0.246&  1.773&  0.279&  0.973&  0.207&	1.99 &  0.22 & 0 &3\\
     B236  &   0.052&  0.012&  0.767&  0.486&  0.318&  0.534&  0.009&  0.613&  0.320&  0.441&	4.43 &  0.46 & 0 &3\\
     B237  &   0.070&  0.033& -0.357&  0.631&  3.251&  0.422&  0.952&  0.485& 99.999& 99.999&	7.60 &  0.30 & 2 &4\\
     B238  &   0.137&  0.006&  3.478&  0.231&  1.714&  0.267&  1.862&  0.301&  0.873&  0.225&	1.81 &  0.24 & 0 &3\\
     B239  &   0.068&  0.026&  1.219&  0.043&  1.183&  0.038& 99.999& 99.999& 99.999& 99.999&	1.67 &  0.04 & 0 &5\\
     B240  &   0.051&  0.007&  0.750&  0.204&  0.742&  0.190&  0.954&  0.208&  0.723&  0.168&	2.05 &  0.19 & 0 &0\\
     B272  &   0.154&  0.033&  2.130&  0.631&  2.300&  0.422&  1.160&  0.485& 99.999& 99.999&	2.05 &  0.30 & 0 &4\\
     B281  &   0.160&  0.033&  1.951&  0.631&  3.956&  0.422&  0.952&  0.485& 99.999& 99.999&	5.73 &  0.30 & 2 &4\\
     B283  &   0.191&  0.033&  3.475&  0.631& -0.140&  0.422&  3.312&  0.485& 99.999& 99.999&  99.99 &  9.99 & 0 &4\\
     B292  &   0.053&  0.016&  0.970&  0.400&  0.950&  0.390&  1.140&  0.450&  0.150&  0.340&	3.14 &  0.32 & 4 &2\\
     B293  &   0.057&  0.021&  0.860&  0.033&  1.093&  0.032& 99.999& 99.999& 99.999& 99.999&	3.55 &  0.03 & 0 &5\\
     B295  &   0.029&  0.033&  1.328&  0.631&  0.633&  0.422&  0.496&  0.485& 99.999& 99.999&	4.94 &  0.30 & 2 &4\\
     B298  &   0.040&  0.019& -0.356&  0.031&  0.981&  0.026& 99.999& 99.999& 99.999& 99.999&	2.30 &  0.03 & 0 &5\\
     B301  &   0.056&  0.015&  1.700&  0.380&  1.430&  0.370&  1.180&  0.420&  0.330&  0.310&	2.94 &  0.33 & 0 &2\\
     B303  &   0.119&  0.033&  1.733&  0.631&  1.561&  0.422& -0.071&  0.485& 99.999& 99.999&	5.95 &  0.30 & 3 &4\\
     B304  &   0.050&  0.015&  1.230&  0.370&  1.400&  0.360&  0.920&  0.420&  0.850&  0.300&	2.52 &  0.30 & 0 &2\\
     B305  &   0.052&  0.033&  2.005&  0.631& -0.892&  0.422&  3.475&  0.485& 99.999& 99.999&	2.77 &  0.30 & 0 &4\\
     B306  &   0.125&  0.033&  1.587&  0.631&  2.200&  0.422&  0.742&  0.485& 99.999& 99.999&	2.47 &  0.30 & 0 &4\\
     B307  &   0.053&  0.033&  2.659&  0.631&  1.798&  0.422&  3.050&  0.485& 99.999& 99.999&	5.93 &  0.30 & 2 &4\\
     B310  &   0.035&  0.015&  0.920&  0.360&  1.260&  0.350&  0.840&  0.400&  0.440&  0.290&	2.56 &  0.30 & 0 &2\\
     B311  &   0.049&  0.012&  0.798&  0.560&  1.100&  0.560&  0.610&  0.560&  0.403&  0.560&	2.80 &  0.52 & 4 &1\\
     B312  &   0.118&  0.010&  1.448&  0.450&  2.150&  0.450&  0.690&  0.450&  0.443&  0.450&	2.94 &  0.43 & 0 &1\\
     B313  &   0.120&  0.014&  2.170&  0.280&  1.690&  0.230&  1.450&  0.260&  0.970&  0.190&	1.51 &  0.21 & 0 &2\\
     B315  &   0.089&  0.011&  0.676&  0.470&  1.930&  0.480&  0.110&  0.500&  0.993&  0.510&	4.75 &  0.40 & 3 &1\\
     B316  &   0.151&  0.033&  2.237&  0.631&  2.433&  0.422&  0.917&  0.485& 99.999& 99.999&	2.64 &  0.30 & 7 &4\\
     B317  &   0.028&  0.030& -0.241&  0.048&  1.539&  0.046& 99.999& 99.999& 99.999& 99.999&	1.67 &  0.05 & 0 &5\\
     B318  &   0.027&  0.004&  0.112&  0.165&  0.586&  0.190&  0.231&  0.222&  0.601&  0.165&	5.49 &  0.12 & 1 &6\\
     B319  &   0.066&  0.033&  0.877&  0.631&  0.668&  0.422&  0.602&  0.485& 99.999& 99.999&	5.54 &  0.30 & 2 &4\\
     B321  &   0.032&  0.016&  0.940&  0.430&  0.800&  0.450&  0.730&  0.520&  0.200&  0.380&	6.85 &  0.32 & 3 &2\\
     B322  &   0.028&  0.015&  0.350&  0.340&  0.630&  0.330&  0.670&  0.380&  0.450&  0.280&	5.06 &  0.24 & 1 &2\\
     B324  &   0.065&  0.014&  1.570&  0.240&  1.660&  0.190&  1.460&  0.220&  0.710&  0.160&	4.69 &  0.14 & 4 &2\\
     B327  &   0.057&  0.014&  0.590&  0.250&  0.830&  0.190&  1.100&  0.220&  0.720&  0.160&	3.78 &  0.14 & 3 &2\\
     B328  &   0.048&  0.016&  0.190&  0.420&  0.890&  0.410&  0.490&  0.480&  0.400&  0.350&	2.58 &  0.35 & 4 &2\\
     B335  &   0.140&  0.033&  2.624&  0.631&  1.186&  0.422&  1.298&  0.485& 99.999& 99.999&	2.31 &  0.30 & 0 &4\\
     B337  &   0.064&  0.013&  1.860&  0.190&  1.480&  0.070&  1.110&  0.090&  0.640&  0.060&	3.23 &  0.07 & 0 &2\\
     B338  &   0.085&  0.009&  1.220&  0.260&  1.640&  0.245&  1.450&  0.273&  0.730&  0.213&	2.10 &  0.24 & 0 &0\\
     B341  &   0.123&  0.033&  2.041&  0.631&  1.254&  0.422&  1.607&  0.485& 99.999& 99.999&	2.05 &  0.30 & 0 &4\\
     B343  &   0.086&  0.015&  1.573&  0.024&  1.824&  0.022& 99.999& 99.999& 99.999& 99.999&	1.48 &  0.02 & 0 &5\\
     B344  &   0.109&  0.007&  1.669&  0.270&  2.161&  0.291&  2.320&  0.330&  1.114&  0.249&	1.88 &  0.26 & 0 &3\\
     B347  &   0.024&  0.014&  0.760&  0.260&  0.510&  0.210&  0.490&  0.240&  0.370&  0.170&	2.87 &  0.17 & 4 &2\\
     B348  &   0.136&  0.008&  2.169&  0.303&  2.221&  0.327&  1.443&  0.383&  1.061&  0.284&	2.55 &  0.28 & 0 &3\\
     B350  &   0.055&  0.015&  1.130&  0.340&  0.980&  0.320&  0.810&  0.370&  0.330&  0.270&	2.80 &  0.27 & 0 &2\\
     B352  &   0.119&  0.026&  2.558&  0.044&  0.201&  0.039& 99.999& 99.999& 99.999& 99.999&	2.82 &  0.04 & 0 &5\\
     B356  &   0.075&  0.008&  0.877&  0.325&  1.133&  0.353&  0.817&  0.405&  0.378&  0.297&	2.47 &  0.31 & 0 &3\\
     B357  &   0.127&  0.022&  1.191&  0.036&  1.780&  0.033& 99.999& 99.999& 99.999& 99.999&	2.12 &  0.03 & 0 &5\\
     B358  &   0.034&  0.007&  0.767&  0.207&  0.810&  0.194&  0.397&  0.187&  0.291&  0.171&	2.63 &  0.20 & 1 &0\\
     B365  &   0.060&  0.014&  1.370&  0.260&  1.330&  0.200&  1.000&  0.240&  0.500&  0.180&	2.72 &  0.17 & 0 &2\\
     B366  &   0.015&  0.033&  0.763&  0.631& -0.339&  0.422&  0.742&  0.485& 99.999& 99.999&	3.07 &  0.30 & 0 &4\\
     B367  &   0.050&  0.033&  0.260&  0.631&  2.466&  0.422&  1.021&  0.485& 99.999& 99.999&	6.38 &  0.30 & 3 &4\\
     B370  &   0.050&  0.015&  0.913&  0.730& -0.080&  0.730&  0.730&  0.740&  0.543&  0.740&	2.71 &  0.71 & 0 &1\\
     B372  &   0.117&  0.016&  1.379&  0.660&  1.930&  0.670&  1.610&  0.670&  0.723&  0.670&	2.36 &  0.63 & 0 &1\\
     B373  &   0.167&  0.016&  2.451&  0.619&  2.169&  0.832&  1.940&  0.810&  1.435&  0.611&	1.58 &  0.53 & 0 &3\\
     B374  &   0.094&  0.033&  1.402&  0.631&  1.932&  0.422&  1.744&  0.485& 99.999& 99.999&	4.24 &  0.30 & 3 &4\\
     B375  &   0.128&  0.025&  0.888&  0.044&  1.539&  0.037& 99.999& 99.999& 99.999& 99.999&	2.35 &  0.04 & 0 &5\\
     B376  &   0.074&  0.038&  2.348&  0.062&  0.799&  0.060& 99.999& 99.999& 99.999& 99.999&	6.40 &  0.06 & 1 &5\\
     B377  &   0.060&  0.030& -1.503&  0.051&  1.160&  0.047& 99.999& 99.999& 99.999& 99.999&	2.07 &  0.05 & 0 &5\\
     B378  &   0.068&  0.033&  0.972&  0.631&  1.357&  0.422&  1.539&  0.485& 99.999& 99.999&	3.09 &  0.30 & 0 &4\\
     B379  &   0.171&  0.020&  1.654&  0.033&  1.516&  0.026& 99.999& 99.999& 99.999& 99.999&	1.48 &  0.03 & 0 &5\\
     B381  &   0.075&  0.006&  1.372&  0.257&  1.766&  0.283&  1.650&  0.323&  0.722&  0.242&	1.70 &  0.25 & 0 &3\\
     B382  &   0.046&  0.033&  1.532&  0.631&  1.152&  0.422&  1.607&  0.485& 99.999& 99.999&	3.09 &  0.30 & 0 &4\\
     B383  &   0.163&  0.013&  3.030&  0.220&  2.120&  0.140&  1.790&  0.170&  1.210&  0.120&	1.75 &  0.13 & 0 &2\\
     B384  &   0.163&  0.014&  1.681&  0.024&  2.084&  0.020& 99.999& 99.999& 99.999& 99.999&	1.20 &  0.02 & 0 &5\\
     B386  &   0.105&  0.013&  1.410&  0.021&  2.019&  0.019& 99.999& 99.999& 99.999& 99.999&	2.07 &  0.02 & 0 &5\\
     B387  &   0.077&  0.020&  0.749&  0.033&  0.686&  0.030& 99.999& 99.999& 99.999& 99.999&	3.44 &  0.03 & 0 &5\\
     B391  &   0.077&  0.033&  1.179&  0.631&  1.323&  0.422&  3.637&  0.485& 99.999& 99.999&	2.75 &  0.30 & 0 &4\\
     B393  &   0.102&  0.014&  1.490&  0.320&  1.930&  0.280&  1.580&  0.330&  0.700&  0.240&	1.90 &  0.24 & 0 &2\\
     B397  &   0.125&  0.025&  1.519&  0.042&  0.868&  0.037& 99.999& 99.999& 99.999& 99.999&	2.30 &  0.04 & 0 &5\\
     B398  &   0.162&  0.014&  3.170&  0.300&  2.310&  0.270&  1.710&  0.310&  1.070&  0.230&	1.60 &  0.24 & 0 &2\\
     B399  &   0.043&  0.004&  0.817&  0.178&  0.854&  0.201&  1.164&  0.229&  0.752&  0.170&	2.89 &  0.16 & 0 &6\\
     B400  &   0.103&  0.033&  1.197&  0.631&  3.348&  0.422& -0.215&  0.485& 99.999& 99.999&	0.65 &  0.30 & 0 &4\\
     B401  &   0.027&  0.014&  0.530&  0.270&  0.630&  0.230&  0.530&  0.270& -0.030&  0.200&	2.84 &  0.19 & 0 &2\\
     B403  &   0.208&  0.057&  3.637&  0.097&  3.985&  0.073& 99.999& 99.999& 99.999& 99.999&  99.99 &  9.99 & 0 &5\\
     B405  &   0.086&  0.009&  1.274&  0.015&  0.663&  0.021& 99.999& 99.999& 99.999& 99.999&	2.14 &  0.01 & 0 &5\\
     B407  &   0.155&  0.017&  2.427&  0.028&  1.317&  0.025& 99.999& 99.999& 99.999& 99.999&	0.59 &  0.03 & 0 &5\\
     B431  &   0.066&  0.030& -1.234&  0.052&  1.714&  0.044& 99.999& 99.999& 99.999& 99.999&	1.74 &  0.04 & 1 &5\\
     B448  &   0.087&  0.033&  0.915&  0.631&  1.083&  0.422& -0.503&  0.485& 99.999& 99.999&	6.87 &  0.30 & 2 &4\\
     B457  &   0.265&  0.000&  4.029&  0.013&  3.851&  0.014&  3.496&  0.016&  2.891&  0.012&	1.44 &  0.01 & 0 &6\\
     B458  &   0.102&  0.033&  0.820&  0.631&  2.333&  0.422&  2.419&  0.485& 99.999& 99.999&	6.36 &  0.30 & 3 &4\\
     B467  &   0.074&  0.033&  0.338&  0.631& -0.033&  0.422&  0.391&  0.485& 99.999& 99.999&	2.33 &  0.30 & 0 &4\\
     B468  &   0.113&  0.007&  2.583&  0.277&  1.472&  0.320&  1.095&  0.367&  0.657&  0.274&	2.50 &  0.25 & 4 &6\\
     B472  &   0.080&  0.004&  3.214&  0.142&  1.378&  0.168&  1.266&  0.192&  0.514&  0.143&	2.26 &  0.15 & 0 &3\\
     B475  &   0.109&  0.033&  0.279&  0.631&  1.083&  0.422&  0.496&  0.485& 99.999& 99.999&	6.13 &  0.30 & 3 &4\\
     B480  &   0.035&  0.033&  1.123&  0.631&  1.697&  0.422&  1.982&  0.485& 99.999& 99.999&	5.36 &  0.30 & 2 &4\\
     B483  &   0.089&  0.033&  0.004&  0.631&  1.014&  0.422& -1.270&  0.485& 99.999& 99.999&	5.75 &  0.30 & 3 &4\\
     B484  &   0.044&  0.033&  1.235&  0.631&  0.980&  0.422&  1.539&  0.485& 99.999& 99.999&	5.87 &  0.30 & 3 &4\\
     B486  &   0.029&  0.038& -1.473&  0.064& -0.057&  0.057& 99.999& 99.999& 99.999& 99.999&	3.22 &  0.06 & 1 &5\\
  G001     &   0.133&  0.003&  2.187&  0.104&  1.866&  0.115&  1.915&  0.131&  0.936&  0.098&	2.37 &  0.10 & 0 &7\\
     G002  &   0.053&  0.016& -0.155&  0.026&  1.138&  0.023& 99.999& 99.999& 99.999& 99.999&	2.12 &  0.03 & 0 &5\\
    B189D  &   0.079&  0.033&  1.217&  0.631&  0.807&  0.422&  2.252&  0.485& 99.999& 99.999&	4.41 &  0.30 & 1 &4\\
    B020D  &   0.092&  0.016&  1.713&  0.653&  0.340&  0.805&  0.897&  0.917&  0.602&  0.696&	3.25 &  0.49 & 0 &3\\
    B103D  &   0.193&  0.033&  3.291&  0.631&  2.631&  0.422&  2.386&  0.485& 99.999& 99.999&	1.74 &  0.30 & 0 &4\\
     G327  &   0.053&  0.033&  0.934&  0.631&  0.876&  0.422&  0.320&  0.485& 99.999& 99.999&	3.00 &  0.30 & 0 &4\\
     VDB0  &   0.031&  0.002&  0.186&  0.088&  0.598&  0.101&  0.568&  0.116&  0.366&  0.087&	4.50 &  0.07 & 1 &6\\
     NB16  &   0.066&  0.013&  1.610&  0.200&  1.180&  0.090&  0.970&  0.110&  0.500&  0.080&	3.34 &  0.08 & 4 &2\\
     NB89  &   0.123&  0.013&  2.430&  0.200&  1.910&  0.090&  1.630&  0.110&  1.020&  0.070&	2.04 &  0.09 & 0 &2\\
    B012D  &   0.076&  0.033&  1.197&  0.631&  1.899&  0.422&  2.016&  0.485& 99.999& 99.999&	7.27 &  0.30 & 2 &4\\
    B025D$^{\ddagger}$  &   0.250&  0.024&  4.182&  0.955&  1.875&  1.094&  2.463&  1.258&  0.035&  1.008&	0.11 &  0.86 & 0 &3\\
    B026D$^{\ddagger}$  &   0.185&  0.033&  2.467&  0.631&  1.764&  0.422&  1.160&  0.485& 99.999& 99.999&	0.02 &  0.30 & 0 &4\\
    B041D  &   0.126&  0.020&  0.685&  0.851&  1.416&  0.945&  1.897&  1.075&  1.019&  0.817&	2.00 &  0.65 & 0 &3\\
    B043D$^{\ddagger}$  &   0.100&  0.033&  1.661&  0.631&  1.697&  0.422&  0.532&  0.485& 99.999& 99.999&	0.04 &  0.30 & 0 &4\\
    B046D$^{\ddagger}$  &   0.230&  0.026&  3.245&  1.091&  2.712&  1.255&  3.013&  1.454&  1.154&  1.160&	2.15 &  0.76 & 0 &3\\
    B090D  &   0.291&  0.007&  4.122&  0.298&  2.604&  0.349&  2.241&  0.408&  1.732&  0.310&	1.54 &  0.23 & 0 &3\\
    B091D  &   0.112&  0.033&  2.431&  0.631&  1.289&  0.422&  2.117&  0.485& 99.999& 99.999&	1.98 &  0.30 & 0 &4\\
    B111D  &   0.070&  0.033&  1.495&  0.631&  1.014&  0.422&  1.160&  0.485& 99.999& 99.999&	5.73 &  0.30 & 2 &4\\
    B215D$^{\ddagger}$  &   0.187&  0.009&  2.449&  0.392&  1.676&  0.463&  1.450&  0.547&  1.061&  0.414&	1.82 &  0.28 & 0 &3\\
    B240D  &   0.043&  0.033&  1.624&  0.631&  0.179&  0.422&  0.602&  0.485& 99.999& 99.999&	2.01 &  0.30 & 7 &4\\
    B248D$^{\ddagger}$  &   0.286&  0.033&  3.558&  0.631&  3.316&  0.422&  4.468&  0.485& 99.999& 99.999&	0.04 &  0.30 & 0 &4\\
    B257D  &   0.040&  0.033&  0.896&  0.631& -1.365&  0.422&  0.036&  0.485& 99.999& 99.999&	5.66 &  0.30 & 2 &4\\
    B289D  &   0.143&  0.033&  2.624&  0.631&  2.565&  0.422& -1.270&  0.485& 99.999& 99.999&	3.54 &  0.30 & 0 &4\\
    B292D  &   0.229&  0.033&  3.934&  0.631&  3.381&  0.422&  0.952&  0.485& 99.999& 99.999&	2.15 &  0.30 & 0 &4\\
   B344D   &   0.121&  0.001&  2.517&  0.022&  2.054&  0.024&  1.679&  0.028&  1.372&  0.021&	2.66 &  0.02 & 0 &6\\
    DAO25$^{\ddagger}$  &   0.060&  0.033&  2.290&  0.631&  1.289&  0.422& -1.122&  0.485& 99.999& 99.999&	1.79 &  0.30 & 0 &4\\
    DAO30  &   0.151&  0.033&  2.273&  0.631&  3.734&  0.422&  3.148&  0.485& 99.999& 99.999&	3.59 &  0.30 & 7 &4\\
    DAO47  &   0.044&  0.033&  1.179&  0.631& -0.033&  0.422&  3.115&  0.485& 99.999& 99.999&	4.20 &  0.30 & 2 &4\\
     V031  &   0.052&  0.033&  1.366&  0.631&  0.319&  0.422&  0.812&  0.485& 99.999& 99.999&	6.01 &  0.30 & 2 &4\\
     BA11  &   0.120&  0.033&  3.054&  0.631&  1.697&  0.422&  0.672&  0.485& 99.999& 99.999&	1.07 &  0.30 & 0 &4\\
  B514     &   0.062&  0.003&  0.300&  0.137&  0.176&  0.154&  1.279&  0.169&  0.282&  0.159&	2.32 &  0.13 & 0 &7\\
  MCGC1    &   0.041&  0.007&  0.566&  0.290&  0.489&  0.327&  0.005&  0.379&  0.397&  0.276&	1.84 &  0.29 & 0 &7\\
   MCGC8   &   0.093&  0.003&  1.334&  0.115&  1.400&  0.128&  1.206&  0.147&  1.041&  0.109&	1.98 &  0.10 & 0 &6\\
   MCGC10  &   0.031&  0.003&  0.395&  0.104&  0.633&  0.118&  0.427&  0.136&  0.677&  0.100&	2.93 &  0.09 & 0 &6\\
\end{longtable}
\end{small}
{\begin{flushleft}{
$^1$ y=BLCC (young cluster) from \cite{blcc} 0-- old cluster; 
1-- color selected; 2-- $H_{\beta}$ selected; 
3-- color and $H_{\beta}$ selected;  
4-- reportedly young objects by other authors and candidates BLCC (table 2) \\ 
7-- classified young by \cite{caldwell}}
\end{flushleft}}
{\begin{flushleft}{
$^2$ Dataset label: 0-- \cite{trager}, 1-- \cite{puziam31}, 2-- \cite{beas},
3-- WHT data, 4-- \cite{perr_cat}, \\
5-- \cite{huchra91}, 6-- TNG data,
7-- LOI data
}
\end{flushleft}}
{\begin{flushleft}{
$^{\ddagger}$ B025D, B026D, B043D, B046D, B215D, B248D and DAO25 are 
classified by \cite{caldwell} to be not-clusters
and in the following analysis are not considered.} 
\end{flushleft}
}

%%%%%%%%%%%%%%%%%%%%%%%%%%%%%%%%%%%%%%%%%%%%%%%%%%%%%%%%%%%%%%%%%%%%%%%%%%%%%%%%%%%%%%%%%%%%%%%

\onecolumn
\clearpage
\begin{small}
\begin{table}[!t]
\caption{Metallicities for M31 globular clusters.  }
\label{meta}
{
\resizebox{\columnwidth}{!}{%
\begin{tabular}{l r r||l r r||l r r ||l r r}
         \hline\hline
Cluster&  [Fe/H]& e[Fe/H]&Cluster& [Fe/H]& e[Fe/H]&Cluster& [Fe/H]& e[Fe/H]&Cluster&  [Fe/H]& e[Fe/H]\\
       &   dex   & dex    &	     & dex   & dex     &        & dex   & dex     &	   &       dex   & dex     \\ 
\hline\hline
    B001  & -0.42 &  0.32  & B085  & -2.10 &  0.26 &  B183  & -0.47 &  0.15&  B344  & -0.80 &  0.21  \\ 
    B003  & -0.99 &  0.48  & B086  & -1.80 &  0.18 &  B184  & -0.01 &  0.22&  B347  & -1.91 &  0.24  \\ 
    B004  & -1.00 &  0.41  & B088  & -1.94 &  0.52 &  B185  & -0.50 &  0.20&  B348  & -0.75 &  0.23  \\ 
    B005  & -0.82 &  0.38  & B090  & -0.17 &  0.26 &  B187  & -1.52 &  0.54&  B350  & -1.54 &  0.31  \\ 
    B006  & -0.59 &  0.41  & B092  & -1.14 &  0.23 &  B188  & -1.51 &  0.51&  B352  & -0.96 &  0.24  \\ 
    B008  & -0.47 &  0.35  & B093  & -0.74 &  0.38 &  B190  & -0.80 &  0.39&  B356  & -1.62 &  0.32  \\ 
    B009  & -1.55 &  0.23  & B094  & -0.35 &  0.30 &  B193  &  0.04 &  0.15&  B357  & -0.88 &  0.20  \\ 
    B010  & -1.64 &  0.68  & B095  & -0.79 &  0.66 &  B197  &  0.02 &  0.21&  B358  & -1.85 &  0.19  \\ 
    B011  & -1.71 &  0.24  & B096  & -0.23 &  0.19 &  B198  & -0.55 &  0.34&  B365  & -1.32 &  0.20  \\ 
    B012  & -1.91 &  0.21  & B097  & -0.95 &  0.42 &  B199  & -1.70 &  0.51&  B366  & -2.14 &  0.39  \\ 
    B013  & -0.74 &  0.51  & B098  & -0.45 &  0.19 &  B200  & -0.43 &  0.32&  B370  & -1.98 &  0.50  \\ 
    B015  &  0.37 &  0.15  & B099  & -0.86 &  0.36 &  B201  & -1.08 &  0.16&  B372  & -1.07 &  0.53  \\ 
    B016  & -0.53 &  0.34  & B103  & -0.43 &  0.15 &  B203  & -0.64 &  0.36&  B373  & -0.59 &  0.47  \\ 
    B017  & -0.82 &  0.24  & B105  & -0.93 &  0.42 &  B204  & -0.39 &  0.15&  B375  & -0.87 &  0.22  \\ 
    B018  & -0.77 &  0.39  & B106  & -0.81 &  0.28 &  B205  & -1.16 &  0.15&  B377  & -1.55 &  0.33  \\ 
    B019  & -0.74 &  0.15  & B107  & -1.20 &  0.15 &  B206  & -1.16 &  0.15&  B378  & -1.38 &  0.51  \\ 
    B020  & -0.83 &  0.15  & B110  & -0.64 &  0.28 &  B207  & -1.10 &  0.44&  B379  & -0.53 &  0.15  \\ 
    B021  & -0.74 &  0.37  & B111  & -0.85 &  0.71 &  B208  & -0.32 &  0.30&  B381  & -1.10 &  0.22  \\ 
    B022  & -1.30 &  0.59  & B112  &  0.13 &  0.15 &  B209  & -0.98 &  0.41&  B382  & -1.16 &  0.44  \\ 
    B023  & -0.91 &  0.15  & B115  &  0.06 &  0.15 &  B211  & -1.92 &  0.29&  B383  & -0.47 &  0.15  \\ 
    B024  & -0.59 &  0.15  & B116  & -0.50 &  0.34 &  B212  & -2.07 &  0.28&  B384  & -0.59 &  0.15  \\ 
    B025  & -1.53 &  0.79  & B117  & -1.78 &  0.23 &  B213  & -0.69 &  0.36&  B386  & -1.09 &  0.15  \\ 
    B026  & -0.09 &  0.25  & B119  & -0.25 &  0.28 &  B214  & -1.00 &  0.47&  B387  & -1.37 &  0.21  \\ 
    B027  & -1.64 &  0.16  & B122  & -1.20 &  0.44 &  B215  & -0.33 &  0.19&  B391  & -0.96 &  0.48  \\ 
    B029  &  0.02 &  0.21  & B125  & -1.99 &  0.34 &  B217  & -0.84 &  0.39&  B393  & -1.03 &  0.24  \\ 
    B030  & -0.14 &  0.26  & B126  & -1.48 &  0.19 &  B218  & -0.71 &  0.18&  B397  & -0.90 &  0.22  \\ 
    B031  & -1.73 &  0.39  & B127  & -0.54 &  0.15 &  B219  & -0.55 &  0.26&  B398  & -0.41 &  0.18  \\ 
    B032  &  0.03 &  0.29  & B129  & -0.69 &  0.36 &  B220  & -1.09 &  0.42&  B399  & -1.63 &  0.18  \\ 
    B033  & -1.12 &  0.47  & B130  & -1.19 &  0.44 &  B221  & -0.83 &  0.39&  B400  & -1.23 &  0.47  \\ 
    B034  & -0.96 &  0.38  & B131  & -0.15 &  0.15 &  B224  & -1.68 &  0.28&  B401  & -1.98 &  0.26\\ 
    B035  & -0.67 &  0.38  & B134  & -0.79 &  0.15 &  B225  & -0.35 &  0.15&  B403  & -0.27 &  0.35\\ 
    B037  & -0.60 &  0.37  & B135  & -1.46 &  0.48 &  B228  & -0.86 &  0.24&  B405  & -1.28 &  0.15\\ 
    B038  & -1.86 &  0.52  & B137  & -1.26 &  0.54 &  B230  & -2.36 &  0.24&  B407  & -0.65 &  0.15\\ 
    B039  & -0.62 &  0.24  & B140  & -0.29 &  0.29 &  B231  & -0.85 &  0.39&  B431$^{\ast}$  & -1.49 &  0.33\\ 
    B041  & -1.14 &  0.47  & B141  & -1.71 &  0.58 &  B232  & -2.01 &  0.23&   B457  &  0.17 &  0.15 \\ 
    B042  & -0.86 &  0.27  & B143  & -0.09 &  0.15 &  B233  & -1.54 &  0.17&   B467  & -2.29 &  0.25 \\  
    B044  & -1.09 &  0.30  & B144  & -0.55 &  0.30 &  B234  & -0.72 &  0.15&   B468  & -0.88 &  0.25 \\  
    B045  & -1.01 &  0.50  & B146  & -0.53 &  0.31 &  B235  & -0.73 &  0.17&   B472  & -0.71 &  0.15 \\
    B046  & -0.99 &  0.36  & B147  &  0.02 &  0.15 &  B238  & -0.43 &  0.17&   B486  & -1.91 &  0.46 \\  
    B047  & -0.28 &  0.22  & B148  & -0.70 &  0.27 &  B239  & -1.46 &  0.28&   B514  & -2.06 &  0.16 \\ 
    B048  & -0.55 &  0.55  & B149  & -1.00 &  0.45 &  B240  & -1.74 &  0.19&   G001  & -0.73 &  0.15 \\  
    B050  & -1.21 &  0.45  & B151  & -0.44 &  0.16 &  B272  & -0.81 &  0.38&   G002  & -1.63 &  0.18  \\  
    B051  & -0.73 &  0.51  & B152  & -1.53 &  0.29 &  B283  & -0.52 &  0.34&   G327  & -1.79 &  0.52 \\  
    B054  & -0.10 &  0.24  & B153  & -0.13 &  0.25 &  B292  & -1.54 &  0.37&   NB16  & -1.27 &  0.15   \\  
    B055  & -0.31 &  0.29  & B154  &  0.03 &  0.15 &  B293  & -1.59 &  0.23&   NB89  & -0.70 &  0.15 \\ 
    B056  &  0.07 &  0.20  & B155  & -0.08 &  0.19 &  B298  & -1.78 &  0.22&  B020D  & -1.52 &  0.59\\ 
    B058  & -1.02 &  0.21  & B156  & -1.30 &  0.34 &  B301  & -1.13 &  0.32&  B041D  & -1.49 &  0.74\\ 
    B059  & -0.75 &  0.40  & B158  & -0.74 &  0.15 &  B304  & -1.38 &  0.33&  B090D  & -0.09 &  0.16  \\ 
    B060  & -1.13 &  0.55  & B159  & -0.77 &  0.38 &  B305  & -1.04 &  0.42&  B091D  & -0.73 &  0.37  \\ 
    B061  & -0.52 &  0.33  & B161  & -0.74 &  0.37 &  B306  & -1.10 &  0.43&  B103D  & -0.22 &  0.27  \\ 
    B063  & -0.76 &  0.17  & B162  &  0.02 &  0.31 &  B310  & -1.57 &  0.34&B240D$^{\ast}$  & -1.74 &  0.53   \\ 
    B064  & -1.37 &  0.21  & B163  & -0.08 &  0.15 &  B311  & -1.71 &  0.53& B289D  & -1.28 &  0.54  \\ 
    B065  & -1.24 &  0.45  & B164  & -0.41 &  0.31 &  B312  & -1.18 &  0.37& B292D  & -0.20 &  0.27  \\ 
    B068  & -0.41 &  0.17  & B165  & -1.66 &  0.29 &  B313  & -0.86 &  0.19& B344D  & -0.64 &  0.15  \\ 
    B070  & -1.42 &  0.43  & B167  & -0.25 &  0.28 &  B316  & -0.79 &  0.38&DAO30$^{\ast}$  & -0.26 &  0.30  \\ 
    B071  &  0.05 &  0.24  & B169  &  0.07 &  0.23 &  B317  & -1.92 &  0.36&   BA11  & -0.82 &  0.40  \\ 
    B072  & -0.28 &  0.29  & B171  & -0.17 &  0.15 &  B328  & -2.17 &  0.30&  MCGC1  & -2.16 &  0.28  \\ 
    B073  & -0.11 &  0.25  & B174  & -1.05 &  0.22 &  B335  & -0.89 &  0.40&  MCGC8  & -1.27 &  0.15 \\ 
    B075  & -1.33 &  0.46  & B178  & -1.16 &  0.21 &  B337  & -1.08 &  0.15& MCGC10  & -2.07 &  0.15  \\ 
    B076  & -0.89 &  0.40  & B179  & -0.98 &  0.27 &  B338  & -1.23 &  0.22&	     &       &  \\ 
    B082  & -0.55 &  0.33  & B180  & -0.75 &  0.18 &  B341  & -0.96 &  0.40&	     &       &  \\
    B083  & -1.73 &  0.46  & B182  & -0.97 &  0.35 &  B343  & -1.28 &  0.15&	     &       &       \\ 
         \hline\hline	     
\end{tabular}					     
}}	
{	
\begin{flushleft}
$^{\ast}$: classified young by \cite{caldwell}.
\end{flushleft}
}										    
\end{table} 										    
\end{small}

%%%%%%%%%%%%%%%%%%%%%%%%%%%%%%%%%%%%%%%%%%%%%%%%%%%%%%%%%%%%%%%%%%%%%%%%%%%%%%%%%%%%%%%%%%%%%%%

\clearpage

%%%%%%%%%%%%%%%%%%%%%%%%%%%%%%%%%%%%%%%%%%%%%%%%%%%%%%%%%%%%%%%%%%%%%%%%%%%%%%%%%%%%%%%%%%%%%%%
\section{Spearman correlation matrix for Lick indices and metallicity \label{correlazione}}

During the first phases of the present study we considered the largest
set of Lick indices to search for the best candidates to be used as
metallicity indicators for our scale. As a useful tool for this choice
we computed the matrix of Spearman rank correlation coefficients  
(\cite{spearman}) for
a set of indices in the T98 system and metallicity for Galactic GCs.
Since this matrix can be of general interest, we present it here as
Table~\ref{corcoef} below. It must be recalled that these correlations
coefficients are computed using a sample of nearly-uniformly {\em old
stellar populations}, hence it refers only to classical old globulars. 

%%%%%%%%%%%%%%%%%%%%%%%%%%%%%%%%%%%%%%%%%%%%%%%%%%%%%%%%%%%%%%%%%%%%%%%%%%%%%%%%%%%%%%%%%%%%%%%
%\onecolumn
\begin{small}
\begin{table}[!h]
  \caption{ Spearman rank correlation coefficients. }

\label{corcoef}
	{\tiny
\resizebox{\columnwidth}{!}{%
\begin{tabular}{lrrrrrrrrrrrrrrrrrrr} 
         \hline\hline
         \noalign{\smallskip}
       &$[Fe/H]_{ZW}$&    CN1&   CN2& Ca4227&  G4300& Ca4455& Fe4531& C24668&H$\beta$&Fe5015&    Mg1&    Mg2&   Mgb&   Fe5270& Fe5335& Fe5709&  NaD&TiO1   &   MgFe\\
\noalign{\smallskip}
\hline
\noalign{\smallskip}
$[Fe/H]_{ZW}$&  1.000&  0.711& 0.620&  0.598&  0.885&  0.684&  0.843&  0.331& -0.875&  0.902&  0.912&  0.958&  0.858&  0.860&  0.703&  0.669&  0.772&  0.613&  0.922\\
CN1        &  0.711&  1.000& 0.961&  0.760&  0.784&  0.537&  0.375&  0.632& -0.676&  0.777&  0.703&  0.672&  0.529&  0.527&  0.765&  0.505&  0.627&  0.618&  0.652\\
CN2        &  0.620&  0.961& 1.000&  0.718&  0.699&  0.566&  0.233&  0.647& -0.554&  0.694&  0.618&  0.556&  0.426&  0.424&  0.681&  0.392&  0.537&  0.613&  0.532\\
Ca4227     &  0.598&  0.760& 0.718&  1.000&  0.630&  0.387&  0.301&  0.311& -0.542&  0.689&  0.630&  0.600&  0.505&  0.583&  0.632&  0.439&  0.583&  0.637&  0.620\\
G4300      &  0.885&  0.784& 0.699&  0.630&  1.000&  0.605&  0.664&  0.390& -0.868&  0.877&  0.873&  0.855&  0.787&  0.696&  0.659&  0.654&  0.662&  0.576&  0.806\\
Ca4455     &  0.684&  0.537& 0.566&  0.387&  0.605&  1.000&  0.569&  0.238& -0.419&  0.755&  0.676&  0.652&  0.667&  0.615&  0.559&  0.279&  0.461&  0.422&  0.630\\
Fe4531     &  0.843&  0.375& 0.233&  0.301&  0.664&  0.569&  1.000&  0.029& -0.679&  0.679&  0.706&  0.863&  0.873&  0.814&  0.493&  0.615&  0.598&  0.230&  0.887\\
C24668     &  0.331&  0.632& 0.647&  0.311&  0.390&  0.238&  0.029&  1.000& -0.328&  0.265&  0.365&  0.279&  0.083&  0.056&  0.431&  0.191&  0.191&  0.453&  0.147\\
H$\beta$   & -0.875& -0.676&-0.554& -0.542& -0.868& -0.419& -0.679& -0.328&  1.000& -0.792& -0.831& -0.873& -0.728& -0.699& -0.659& -0.809& -0.811& -0.495& -0.792\\
Fe5015     &  0.902&  0.777& 0.694&  0.689&  0.877&  0.755&  0.679&  0.265& -0.792&  1.000&  0.853&  0.887&  0.838&  0.799&  0.775&  0.507&  0.735&  0.657&  0.885\\
Mg1        &  0.912&  0.703& 0.618&  0.630&  0.873&  0.676&  0.706&  0.365& -0.831&  0.853&  1.000&  0.885&  0.745&  0.721&  0.713&  0.681&  0.752&  0.505&  0.787\\
Mg2        &  0.958&  0.672& 0.556&  0.600&  0.855&  0.652&  0.863&  0.279& -0.873&  0.887&  0.885&  1.000&  0.922&  0.848&  0.672&  0.647&  0.792&  0.510&  0.946\\
Mgb        &  0.858&  0.529& 0.426&  0.505&  0.787&  0.667&  0.873&  0.083& -0.728&  0.838&  0.745&  0.922&  1.000&  0.814&  0.596&  0.569&  0.689&  0.444&  0.951\\
Fe5270     &  0.860&  0.527& 0.424&  0.583&  0.696&  0.615&  0.814&  0.056& -0.699&  0.799&  0.721&  0.848&  0.814&  1.000&  0.502&  0.502&  0.806&  0.569&  0.892\\
Fe5335     &  0.703&  0.765& 0.681&  0.632&  0.659&  0.559&  0.493&  0.431& -0.659&  0.775&  0.713&  0.672&  0.596&  0.502&  1.000&  0.608&  0.593&  0.537&  0.708\\
Fe5709     &  0.669&  0.505& 0.392&  0.439&  0.654&  0.279&  0.615&  0.191& -0.809&  0.507&  0.681&  0.647&  0.569&  0.502&  0.608&  1.000&  0.689&  0.208&  0.623\\
NaD        &  0.772&  0.627& 0.537&  0.583&  0.662&  0.461&  0.598&  0.191& -0.811&  0.735&  0.752&  0.792&  0.689&  0.806&  0.593&  0.689&  1.000&  0.480&  0.762\\
TiO1       &  0.613&  0.618& 0.613&  0.637&  0.576&  0.422&  0.230&  0.453& -0.495&  0.657&  0.505&  0.510&  0.444&  0.569&  0.537&  0.208&  0.480&  1.000&  0.527\\
MgFe       &  0.922&  0.652& 0.532&  0.620&  0.806&  0.630&  0.887&  0.147& -0.792&  0.885&  0.787&  0.946&  0.951&  0.892&  0.708&  0.623&  0.762&  0.527&  1.000\\
         \hline\hline
         \noalign{\smallskip}
\end{tabular}
}
}											    
\end{table} 										    
\end{small}
\end{appendix}
\end{document}